\documentclass[twocolumn]{aastex631}
\usepackage{amssymb,amsmath,amsopn}
\usepackage{bm}
\usepackage{dcolumn}
\usepackage{enumitem}
\usepackage{mathtools}
\usepackage{hyperref}
\usepackage{soul}
\usepackage{color}
\usepackage{comment}
\usepackage{rotating}
\usepackage[utf8]{inputenc}
\usepackage{newunicodechar,graphicx} 

\DeclareRobustCommand{\okina}{%
	\raisebox{\dimexpr\fontcharht\font`A-\height}{%
		\scalebox{0.8}{`}%
	}%
}
\newunicodechar{ʻ}{\okina}

\newcommand{\vect}[1]{\boldsymbol{#1}}
\newcommand{\Matrix}[1]{\mathbb{#1}}
\newcommand{\code}[1]{\textsf{#1}}

\newcommand{\cfast}{c_{f}}
\newcommand{\alfven}{c_{a}}
\newcommand{\cslow}{c_{s}}
\newcommand{\csound}{a}

\newcommand{\lare}{\code{LaRe3D}}
\newcommand{\charcode}{\code{CHAR}}

\newcommand{\Linc}{\vect{L}_\ensuremath{\mathcal{I}}} 
\newcommand{\Lsiginc}{L_{\sigma,\ensuremath{\mathcal{I}}}} 
\newcommand{\Lout}{\vect{L}_\ensuremath{\mathcal{O}}}
\newcommand{\Lsigout}{L_{\sigma,\ensuremath{\mathcal{O}}}} 

\defcitealias{TarKee24}{Paper I}

\definecolor{dylan}{rgb}{0, 0.5, 1.0}

\definecolor{lucas}{rgb}{0.5, 0.0, 0.2}

\definecolor{markl}{rgb}{0.5, 0.5, 1.0}

\definecolor{JEL}{rgb}{1.0, 0.0, 1.0}

\newcommand{\change}[1]{#1}

\shorttitle{NRBC implementation in \lare{}}
\shortauthors{Kee et al.}

\graphicspath{{./}{figures/}}

\begin{document}

\title{Simulating the photospheric to coronal plasma using magnetohydrodyanamic characteristics II: reflections on  non-reflecting boundary conditions}
\correspondingauthor{N. Dylan Kee}
\email{dylan.kee@nasa.gov}
\author[0000-0001-7322-5401]{N. Dylan Kee}
\altaffiliation{Currently at NASA Goddard Space Flight Center, 8800 Greenbelt Rd., Greenbelt, MD 20771, USA}
\affil{National Solar Observatory, 22 Ohi\okina a Ku St, Makawao, HI 96768, USA}

\author[0000-0002-8259-8303]{Lucas A. Tarr}
\affiliation{National Solar Observatory, 22 Ohi\okina a Ku St, Makawao, HI 96768, USA}

\author[0000-0003-1522-4632]{Peter W. Schuck}
\affil{Goddard Space Flight Center, 8800 Greenbelt Rd, Greenbelt, MD 20771, USA
}

\author[0000-0002-4459-7510]{Mark G. Linton}
\affil{Naval Research Laboratory, 4555 Overlook Ave SW, Washington, DC 20375, USA
}

\author[0000-0002-6936-9995]{James E. Leake}
\affil{Goddard Space Flight Center, 8800 Greenbelt Rd, Greenbelt, MD 20771, USA
}



\begin{abstract}

We present our implementation of non-reflecting boundary conditions in the magnetohydrodynamics (MHD) code \lare{}.
This implementation couples a characteristics-based boundary condition with a Lagrangian remap code, demonstrating the generality and flexibility of such non-reflecting boundary conditions for use with arbitrary grid-based MHD schemes.
To test this implementation 
for perturbations on a background state, we present simulations of a hot sphere in an angled magnetic field.
We then examine a series of simulations where we advect a spheromak through a non-reflecting boundary condition at four speeds related to the fast and slow magnetosonic speeds and the Alfv\'en speed.
We compare the behavior of these simulations to ground truth simulations run from the same initial condition on an extended grid that keeps the spheromak in the simulation volume at all times.
We find that the non-reflecting boundary condition can lead to severe, physical differences developing between a simulation using a non-reflecting boundary and a ground truth simulation using a larger simulation volume.
We conclude by discussing the origins of these differences.

\end{abstract}




\section{Introduction} \label{sec:intro}



The formulation of general, open boundary conditions for numerical simulations when the region outside the simulation is not simply described by the region inside the simulation, and especially when we do not know how the region outside the simulation reacts to what is happening inside the simulation, is a highly non-trivial but universally important task. 
The role of a boundary condition is to approximate the impact of the external universe on the internal state of a simulation and vice versa.
When this approximation is poor, as it often is, the full internal state of the simulation can be corrupted by the poor approximation, thus rendering the simulation difficult to interpret at best, or numerically unstable or flatly wrong at worst.
Unfortunately, this issue is often not adequately discussed in the literature.

Standard boundary conditions for ideal magneto-hydrodynamics (MHD) simulations, namely those that are implemented in terms of the MHD variables themselves, are especially prone to these types of numerical instabilities and errors.
An appropriately specified boundary condition specifies only the information which is entering from the external universe, and utilizes information from the simulation interior to set the remainder of the properties at the boundary.
As we highlight in this paper, mapping between the information entering the simulation through the boundary and the MHD variables is hardly one-to-one, such that standard MHD boundary conditions generally either severely over- or under-specify the problem.
Moreover, it is often the case that these standard boundary conditions simultaneously over-specify information for the time update of one MHD quantity while under-specifying the information needed for the time update of another.

As we discuss in \cite{TarKee24} (hereafter \citetalias{TarKee24}), there is a mathematically and physically rigorous solution to this separation of incoming and outgoing information for the ideal MHD equations.
Specifically, it is possible to formulate a boundary condition by casting the MHD equations in terms of characteristic derivatives and then setting these derivatives individually based on whether they represent incoming (entering the simulation volume from outside) or outgoing (exiting the simulation volume) information.
Such a solution is particularly beneficial for the data-driving scenario discussed in \citetalias{TarKee24}, as this is a case when the state of the system is well defined on a surface and the simulation must evolve in a way that is consistent with the observed evolution of that boundary.
In that case, the boundary is developed to get information into the system from an external universe whose state is partially or completely known.
Characteristics-based boundaries are also highly effective in the extreme opposite limit when the outside universe is assumed to be in a steady state or inert as regards the simulation domain such that the boundary conditions evolve only due to the impact of waves exiting the simulation domain without reflection.
In this case, the boundary is designed to get information out of the system.
In this paper we implement and test a characteristic based boundary condition in this latter, non-reflecting boundary condition (NRBC) scenario.

The idea of such non-reflecting boundary conditions is not new in the solution of hyperbolic systems of equations, nor in MHD in particular.
These were first implemented by \citet{Hed79} and have since been used by various other authors
\citep[e.g.,][among others]{GraLeo00,GraAul08,LanVel05,JiaFen11,GudCar11,LioDow13}.
In this paper we discuss our implementation of such boundary conditions into the Staggered Grid, Lagrangian–Eulerian Remap Code for 3-D MHD Simulations, \lare{} \citep{ArbLon01}.

In addition to discussing our implementation, in this paper we also want to remind the reader that the term ``non-reflecting boundary conditions'' can be deceptively broad, as various implementations carrying that name treat the incoming characteristic derivatives substantially differently.
These different implementations then imply different assumptions about the impact of the external universe on the simulation interior evolution.
As such, we dedicate much of the discussion in this paper to 
emphasizing that mismatches between the evolution of ground truth simulations and those using non-reflecting boundary conditions can be dominated by the implicit assumptions encoded in the implementation of a non-reflecting boundary. 
While careful consideration of the problem being solved can ameliorate these effects, as we discuss throughout the paper, some are unavoidable as we do not know the state of the external universe.
Ultimately, you can't always get what you want, and sometimes even trying can prove insufficient to get you what you need.

One of the central issues here is that the boundary condition that is commonly sought when using a non-reflecting boundary is one that unobtrusively allows waves, plasma, magnetic fields, etc. out of the simulation volume.
Ideally such a boundary condition would minimally alter anything leaving the simulation volume such that a simulation with non-reflecting boundaries resembles a larger simulation with the boundaries moved far away.
However, it is not clear how such a boundary condition should be implemented, or even what the appropriate physical quantities and properties are that the boundary condition must respect in order to achieve this goal.
This is because specifying such a boundary condition requires knowledge of the external universe, namely the region which we are not simulating.
Without knowledge of the state of the external universe, it is not possible to determine the reaction of the external universe to the action of the simulation.
In turn, if one cannot simulate the reaction of the external universe on the simulation volume it becomes impossible to determine which information is the ``right'' information to allow to leave the simulation volume.
Therefore, one often falls back on standard implementations of non-reflecting boundaries which only achieve this goal in very special circumstances.


In Section~\ref{sec:formulation} we review the formulation of a characteristics-based boundary condition as laid out in \citetalias{TarKee24}.
This section also presents the numerical specifications of our implementation in \lare{}.
We then continue in Section~\ref{sec:validation} by presenting a simple test case of a hot sphere in an angled magnetic field to demonstrate that our implementation of a characteristics-based non-reflecting boundary is operating as expected.
In Section~\ref{sec:advect_spheromak}, we introduce a more complex test case used throughout the remainder of the paper, namely the advection of a balanced spheromak.
We then analyze the evolution of simulations of this test case using two common implementations of non-reflecting boundary conditions, and compare these to ground truth simulations in the second portion of this section, with a key result being that simulations with non-reflecting boundaries do not accurately represent the impact of the external universe.
In Section~\ref{sec:discussion}, we break down why simulations including a non-reflecting boundary condition in general cannot recover the behavior of ground truth simulations which treat a larger volume.
Finally we summarize the overall conclusions of the paper in Section~\ref{sec:conclusions}.

\newpage
\section{NRBC boundary condition formulation}\label{sec:formulation}

\subsection{Summary of the mathematical formulation}\label{sec:formulation_math}


Before presenting the results of our implementation of non-reflecting boundary conditions (NRBCs) into \lare{}, we summarize some information from \citetalias{TarKee24}, \S 2-3 regarding a characteristics-based formulation of MHD which is necessary for the discussion in this paper.
For the full discussion, we refer the reader to \citetalias{TarKee24}.
To facilitate the reader moving back and forth between this section and the associated discussion in \citetalias{TarKee24}, the first several equations here have numbers P1.xx where xx is the equation number in \citetalias{TarKee24}.

To derive a characteristics-based formulation of MHD, we begin by casting the MHD equations
\begin{subequations}
  \label{eq:mhd}
  \begin{gather}
    \frac{\partial\rho}{\partial t} + \nabla\cdot(\rho\vect{v})=0 \tag{P1.1a}\\
    \frac{\partial\vect{v}}{\partial t} + (\vect{v}\cdot\nabla)\vect{v} + \frac{1}{\rho}\nabla P - \frac{1}{\mu_0\rho}(\nabla\times\vect{B})\times\vect{B} - \vect{g}=\vect{0} \tag{P1.1b}\label{eq:momentum}\\
    \frac{\partial\epsilon}{\partial t} + \vect{v}\cdot\nabla\epsilon + \frac{P}{\rho}\nabla\cdot\vect{v}=0 \tag{P1.1c}\\
    \label{eq:induction}
    \frac{\partial\vect{B}}{\partial t} +(\vect{v}\cdot\nabla)\vect{B} - (\vect{B}\cdot\nabla)\vect{v} + \vect{B}(\nabla\cdot\vect{v}) = \vect{0} \tag{P1.1d}\;,
    \intertext{closed using the ideal equation of state}
    P = (\gamma-1)\rho\epsilon \tag{P1.1e}\;,
  \end{gather}
\end{subequations}
in matrix form in terms of the primitive variable MHD state vector $\vect{U}^T~=~(\rho,\epsilon,v_x,v_y,v_z,B_x,B_y,B_z)$ as
\begin{equation} \tag{P1.2}
  \label{eqn:mhd-matrix}
  \partial_t \vect{U} + \Matrix{A}_{x}\cdot\partial_x\vect{U} + \Matrix{A}_{y}\cdot\partial_y\vect{U}+ \Matrix{A}_{z}\cdot\partial_z\vect{U} + \vect{D} = \vect{0}\;.
\end{equation}
Eqn.~\ref{eq:momentum} uses $\vect{J} = \nabla\times \vect{B}/\mu_0$.
\change{We use internal energy per unit mass ($\epsilon$) to match the MHD variables used by \lare{}, but note that $\epsilon$ could be replaced by another thermodynamic variable, for instance pressure, with minor changes to the following equations and without altering the overall discussion.}
The vector $\vect{D}$ contains inhomogeneous terms (e.g., gravity in Eqn.~\ref{eq:momentum}).
The coefficient matrices $\Matrix{A}$ can be diagonalized, although not in general simultaneously in all three directions ($\hat{x},\hat{y},\hat{z}$) \citep{RoeBal96}.
As we are focusing on a boundary condition, and can select the direction perpendicular to the boundary surface, we denote the boundary-normal direction as $\hat{z}$ and proceed by diagonalizing $\Matrix{A}_z = \Matrix{S}_z \Matrix{D}_z \Matrix{S}_z^{-1}$ with $\Matrix{S}_z$ and $\Matrix{S}_z^{-1}$ the right and left eigenmatrices of $\Matrix{A}_z$, respectively.
$\Matrix{D}_z$ is the diagonal matrix of the eigenvalues $\lambda$ of $\Matrix{A}_z$, which are the projected advection speed $v_z$ and this speed plus or minus the absolute value of the projected Alfv\'en speed $\alfven$, the magnetosonic slow speed $\cslow$, or the magnetosonic fast speed $\cfast$.
These speeds are defined as

\begin{subequations}
    \begin{gather}
    \alfven = \lvert b_z\rvert \label{eq:ca}, \tag{P1.7i}\\
    \cfast^2 = \frac{1}{2}(a^2+b^2) + \frac{1}{2}\sqrt{(a^2+b^2)^2 - 4a^2b_z^2}\label{eq:cf}, \tag{P1.7j}\\
    \cslow^2 = \frac{1}{2}(a^2+b^2) - \frac{1}{2}\sqrt{(a^2+b^2)^2 - 4a^2b_z^2}\label{eq:cs}, \tag{P1.7k}\\
    a^2 = \gamma(\gamma-1)\epsilon, \tag{P1.7l}\\
    b^2 = \sum b_n^2,\text{ for } b_n = B_n/\sqrt{\rho},\ n\in(x,y,z)\;, \tag{P1.7m}
    \end{gather}
\end{subequations}
which further introduces the sound speed $a$ and the un-projected Alfv\'en speed $b$.
The resulting seven fundamental speeds are crucial for the problem of characteristics-based MHD as they denote the speeds of information propagation in MHD.
The information which is transported in the $z$-direction at each of these speeds is given by the $z$-direction characteristic derivative vector

\begin{equation} \tag{P1.9c}
    \vect{L}_z = \Matrix{D}_z \Matrix{S}^{-1}_z \partial_z \vect{U}\;.
\end{equation}
The full set of components $L_\sigma$ of $\vect{L}_z$, hereafter referred to as the boundary-normal characteristic derivatives, are explicitly given by 
\begin{subequations}
  \label{eqn:li}
  \begin{align*}
    L_1 & = v_z \biggl[B_z^\prime \biggr]\tag{P1.14a} \label{eqn:li1}\\
    L_2 & = v_z\biggl[ \frac{1-\gamma}{\gamma \rho}\rho^\prime +\frac{1}{\gamma\epsilon}\epsilon^\prime \biggr] \tag{P1.14b} \label{eqn:li2}\\
    L_3 & = \frac{v_z-c_a}{2}\biggl[-\beta_yv_x^\prime + \beta_x v_y^\prime \\
    &\hspace{0.21\linewidth}-\beta_y\sqrt{\rho}s_zB_x^\prime +\beta_x\sqrt{\rho}s_zB_y^\prime \biggr] \tag{P1.14c} \label{eqn:li3}\\
    L_4 & = \frac{v_z+c_a}{2}\biggl[\beta_yv_x^\prime - \beta_x v_y^\prime \\
    &\hspace{0.21\linewidth}- \beta_y\sqrt{\rho}s_zB_x^\prime +\beta_x\sqrt{\rho}s_zB_y^\prime \biggr] \tag{P1.14d} \label{eqn:li4}\\
    L_5 & = \frac{v_z-\cslow }{2}\biggl[\frac{\alpha_s}{\gamma\rho}\rho^\prime +\frac{\alpha_s}{\gamma\epsilon}\epsilon^\prime \\
    &\hspace{0.21\linewidth}- \frac{\beta_x\alpha_f \cfast}{\csound^2}s_zv_x^\prime - \frac{\beta_y\alpha_f \cfast}{\csound^2}s_zv_y^\prime - \frac{\alpha_s \cslow}{\csound^2}v_z^\prime \\
    &\hspace{0.21\linewidth}-\frac{\beta_x\alpha_f}{\csound\sqrt{\rho}}B_x^\prime - \frac{\beta_y\alpha_f}{\csound\sqrt{\rho}}B_y^\prime \biggr] \tag{P1.14e} \label{eqn:li5}\\
    L_6 & = \frac{v_z+\cslow}{2}\biggl[\frac{\alpha_s}{\gamma\rho}\rho^\prime +\frac{\alpha_s}{\gamma\epsilon}\epsilon^\prime \\
    &\hspace{0.21\linewidth}+ \frac{\beta_x\alpha_f \cfast}{\csound^2}s_zv_x^\prime + \frac{\beta_y\alpha_f \cfast}{\csound^2}s_zv_y^\prime + \frac{\alpha_s \cslow}{\csound^2}v_z^\prime \\
    &\hspace{0.21\linewidth}-\frac{\beta_x\alpha_f}{\csound\sqrt{\rho}}B_x^\prime - \frac{\beta_y\alpha_f}{\csound\sqrt{\rho}}B_y^\prime \biggr] \tag{P1.14f} \label{eqn:li6}\\
    L_7 & = \frac{v_z-\cfast }{2}\biggl[\frac{\alpha_f}{\gamma\rho}\rho^\prime +\frac{\alpha_f}{\gamma\epsilon}\epsilon^\prime \\
    &\hspace{0.21\linewidth}+ \frac{\beta_x\alpha_s \cslow}{\csound^2}s_zv_x^\prime + \frac{\beta_y\alpha_s \cslow}{\csound^2}s_zv_y^\prime - \frac{\alpha_f \cfast}{\csound^2}v_z^\prime \\
    &\hspace{0.21\linewidth}+\frac{\beta_x\alpha_s}{\csound\sqrt{\rho}}B_x^\prime + \frac{\beta_y\alpha_s}{\csound\sqrt{\rho}}B_y^\prime \biggr] \tag{P1.14g} \label{eqn:li7}\\
    L_8 & = \frac{v_z+\cfast}{2}\biggl[\frac{\alpha_f}{\gamma\rho}\rho^\prime +\frac{\alpha_f}{\gamma\epsilon}\epsilon^\prime \\
    &\hspace{0.21\linewidth}- \frac{\beta_x\alpha_s \cslow}{\csound^2}s_zv_x^\prime - \frac{\beta_y\alpha_s \cslow}{\csound^2}s_zv_y^\prime + \frac{\alpha_f \cfast}{\csound^2}v_z^\prime \\
    &\hspace{0.21\linewidth}+\frac{\beta_x\alpha_s}{\csound\sqrt{\rho}}B_x^\prime + \frac{\beta_y\alpha_s}{\csound\sqrt{\rho}}B_y^\prime \biggr] \tag{P1.14h} \label{eqn:li8}\;.
  \end{align*}
\end{subequations} 
We also introduce a set of convenient auxiliary variables to keep the notation more compact, namely $\beta_x \equiv B_x/\sqrt{B_x^2 + B_y^2}$, $\beta_y \equiv B_y/\sqrt{B_x^2 + B_y^2}$, $s_z \equiv {\rm sign}(B_z)$, $\alpha_f^2 \equiv (a^2 - c_s^2)/(c_f^2 - c_s^2)$, and $\alpha_s^2 \equiv (c_f^2 - a^2)/(c_f^2 - c_s^2)$.
In terms of these definitions, we can now write the MHD equations in the desired form 

\begin{equation} \tag{P1.12}
    \label{eqn:mhd-char-derivs}
    \partial_t \vect{U} + \Matrix{S}_z\cdot \vect{L}_z + \vect{C} = \vect{0},
\end{equation}
with $\vect{C} = \Matrix{A}_x \partial_x \vect{U} + \Matrix{A}_y \partial_y \vect{U} + \vect{D}$ consolidating all inhomogeneous terms, such as gravity, and those terms transverse to the boundary condition.
An equivalent, expanded version of this equation which is also useful to the discussion here is 

\begin{subequations}
  \label{eq:mhd-chars}
  \begin{align}
    \partial_t \rho &-\rho [L_2] + \alpha_s\rho[L_5+L_6] & \notag \\
    &+ \alpha_f\rho[L_7+L_8]+  C_\rho = 0\label{eq:rho_char} \tag{P1.18a} \\
    \partial_t \epsilon &+ \epsilon[L_2] +\alpha_s\frac{a^2}{\gamma}[L_5+L_6] \notag\\
    &+\alpha_f\frac{a^2}{\gamma}[L_7+L_8] + C_\epsilon = 0\label{eq:eps_char} \tag{P1.18b} \\
    \partial_t v_x &- \beta_y[L_3-L_4]-\alpha_f\beta_xc_fs_z[L_5-L_6] \notag\\
    &+ \alpha_s\beta_xc_ss_z[L_7-L_8] + C_{v_x} = 0\label{eq:vx_char} \tag{P1.18c} \\
    \partial_t v_y &+ \beta_x[L_3-L_4] - \alpha_f\beta_yc_fs_z[L_5-L_6] \notag\\
    &+\alpha_s\beta_yc_ss_z[L_7-L_8]+ C_{v_y} = 0\label{eq:vy_char} \tag{P1.18d} \\
    \partial_t v_z &- \alpha_sc_s[L_5-L_6] \notag\\
    &- \alpha_fc_f[L_7-L_8]+ C_{v_z} = 0\label{eq:vz_char} \tag{P1.18e}\\
    \partial_t B_x &-\beta_y\sqrt{\rho}s_z[L_3+L_4] - \alpha_f\beta_x\sqrt{\rho} a[L_5+L_6] \notag\\
    &+ \alpha_s\beta_x\sqrt{\rho}a[L_7+L_8] + C_{B_x} = 0\label{eq:Bx_char} \tag{P1.18f} \\
    \partial_t B_y &+\beta_x\sqrt{\rho}s_z[L_3+L_4] - \alpha_f\beta_y\sqrt{\rho}a[L_5+L_6] \notag\\
    &+\alpha_s\beta_y\sqrt{\rho}a[L_7+L_8]+ C_{B_y} = 0\label{eq:By_char} \tag{P1.18g} \\
    \partial_t B_z &+[L_1] + C_{B_z} = 0\label{eq:Bz_char}\;. \tag{P1.18h}
  \end{align}
\end{subequations}
The $C_\zeta$ terms are the elements of the vector $\vect{C}$ appearing in each of the eight MHD equations.
The mathematical expressions for these combined inhomogeneous and transverse (to $\hat{z}$) terms are given in \citetalias{TarKee24}, Appendix D.

\begin{deluxetable*}{ccl||ccc}
  \tablecaption{Characteristic Derivatives at a $z_{min}$ Boundary\label{tab:incoming}}
  \tablehead{\colhead{Velocity} & \colhead{Incoming $L_\sigma$ ($\Lsiginc$)} & \colhead{Dependent Derivatives} & \colhead{Outgoing $L_\sigma$ ($\Lsigout$)} & \colhead{P} & \colhead{Q}}
  \startdata
  $v_z\leq-c_f$           &  -  & - & $L_7,L_3,L_5,L_1,L_2,L_6,L_4,L_8$ & 0 & 0\\
  $-c_f\leq v_z \leq -c_a$& $L_8$ & $\rho^\prime,\epsilon^\prime, v_x^\prime, v_y^\prime,v_z^\prime,B_x^\prime,B_y^\prime$ & $L_7,L_3,L_5,L_1,L_2,L_6,L_4$ & 1 & 7\\
  $-c_a\leq v_z \leq -c_s$& $L_4,L_8$ & $\rho^\prime,\epsilon^\prime, v_x^\prime, v_y^\prime,v_z^\prime,B_x^\prime,B_y^\prime$ & $L_7,L_3,L_5,L_1,L_2,L_6$ & 2 & 7 \\
  $-c_s\leq v_z \leq 0$   & $L_6,L_4,L_8$ & $\rho^\prime,\epsilon^\prime, v_x^\prime, v_y^\prime,v_z^\prime,B_x^\prime,B_y^\prime$ & $L_7,L_3,L_5,L_1,L_2$ & 3 & 7 \\
  \hline
  $0\leq v_z \leq c_s $   & $L_1,L_2,L_6,L_4,L_8$ & $\rho^\prime,\epsilon^\prime, v_x^\prime, v_y^\prime,v_z^\prime,B_x^\prime,B_y^\prime,B_z^\prime$ & $L_7,L_3,L_5$ & 5 & 8 \\
  $c_s\leq v_z \leq c_a$  & $L_5,L_1,L_2,L_6,L_4,L_8$ & $\rho^\prime,\epsilon^\prime, v_x^\prime, v_y^\prime,v_z^\prime,B_x^\prime,B_y^\prime,B_z^\prime$ & $L_7,L_3$ & 6 & 8 \\
  $c_a\leq v_z \leq c_f$  & $L_3,L_5,L_1,L_2,L_6,L_4,L_8$ & $\rho^\prime,\epsilon^\prime, v_x^\prime, v_y^\prime,v_z^\prime,B_x^\prime,B_y^\prime,B_z^\prime$ & $L_7$ & 7 & 8\\
  $c_f\leq v_z$        & $L_7,L_3,L_5,L_1,L_2,L_6,L_4,L_8$ & $\rho^\prime,\epsilon^\prime, v_x^\prime, v_y^\prime,v_z^\prime,B_x^\prime,B_y^\prime,B_z^\prime$ & - & 8 & 8 \\
  \enddata
  \tablecomments{Boundary-normal characteristic derivatives (second and fourth columns) ordered by boundary--normal velocity (first column) for the bottom, $z_{min}$ boundary of some simulation.
  The horizontal line demarcates $v_z\leq0$ (above the line) and $v_z\geq 0$ (below). When the bulk velocity equals one of the eigenvalues of the system, the corresponding mode is non-propagating, and has zero amplitude, so we include an equals sign on both sides of all divisions.
  For reference, the fourth column shows the outward characteristics, which is simply the reverse ordering of the inward characteristics, and $P$ and $Q$ refer to the number of inward characteristic derivatives and the number of MHD equations in which they appear, respectively. Table reproduced from \citetalias{TarKee24}.}
\end{deluxetable*}

With the MHD equations cast in this form, the task of formulating a characteristic-based boundary comes down to setting the vector of boundary-normal characteristic derivatives, $\vect{L}_z$, at the boundary.
Focusing on a single boundary, here at the minimum coordinate position in the $\hat{z}$ direction $z_{\rm min}$, a natural distinction can be made between $L_\sigma$ associated with eigenvalues implying an \emph{outward} directed information propagation (i.e.,~$\lambda_\sigma<0$ at $z = z_{\rm min}$) which we denote as $\Lout$ when referring to the full set and $\Lsigout$ when referring to individual characteristic derivatives, and $L_\sigma$ associated with eigenvalues implying an \emph{inward} directed information propagation (i.e.,~$\lambda_\sigma>0$ at $z = z_{\rm min}$) which we denote as $\Linc$  when referring to the full set and $\Lsiginc$ when referring to individual characteristic derivatives.
All information necessary to compute outgoing characteristic derivatives using Eqn.~P1.14 (as well as $\vect{C}$) is available in the simulation volume, and therefore all $\Lsigout$ are set using their corresponding subsets of Eqn.~P1.14.
The same is not true for the $\Linc$, which represent the action and response of the external universe, and therefore alternate definitions are required.
Table \ref{tab:incoming} summarizes which $L_\sigma$ are outgoing and which are incoming as a function of the value of $v_z$, thereby itemizing which characteristics, respectively, can be set using Eqn.~P1.14 and which must be set in other ways.


In formulating a non-reflecting boundary condition, the goal in setting the $\Lsiginc$ is to ensure that no outward propagating disturbance reflects off the boundary condition and back into the simulation domain.
If a reflection were to occur that would mean that information would be returned from the location where the reflection occurred, with that information representing the nature of the reflection.
In the language of characteristics, the amplitude of the characteristic derivatives is what carries this type of information, and the information is carried at the propagation speed of the associated eigenvalue, or characteristic speed, of the MHD system.
Therefore, if a reflection occurred, it would be represented by a change in the amplitude of at least one characteristic derivative carrying information into the simulation.
Based on this, imposing that none of the outgoing characteristic derivatives are reflected is equivalent to imposing that outgoing characteristic derivatives do not contribute to a change in time of the amplitude of incoming characteristic derivatives.

In one-dimensional systems (1D), where all characteristic derivatives are guaranteed to be either purely incoming or outgoing without transverse components (i.e., $\Matrix{A}_x \cdot \partial_x \vect{U} = \Matrix{A}_y \cdot \partial_y \vect{U} =0$), this can be imposed exactly by setting the set of $\Linc$ equal to their initial values. \change{These are possible to compute from Eqn.~P1.14 for the special case of analytically or numerically defined initial conditions that extend beyond the numerically simulated region; otherwise, $\vect{U}'$ immediately beyond the boundary condition is unknown.
As the simplest example, all $\Linc$ are zero for a locally homogeneous system because all boundary normal spatial derivatives are zero in this region.}
In two- and three-dimensional systems (2D and 3D), however, the presence of transverse derivatives means that information propagation is in general no longer perpendicular to the boundary, namely the transverse characteristic derivatives may have non-zero amplitude, nor are the eight different types of MHD eigenmodes represented by the characteristic derivatives guaranteed to be aligned with one another.
As a simple example, a magnetic field-aligned flow with a gradient in density along the field and a gradient in magnetic field strength perpendicular to the field gives a non-zero amplitude to the Alfv\'en modes perpendicular to the field and a non-zero amplitude to the entropy mode in the orthogonal direction along the field.
Therefore, the meaning of non-reflecting in multi-D is no longer so straightforward.
Due to this ambiguity, individual authors refer to different methods of setting the $\Lsiginc$ in multi-D as imposing a ``non-reflecting'' boundary condition.
As an example, we now compare the three different methods used by \citet{GraLeo00,GraAul08}, which we refer to as ``Fixed'' NRBCs, \citet{JiaFen11}, which we refer to as ``Cancellation'' NRBCS, and \citet{GudCar11}, which is a variation on the cancellation NRBCs.
Importantly, all of these methods are referred to as ``non-reflecting''.

\subsubsection{Fixed NRBCs}
One common interpretation of non-reflecting in multi-D, which is analogous to the 1D case, is to again set each $\Lsiginc$ to its initial function of space.
For any $L_\sigma$ which is initially incoming and switches to outgoing, its value is computed from the simulation while outgoing like all other $\Lsigout$.
If this $L_\sigma$ is later incoming again, its value is set back to $\Lsiginc(t=0)$.
For any $L_\sigma$ which is initially outgoing and is later switched to incoming, $\Lsiginc=0$ while it is incoming, as no information was initially entering the simulation via this mode.
As mentioned above, frequently $\Lsiginc(t=0)=0$ in which case all $\Lsiginc=0$ at all times independent of their initial status.
This is the case for all the simulations we perform later in this paper.

Conceptually, this imposes the condition that no information in the simulation is allowed to alter $\Linc$.
As such, the boundary-perpendicular, incoming characteristic derivatives are independent of either the boundary-perpendicular, outgoing characteristic derivatives or the transverse characteristic derivatives.
Consulting Eqn.~\ref{eqn:mhd-char-derivs}, this implies that the evolution in time of $\vect{U}$ at the boundary is driven by $\Lout$, the transverse characteristic derivatives, and inhomogeneous terms in $\vect{C}$, potentially with an additional contribution from whatever function each $\Lsiginc$ had initially.
This is, for instance, the method chosen by \citet{GraLeo00,GraAul08}.
We will explore this case in detail below, and refer to it as Fixed NRBCs.

\subsubsection{Cancellation NRBCs}
Another common, alternative interpretation of how to generalize a non-reflecting boundary condition from 1D to multi-D arises from examining Eqn.~\ref{eqn:mhd-char-derivs}.
Solving this for $\vect{L}_z$ yields
\setcounter{equation}{0}
\begin{equation}
    \vect{L}_z = -\Matrix{S}^{-1} \partial_t \vect{U} -\Matrix{S}^{-1}\vect{C}\;.
\end{equation}
In a 1D setup where the plasma near the boundary is initially spatially uniform, and in the absence of inhomogeneous terms (i.e., $\vect{C} = \vect{0}$), this reduces to 
\begin{equation}
    L_\sigma = -\sum_\zeta S_{\sigma,\zeta}^{-1}\partial_t U_\zeta\;.
\end{equation}
In this limit, $\Linc = 0$ because the plasma is spatially uniform and an NRBC is defined such that outgoing characteristics can not change the value of the incoming characteristics.
This then means $\sum_\zeta S_{\sigma,\zeta}^{-1}\partial_t U_\zeta = 0$ if $L_\sigma$ is incoming.
This property is then applied to the multi-D case, such that each $\Lsiginc$ must be set to exactly cancel its associated transverse and inhomogeneous terms, namely $\Lsiginc = -\sum_\zeta S_{\sigma,\zeta}^{-1} C_\zeta$.
This imposes that $\Linc$ can not be changed by $\Lout$, but can be changed by other information at the boundary in the simulation volume through the transverse and inhomogeneous terms $C_\zeta$.
Additionally, this method implies that time derivatives of the primitive variables approach zero at the boundary with increasing numbers of incoming characteristic derivatives because the incoming characteristics cancel out increasing portions of the time update imposed by transverse terms.
``Fixed'' NRBCs have the same behavior with increasing numbers of incoming characteristics for an initially spatially uniform plasma because the transverse terms are zero and all $\Lsiginc=0$.
This second method is used, for instance, by \citet{JiaFen11} and underpins the \code{Bifrost} boundary conditions\footnote{\code{Bifrost} boundary conditions actually satisfy $\Lsiginc = -\sum_\zeta S_{\sigma,\zeta}^{-1} C_\zeta\biggr|_{(\vect{v}=\vect{0})}$.} described by \citet{GudCar11}.
We will also explore this case in detail below, and refer to it as Cancellation NRBCs.

In order to discuss the differences introduced by these two methods, we present simulations using both prescriptions for $\Linc$ in the following sections.
Before moving on, however, it is important to highlight a mismatch between what non-reflecting boundary conditions do and how they are commonly deployed.
A non-reflecting boundary condition removes the impact of $\Lout$ on $\Linc$.
Meanwhile, non-reflecting boundary conditions are often utilized as though they will mimic the results of a larger simulation by minimizing the action of the boundary condition on the simulation interior.
As we discuss in more depth in Section \ref{sec:discussion}, this is explicitly \emph{not} what a non-reflecting boundary condition does.
To illuminate why this is, consider the simple case of a zero-velocity, force-free plasma configuration, namely a configuration in which all the terms in the three momentum equations sum to give $\partial_t \vect{v} = \vect{0}$.
Consulting Equations \ref{eq:vx_char}, \ref{eq:vy_char}, and \ref{eq:vz_char}, we see that we can conceptualize a force-free configuration in terms of the characteristic derivatives as a superposition of standing waves at every location imposed by oppositely directed Alfv\'en (via $L_3$ and $L_4$) and magnetosonic slow (via $L_5$ and $L_6$) and fast (via $L_7$ and $L_8$) modes.
This superposition must cancel out the contribution of $\vect{C}$ in some non-trivial way to enforce $\partial_t \vect{v} = \vect{0}$.
Therefore, one might imagine that altering any one $L$ would upset the delicate balance.
As we discuss in Section \ref{sec:discussion}, this intuition is correct.
If we were to place a simple non-reflecting boundary condition at some point in this force-free configuration, then the Fixed method, where all $\Lsiginc$ are set to their initial value, will preserve the force-free nature of the plasma configuration because every $L$ has its initial well balanced value at every time. 
Meanwhile, it is not clear that the Cancellation method, where $\Lsiginc = -\sum_\zeta S_{\sigma,\zeta}^{-1} C_\zeta$ rather than its well balanced value, can maintain a force-free balance outside of trivial scenarios where all the characteristic derivatives are zero initially.
Further, as soon as even spatially constant advection is introduced, as we do in Section \ref{sec:advect_spheromak}, the two methods differ from one another and from a larger simulation.
To understand why this is, it is important to recognize that none of the characteristic derivatives are likely to be uniform in space for any arbitrary plasma configuration. 
Advection moves this non-uniform distribution through the boundary condition, such that imposing force-balance at the boundary requires a time varying description of $\Linc$ to perform the delicate cancellation required by Equations \ref{eq:vx_char}, \ref{eq:vy_char}, and \ref{eq:vz_char}.
The Fixed NRBC is explicitly not varying $\Linc$ in time, so it can not correctly enforce force balance in this simple advection case. 
The Cancellation NRBC does vary $\Linc$ in time, but this time varying description of $\Linc$ depends only on the transverse terms, while enforcing force balance depends on $\vect{C}$, $\Linc$, and $\Lout$.
This conceptual example illustrates that, while it is easy to remove information from the simulation volume, it is very complex to \emph{correctly} remove information from the simulation volume.

\subsection{Numerical details of the implementation}\label{sec:formulation_numerics}

From a numerical perspective, our implementation of the conditions laid out in the prior subsection is as a new boundary condition for the Lagrangian remap MHD code \lare{} \citep{ArbLon01}.
Because \lare{} is not a characteristics-based code, and our implementation must also be able to solve the data-driving problem discussed in \citetalias{TarKee24}, this requires writing a second MHD code based on the characteristic formulation, referred to hereafter as \charcode{}.
This new MHD code is only active in a thin grid encapsulating the ghost cells of \lare{}.
For the simulations in this paper, we focus on a boundary in the $\hat{z}$-normal direction, so this grid spans the full simulation in the $x$- and $y$-directions, but is thin in the $z$-direction.
We then exchange information at each Courant timestep back and forth between the main MHD code \lare{} and the characteristics-based boundary condition MHD code \charcode{} to evolve both \charcode{} and the \lare{} ghost cells.
Specifically, information on all primitive variables is transferred from \lare{} to \charcode{} at positions corresponding to the first two layers of active cells of \lare{}, \charcode{} is run, and information on all primitive variables is then transferred back from \charcode{} to \lare{} at positions corresponding to all the ghost cells of \lare{}.
Using this very general approach makes \charcode{}, and the characteristics-based boundary conditions it provides, easily implementable into any MHD code independent of the underlying grid structure or numerical methods the base code may utilize in its own solution of MHD.
This process is also described in detail in \S6 of \citetalias{TarKee24}.

Practically speaking, this transfer of information must itself be done carefully.
In order to facilitate its Lagrangian remap \lare{} stores its variables in several staggered locations in the grid cell, namely $\rho$ and $\epsilon$ are stored at cell centers, $B_n$ are each stored at the right cell face perpendicular to the $\hat{n}$ direction (e.g., $B_x$ is stored on the $\hat{x}$ face with the maximum $x$ value for the cell), and $v_n$ are all stored at the cell vertex where the right faces of the cell in all three directions meet.
Meanwhile, the characteristics-based code \change{performs the time evolution of} all its variables at a single location in each cell, which we chose to be cospatial with the \lare{} cell center where densities and energies are stored.
\change{To perform this time evolution, spatial derivatives for the characteristics interior to the \charcode{} grid are therefore computed between neighboring cell centers, while $\Matrix{S}$, $\Matrix{D}_z$, and $\Matrix{S}^{-1}$ are computed at all $\hat{z}$ cell faces using $\vect{U}$ averaged from the neighboring cell centers.\footnote{\change{The corresponding matrices for the $\hat{x}$ and $\hat{y}$ direction characteristics are accordingly computed at the $\hat{x}$ and $\hat{y}$ cell faces, respectively.}}
For the boundary condition it is no longer possible to compute a spatial average to obtain a face-centered $\vect{U}$, as $\vect{U}$ is not well defined beyond this surface, so $\Matrix{S}$, $\Matrix{D}_z$, and $\Matrix{S}^{-1}$ are approximated by their values at the nearest cell center.
See the end of Section 5 of \citetalias{TarKee24} for further discussion of this approximation.}

The mapping between the two codes is then done in the \lare{}$\rightarrow$\charcode{} direction by averaging \lare{} B-fields from the two cell faces on either side of a cell center and averaging \lare{} velocities from the eight cell vertices surrounding each cell center.
In the \charcode{}$\rightarrow$\lare{} direction we take control volume averages of the \charcode{} cell center variables immediately surrounding each location where a \lare{} variable is required.
The only exception to this is $B_z$; to preserve $\nabla\cdot\vec{B}=0$ to machine precision on the staggered grid that \lare{} uses, $\nabla\cdot\vec{B}=0$ is solved for $B_z$ at each $\hat{z}$-normal cell face in the boundaries using the \lare{} numerical derivative stencil for magnetic fields.
Note that this \lare{} specific choice does not limit the applicability of \charcode{} to arbitrary other MHD codes, as this is part of the remapping step which already needs to be tailored to the specific grid of the base MHD code, and does not replace the \charcode{} internal computations for $B_z$.

Beyond the choices of where to store the characteristic code variables and perform their time update and how to map back and forth between \charcode{} and \lare{}, there is also the issue of where on the grid to locate derivatives and perform the diagonalization of the $\Matrix{A}$ matrices.
We choose to follow an upwind Finite Volume method, namely we perform the diagonalization and locate the spatial derivatives for each $L_\sigma$ on cell faces upwind of the cell center in the direction of its associated eigenvalue.
While such a method introduces higher order diffusion terms, this is an unavoidable consequence of an upwind scheme \citep{HarLax83}.
Moreover, this method conserves mass, momentum, energy, and magnetic flux, which is not guaranteed for other combinations of diagonalization and derivative locations.


\section{Validation}\label{sec:validation}

For all following simulations, we work in dimensionless units $\vect{U} = \vect{U}^\ast/\vect{U}_N$, where $\vect{U}^\ast$ are the dimensional units, and $\vect{U}_N$ the normalizations.
We proceed with the normalizations chosen in length $L_N = 1$ m, density $\rho_N = 1$ kg m$^{-3}$, and magnetic field $B_N = \sqrt{4 \pi 10^{-7}}$ T. 
Note that the normalization of magnetic field here absorbs a factor of the permeability of free space, $\mu_0$.
The normalization of energy density and velocity can be derived, respectively, as $\epsilon_N = B_N^2/\mu_0\rho_N$ and $v_{N} = B_N^2/\mu_0\rho_N$.
Finally time is normalized in units of $t_N = L_N/v_N$.

First, as a test of the non-reflecting boundary implementation, we consider a hot sphere in an angled magnetic field with no inhomogeneous terms (i.e., $\vect{D}=\vect{0}$).
Specifically, for this test we initialize a uniform volume \change{of plasma with} $\rho = 1$, $\epsilon = 1$, $v_x=v_y=v_z = 0$ and $B_x = B_y = B_z = 1$.
Into this, we insert a sphere with a 10\% perturbation in internal energy such that a sub-volume of the plasma is over-pressurized compared to the background and will expand.
Due to the presence of the magnetic field and the choice of a $\beta \equiv P_{gas}/P_{mag} = 2 \gamma \rho \epsilon/|\vect{B}|^2 \sim 1$ plasma\footnote{Plasma $\beta$ as the ratio of gas and magnetic pressures is not to be confused with the auxiliary variables $\beta_x$ and $\beta_y$.}, this expansion will be channeled along the magnetic field, resulting in a situation where the number of $\Lsiginc$ is neither constant in space nor time along the non-reflecting boundary as each of the perturbations initialized by this expansion passes through it.
All simulations are run with periodic side ($x$ and $y$) boundaries to allow perturbations to wrap around and remain inside the simulation volume.
The top boundary in all three cases is a simplified ``open'' boundary condition built into \lare{}, while the bottom boundary is either a version of our newly implemented NRBC (applied at $z=0$) or a more distant ($z<0$) \lare{} ``open'' boundary for the Ground Truth simulation.
\change{The use of periodic side boundaries} means perturbations propagating into the upper portion of all the simulations, as well as perturbations propagating into the lower portion of the ground truth simulation where it extends beyond the location of the non-reflecting boundary in the other two simulations, will interact with one another and generate contaminating diffracted waves which continue to bounce around the simulation until they leak out the top or bottom of the box through the NRBC or ``open'' boundary. 
As we discuss in more depth later, these result in unavoidably different interference patterns in the various simulations we compare due to the different treatments of the boundary or simulation at $z=0$.

As for the numerical specifications, we run this test in a cube of $N_x = N_y = N_z = 128$ grid cells and dimensionless length $\ell_x = \ell_y = \ell_z = 1$ in each direction.
The hot sphere is centered $z = 0.3$ units above the non-reflecting boundary in the middle of the grid in the $x$- and $y$-directions, and has radius $0.1\sqrt{2}$.

The test is run for two scenarios: using non-reflecting boundaries with $\Lsiginc=0$ (Fixed NRBCs) and with $\Lsiginc=-\sum_\zeta S_{\sigma,\zeta}^{-1} C_\zeta$ (Cancellation NRBCs). 
As a comparison, we run a ground truth simulation which extends one spatial unit beyond the non-reflecting boundary location in the $z$-direction, and as such has $N_z=256$ in order to keep the same spatial resolution.

\begin{figure*}[!htbp]
	\centering

        \hspace{0.1125\textwidth}
        \includegraphics[width=0.225\textwidth, viewport=250 350 750 790, clip=true]{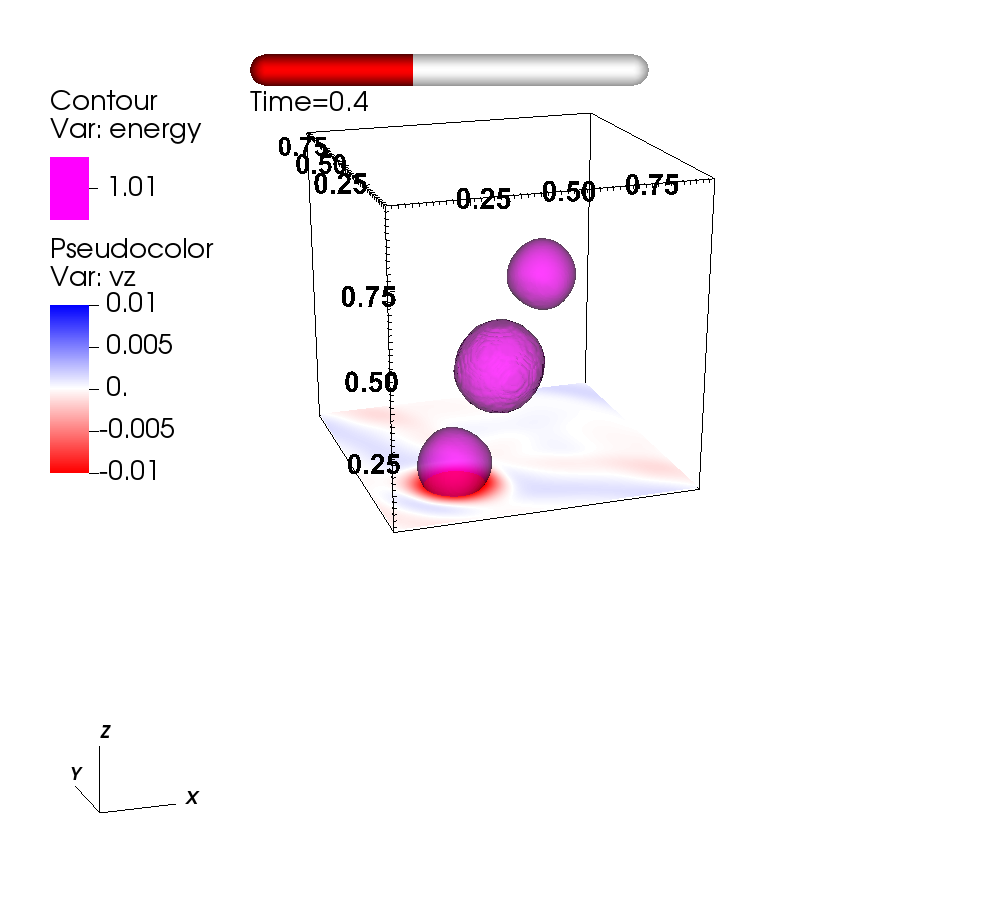}
        \includegraphics[width=0.225\textwidth, viewport=250 350 750 790, clip=true]{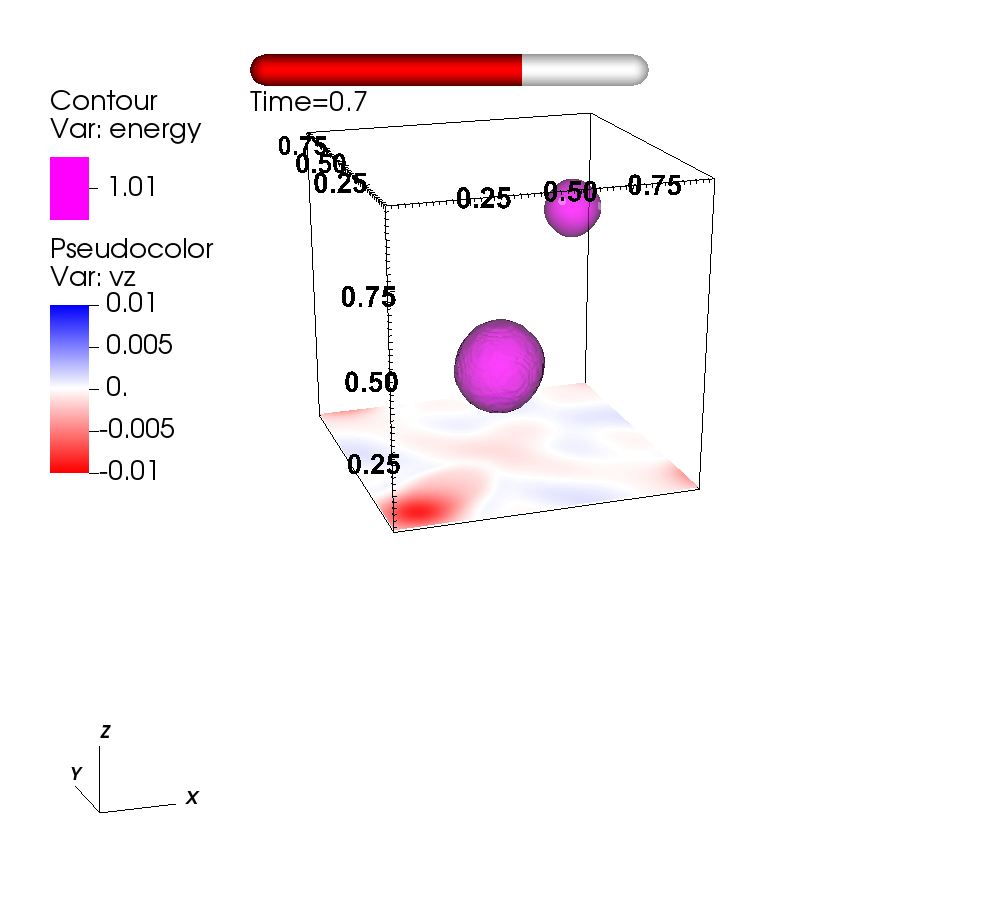}
        \includegraphics[width=0.225\textwidth, viewport=250 350 750 790, clip=true]{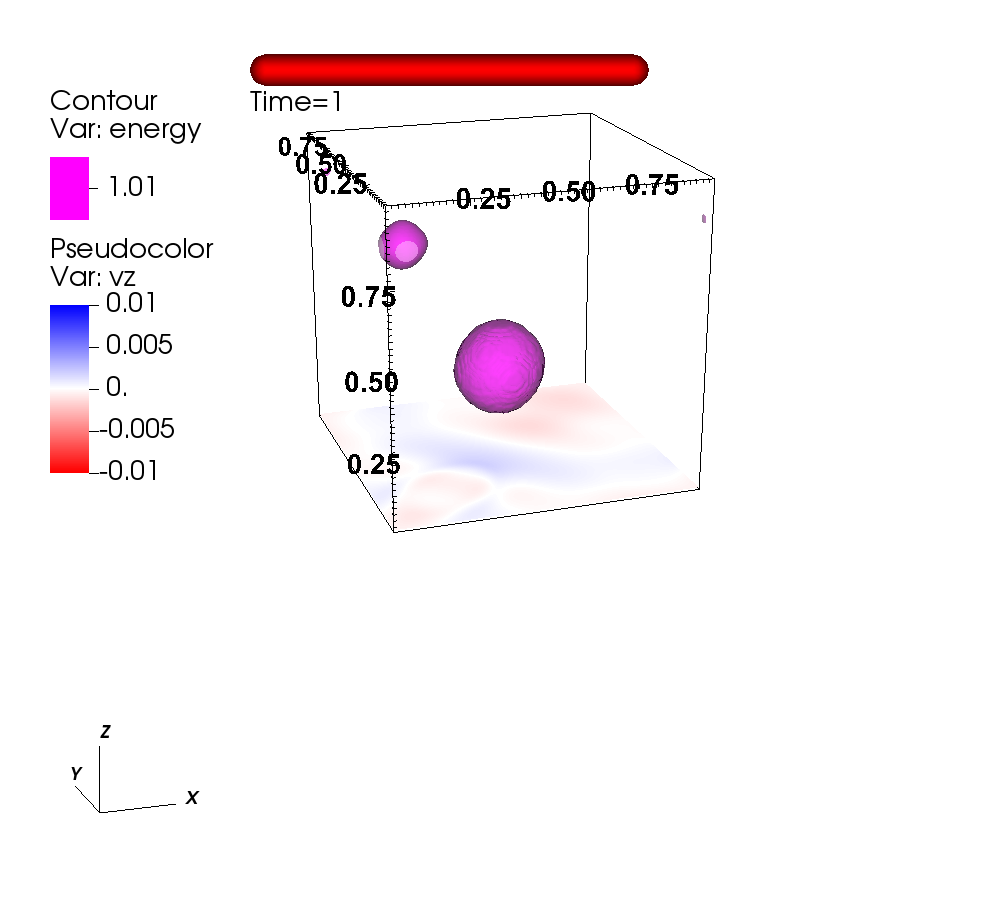}

        a) Fixed NRBC
        
        \includegraphics[width=0.3375\textwidth, viewport=0 75 750 903, clip=true]{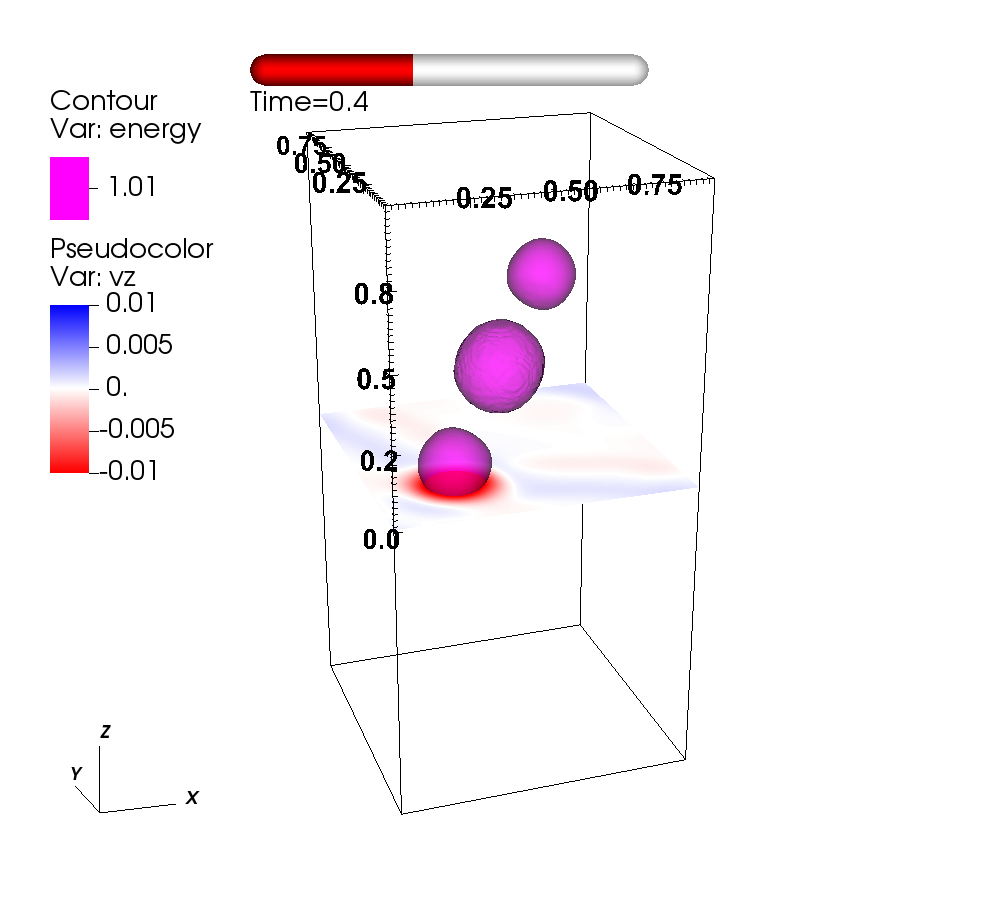}
        \includegraphics[width=0.225\textwidth, viewport=250 75 750 903, clip=true]{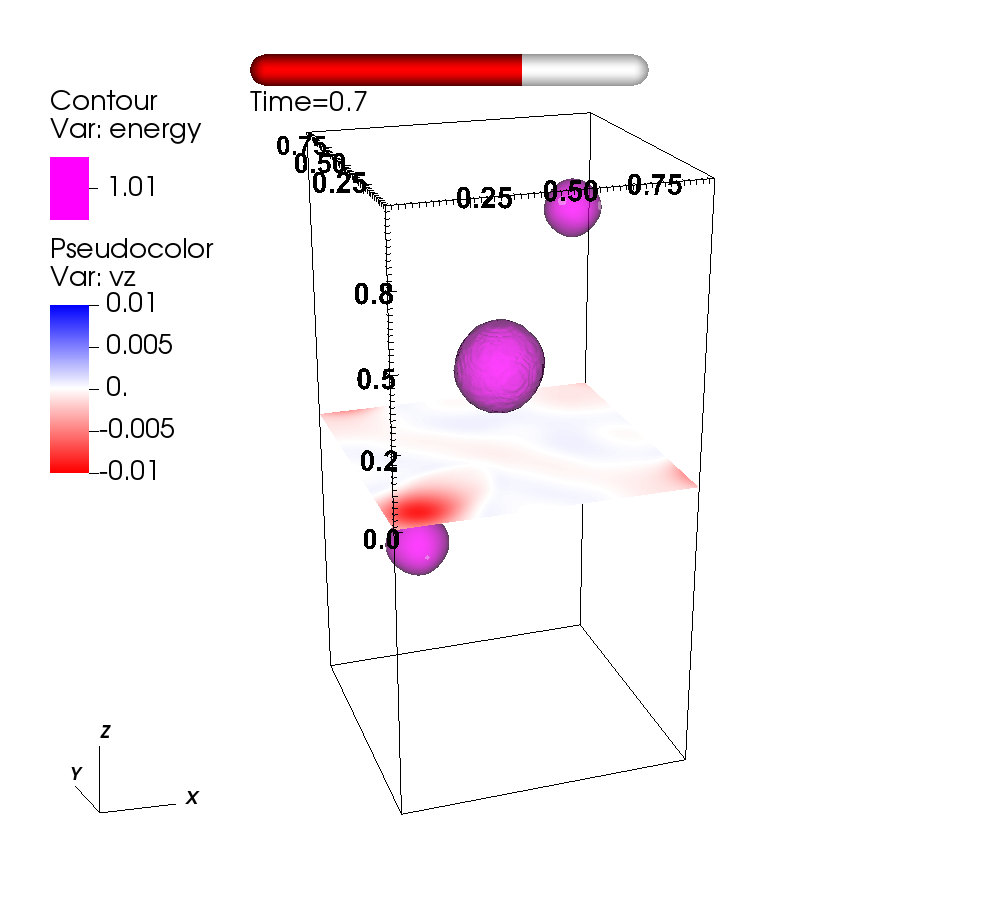}
        \includegraphics[width=0.225\textwidth, viewport=250 75 750 903, clip=true]{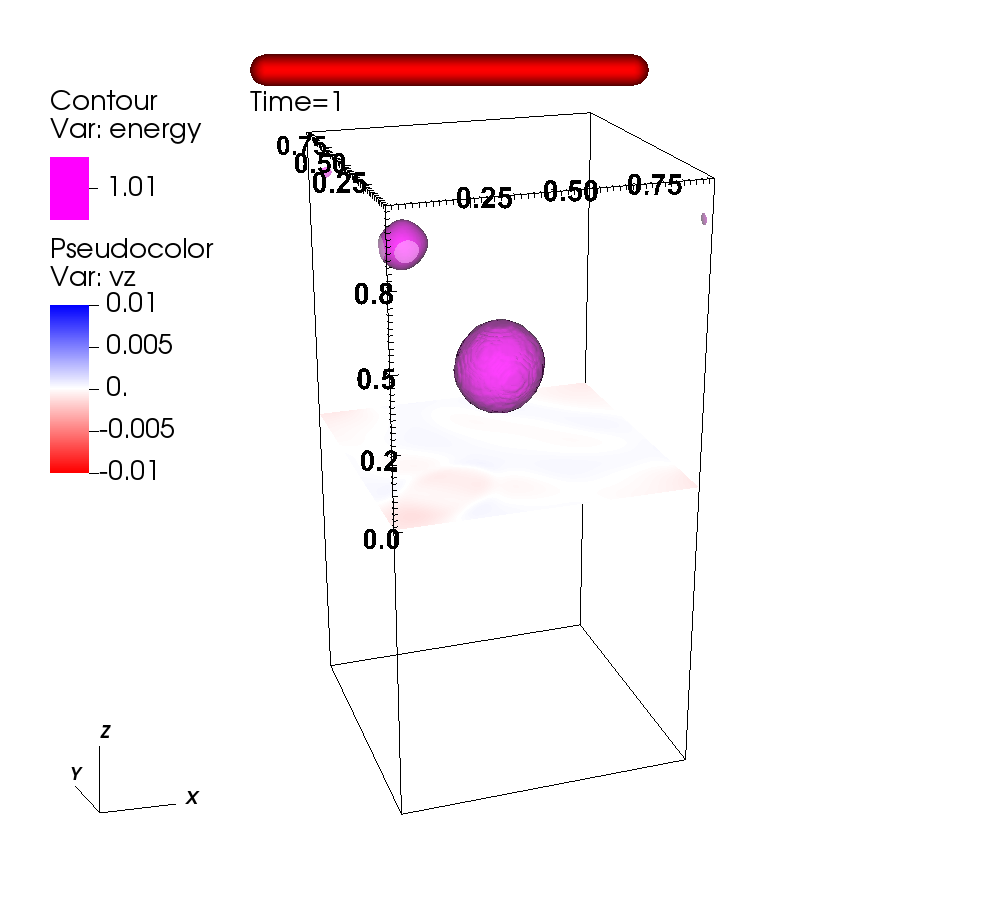}

        b) Ground Truth
        \vspace{10pt}

        \hspace{0.1125\textwidth}
        \includegraphics[width=0.225\textwidth, viewport=250 350 750 790, clip=true]{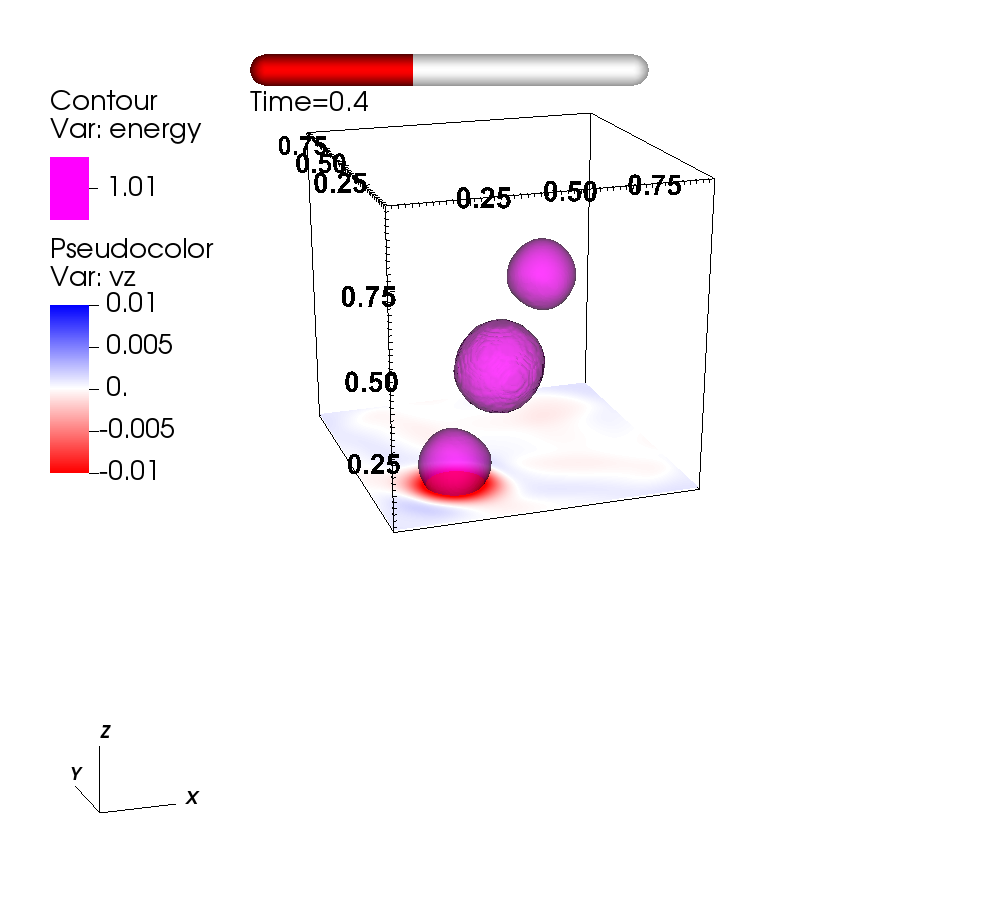}
        \includegraphics[width=0.225\textwidth, viewport=250 350 750 790, clip=true]{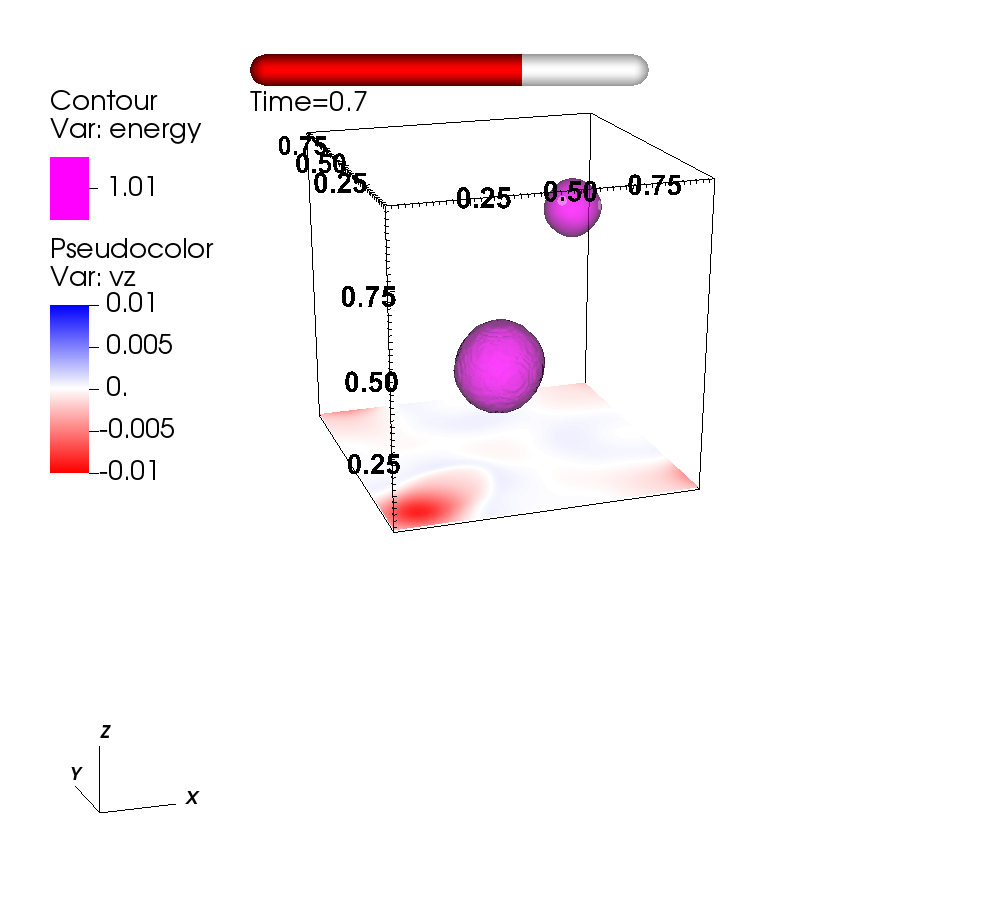}
	\includegraphics[width=0.225\textwidth, viewport=250 350 750 790, clip=true]{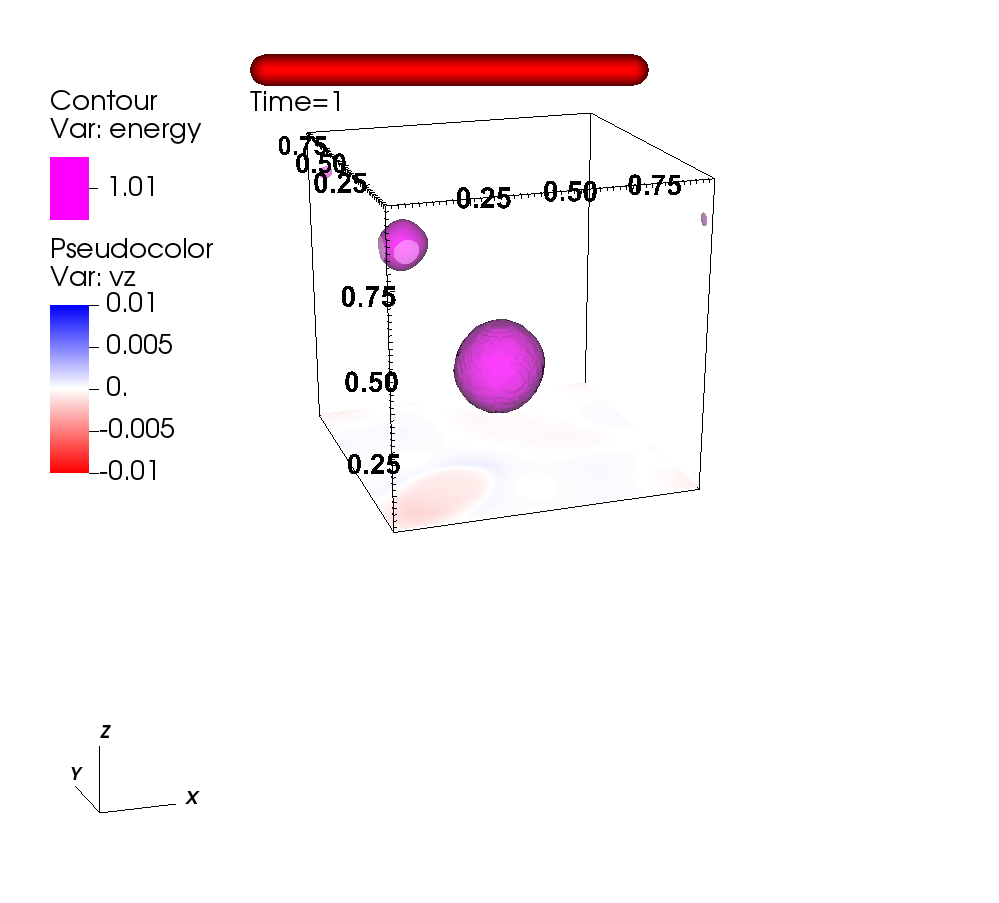}
 
         c) Cancellation NRBC
	
	\caption{Snapshots of the hot sphere with angled $\vect{B}$ test of the non-reflecting boundary conditions. Each panel shows an isocontour of internal energy 1\% above the background and a map of the vertical velocity at the layer of the non-reflecting boundary. The results from a simulation with a Fixed NRBC (here $\Lsiginc = 0$) are in the top row, and those from a simulation with a Cancellation NRBC ($\Lsiginc = -\sum_\zeta S_{\sigma,\zeta}^{-1} C_\zeta$) are in the bottom row, while the ground truth simulation is in the middle row. The left column is at a time shortly after the perturbation from the hot sphere has impacted the non-reflecting boundary ($t=0.4$), and the right column is the final state of the simulation at $t=1$ in normalized code units. An animated version of this Figure is available as Supplementary Video 1.}
	\label{fig:hot_sphere}
\end{figure*}

In Figure \ref{fig:hot_sphere} we present snapshots of all three simulations at a time ($t = 0.4$) shortly after the expansion of the hot sphere has encountered the layer corresponding to the non-reflecting boundary (left column), a time ($t=0.7$) shortly after the downward propagating perturbation leaves the simulations with NRBCs (middle column), as well as the final state of the plasma at time $t=1$ in normalized code units (right column).
By $t=1$ when the simulations are terminated, fast, slow, and Alfv\`enic perturbations have had time to fully interact with the NRBC.
Moreover, if any fast or Alfv\`enic perturbations were reflected from the NRBC they would have time to propagate through the full simulation volume.
All panels of the figure show the vertical velocity at the non-reflecting boundary (or at the corresponding plane in the ground truth simulation) and an isocontour of energy at $\epsilon=1.01$.
The top row shows the simulation with a Fixed NRBC ($\Lsiginc=0$), the bottom row shows the simulation with a Cancellation NRBC ($\Lsiginc=-\sum_\zeta S_{\sigma,\zeta}^{-1} C_\zeta$), and the middle row shows the ground truth simulation.
A movie of the full evolution of the three simulations is also available as online supplementary material (Supplementary Video 1).


Generally, the evolution of all three simulations can be outlined by following the propagation and expansion or contraction of the three spheres of hot plasma visible \change{in} the left column of Figure \ref{fig:hot_sphere}.
As can be seen from the animated version of this figure, all three originate from the initial hot sphere centered at $\{x,y,z\} = \{0,0,0.3\}$ at time $t=0$.
The higher pressure inside this original hot sphere causes it to expand.
As a result the density inside the heated region drops, and the hot sphere enters pressure equilibrium with the background plasma.
Meanwhile, the perturbation to the background plasma initiated by this expansion
gets channeled by the magnetic field and creates the two additional propagating spheres.
\change{
The appearance of these three spheres as separated from one other is due to the specific choice of energy density contour.
At a lower energy density closer to the initial background the three remain linked as a tube of material with energy density above the background develops along the magnetic field-lines passing through the initial hot sphere.
}
The upward propagating perturbations pass through the periodic side boundaries in all three simulations after $t=0.7$, reappearing near the top boundary and the front edge of the simulation volume, namely the edge where $x=0$ and $y=0$, by $t=1$.
These perturbations become weaker as they propagate, and as such the upward propagating spheres contract.
The downward propagating perturbation leaves the simulation volume through the NRBC in both simulations with such boundaries. 
For the ground truth simulation, the downward propagating perturbation passes through the periodic boundaries in the lower portion of the ground truth simulation, again between $t=0.7$ and $t=1$, but ends the simulation obscured from view behind the cut showing vertical velocity near the back edge of the simulation volume, namely the edge where $x=1$ and $y=1$.
All three simulations show the interference of the various waves kicked off by the expansion of the hot sphere which we mentioned a few paragraphs earlier in this section.
The resulting interference pattern that can be seen in Figure \ref{fig:hot_sphere} in the vertical velocity shown in the $z=0.0$ plane is noticeably different between the three simulations, but this is to be expected as this layer of the simulation is treated substantially differently in the three simulations.
Moreover, the amplitude of the differences and the vertical velocity itself diminishes in time as information \change{is} lost from the volume at the depicted surface (for the two cases with non-reflecting boundary conditions) or as the perturbation moves away from the $z=0.0$ plane and damps out (in the ground truth case).

\change{
Returning to the discussion in the introduction, the perilous hope in applying a non-reflecting boundary is frequently that its presence will allow for a small simulation to evolve equivalently to a simulation including the region beyond the non-reflecting boundary.
While the bulk evolution of the simulations are all quite similar to one another by eye, there are differences in the details as we have just discussed.
Therefore, it is useful to quantify the difference between the ground truth simulation and the simulations with non-reflecting boundary conditions.
Any single metric can and will obscure important similarities and differences between two simulations, but our primary goal here is to quantify a point-by-point difference between two simulations throughout their complete simulation volumes.
To this end, we utilize a weighted mean squared difference which is defined as
}

\begin{multline} \label{eq:wMSD}
    {\rm wMSD} =\\ \Big\langle \Bigl( \vect{N}(\vect{x},t) - \vect{G}(\vect{x},t) \Bigr) \Matrix{K}^{-1} \Bigl( \vect{N}(\vect{x},t) - \vect{G}(\vect{x},t) \Bigr)\Big\rangle\;,
\end{multline}
\change{
where $\vect{N}$ is the vector of primitive variables in the simulation with a non-reflecting boundary, $\vect{N}^T = (\rho,\epsilon,v_y,v_y,v_z,B_x,B_y,B_z)$.
$\vect{G}$ is the same vector for the ground truth simulation.
The expectation value $\langle\cdots\rangle$ is taken over the total number of cells $N = N_x N_y N_z$ in the simulations with non-reflecting boundary conditions.
$\Matrix{K}$ is the covariance matrix of the vector of primitive variables computed for the ground truth simulation at $t=0.1$\change{, by which time }all the perturbations driven by the expansion of the hot sphere have had time to develop but before any of these have reached the sides of the simulation volume in the lateral direction or $z=0$\change{, the location of the NRBC in the Fixed and Cancellation simulations}.
For reference, the \{$\chi,\zeta$\} element of the covariance matrix, where both $\chi$ and $\zeta$ vary over $\{\rho,\epsilon,v_x,v_y,v_z,B_x,B_y,B_z\}$, is defined as
}

\begin{multline} \label{eq:covariance}
	K_{\chi,\zeta} = \Big\langle\Bigl(G_\chi \left(\vect{x},\tau_{ad}\right) - \langle G_\chi \left(\tau_{ad}\right) \rangle \Bigr)\\
 \Bigl(G_\zeta\left(\vect{x},\tau_{ad}\right) - \langle G_\zeta \left(\tau_{ad}\right)\rangle\Bigr)\Big\rangle\;.
\end{multline}
\change{
$\Matrix{K}$ is computed over the subvolume of a ground truth simulation which overlaps the volumes of the simulations with non-reflecting boundary conditions. 
}

\begin{figure}[!htbp]
    \centering
    \includegraphics[width=0.4\textwidth]{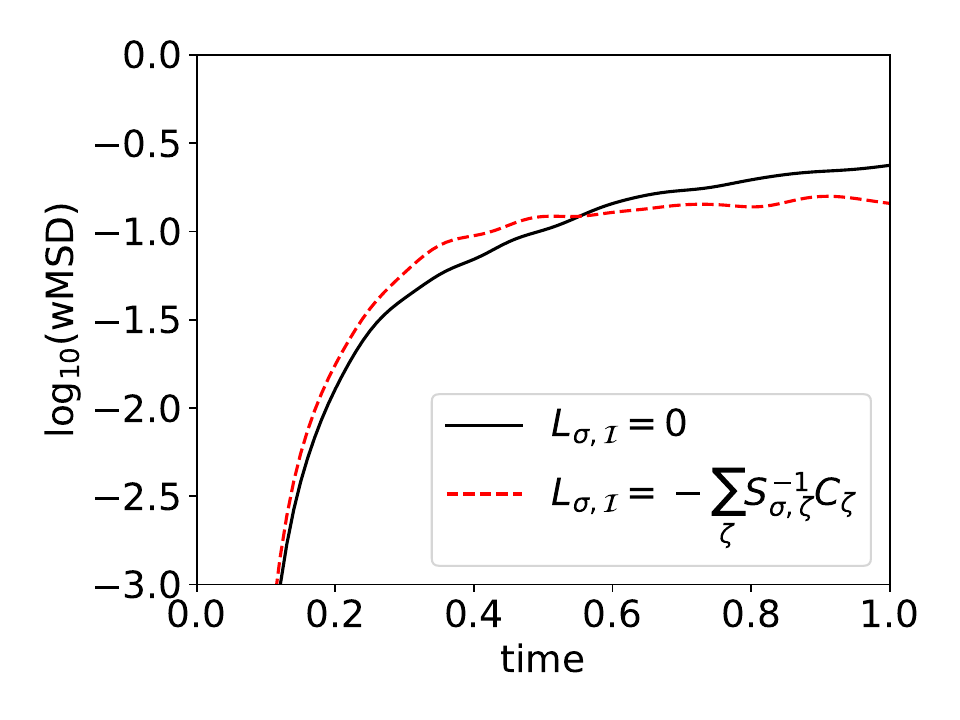}	
	\caption{Mean squared difference between the ground truth hot sphere simulation and the simulations with non-reflecting boundary conditions. The solid black curve denotes the simulation with a Fixed NRBC (here $\Lsiginc = 0$) and the dashed red curve the simulation with a Cancellation NRBC ($\Lsiginc = -\sum_\zeta S_{\sigma,\zeta}^{-1} C_\zeta$).}
	\label{fig:hot_sphere_mean_sq_error}
\end{figure}

\change{
Figure \ref{fig:hot_sphere_mean_sq_error} plots this mean squared difference for the simulations with a Fixed NRBC (here $\Lsiginc = 0$) and with a Cancellation NRBC ($\Lsiginc = -\sum_\zeta S_{\sigma,\zeta}^{-1} C_\zeta$).
Both simulations show maximum differences on the order of $wMSD = 0.1 - 0.3$.
To give a sense of the level of difference that this implies, we note that the maximum element of $\Matrix{K}$ is the standard deviation of $\epsilon$, which is $6\times10^{-5}$, and several elements of the covariance matrix are order $10^{-7} - 10^{-6}$, while $\rho$, $\epsilon$, and $\vect{B}$ are order unity and $\vect{v}$ is order $0.01-0.1$ depending on the time in the simulation.
This means that on average, individual cells in the simulation agree between the ground truth and the simulations with non-reflecting boundaries in all MHD quantities at the level of fractions of a percent.
This reinforces the point that the perturbations shown by the contours of $\epsilon$ in Figure \ref{fig:hot_sphere} all are indistinguishable by eye between the three simulations.
}

\change{Overall,} this test case shows NRBCs behaving as they are hoped to in the literature.
First, both implementations of NRBCs pass perturbations out of the simulation without reflection, thereby fulfilling the purpose for which they are designed.
Second, the simulations with NRBCs look for the most part like cut outs of the larger ground truth simulation, as is often the hope when running simulations with NRBCs.
The three simulations do show subtle differences from one another, here most evident in the interference pattern of non-zero vertical velocity remaining at $z=0$ after the expansion of the hot sphere passes this layer.
However, these can be explained by the loss of information in the simulations with a non-reflecting boundary, as a non-reflecting boundary prevents the external universe from communicating back to the simulation.
In contrast, the propagating information in the fully periodic ground truth simulation wrapped through the periodic boundaries in the bottom portion of the expanded volume\footnote{Recall that the contour of this downward propagating perturbation in energy is obscured behind the cut of $v_z$ at $z=0$ in the final snapshot of the simulation.} and information about its continuing propagation is allowed to enter the top portion of the simulation.
As such, the different phase patterns in the $z=0$ plane are due to the implicitly different models of the external universe below that plane for the three simulations.
These differences are small and do not lead to meaningful differences in the evolution of the simulations, and as such it is reasonable to argue that both non-reflecting boundaries have done their job and ``correctly'' passed information out of the simulation with minimal reflections.

To further verify the implementation, we have run a variety of other one-, two- and three-dimensional test cases including the Sod \citep{Sod78} and Brio-Wu \citep{BriWu88} shock tubes, pure Alfv\'en waves, a hydrodynamic equivalent of the hot sphere test, and initially sinusoidal velocity pulses in hydrodynamic and magnetohydrodynamic simulations.
\change{We present a sample of familiar, 1D test problems, namely a linear Alfv\'en wave, the Sod shock tube, and the Brio-Wu shock tube, in Appendix \ref{app:waves}.}
We find comparable results in all cases.
Therefore, we can confidently assert that this non-reflecting boundary implementation is working as specified.

\section{A surprisingly more complicated test case: advecting a spheromak}\label{sec:advect_spheromak}

Moving on, we now want to consider a different type of test case where we advect a topologically complex magnetic structure through the boundary condition.
This situation is much more in line with the type of simulations one expects in a solar physics setting.
Specifically, for a simulation of the solar atmosphere, we expect energy, matter, and magnetic fields to enter the simulation through a boundary cospatial with the solar photosphere, and in many cases to exit the simulation through another boundary.
For instance, consider the case of an active region coronal mass ejection (CME), where the simulation begins with magnetic flux emergence through the photosphere and ends with the CME plasma and its associated magnetic field exiting the simulation through the top boundary of the simulation.
As we have discussed in \citetalias{TarKee24}, the magnetic flux emergence at the photosphere can be readily handled by data-driven boundary conditions, as this is a case where we have invaluable information about the state of the external universe through observations of the solar photosphere.
As we will show in this section, the latter case where the CME plasma and magnetic field must leave the simulation again is very difficult, as we have very little or no knowledge of the state of the external universe.

As our test case for this scenario, we initialize a spheromak in the center of the simulation, allow the simulation to adjust to an asymptotic equilibrium configuration in the presence of the spheromak, and then advect the spheromak through the NRBC by reinitializing $v_z(x,y,z) = v_{\rm ad}$, for $v_{\rm ad}$ one of a chosen set of advection speeds discussed in more depth later in this section.
To provide ground truths against which to compare these simulations, we also run a second suite of simulations\change{ having domains extended in the advection ($z$) direction, with the $z_{min}$ boundary placed }beyond $z = v_{\rm ad}t_{\rm tot}$, where $t_{\rm tot}$ is the total time for which the simulations are run.
\change{The ground truth simulations use a simple zero-gradient boundary condition on $\rho$, $\epsilon$, $\vect{v}$, $B_x$, and $B_y$, while $B_z$ is set to preserve $\nabla\cdot\vect{B} = 0$.  This analytic solution to the equations is also numerically well behaved up until the leading edge of the spheromak reaches the $z_{min}$ boundary.
Therefore we halt the simulations, and our analysis, before that occurs.}

The analytic description of a spheromak was developed by \cite{RosBus79}.
For our purposes, an important feature of the spheromak is that it is a solution to the linear force-free equation, $\vect{J} = \kappa \vect{B}$ with $\kappa$ a scalar.
The classical spheromak is an axisymmetric solution to this equation with a closed magnetic surface at $\kappa r = 4.493\ldots$, the first zero of the the spherical Bessel function.
We set $\vect{B} = \vect{0}$ outside this surface which results in a tangential discontinuity within the surface (i.e, a surface current)\footnote{As discussed in \citetalias{TarKee24}, for numerical stability we smooth out this tangential discontinuity so that it spans a few grid cells in our simulation.}.
Therefore, the spheromak is a self-contained magnetic structure embedded in an initially homogeneous field-free plasma, and, after the brief relaxation discussed below, is in stable equilibrium.
Our magnetic field implementation of the spheromak in \lare{} is fully described in \citetalias{TarKee24}, Appendix F. 
In contrast to \citetalias{TarKee24}, however, we wish to construct a spheromak initially as close as possible to a steady-state equilibrim instead of the pressure-driven expanding spheromak considered in that work.  We therefore use a uniform pressure and density for \change{all simulations in this paper}.

\begin{figure*}
    \centering
    \includegraphics[width=0.45\textwidth]{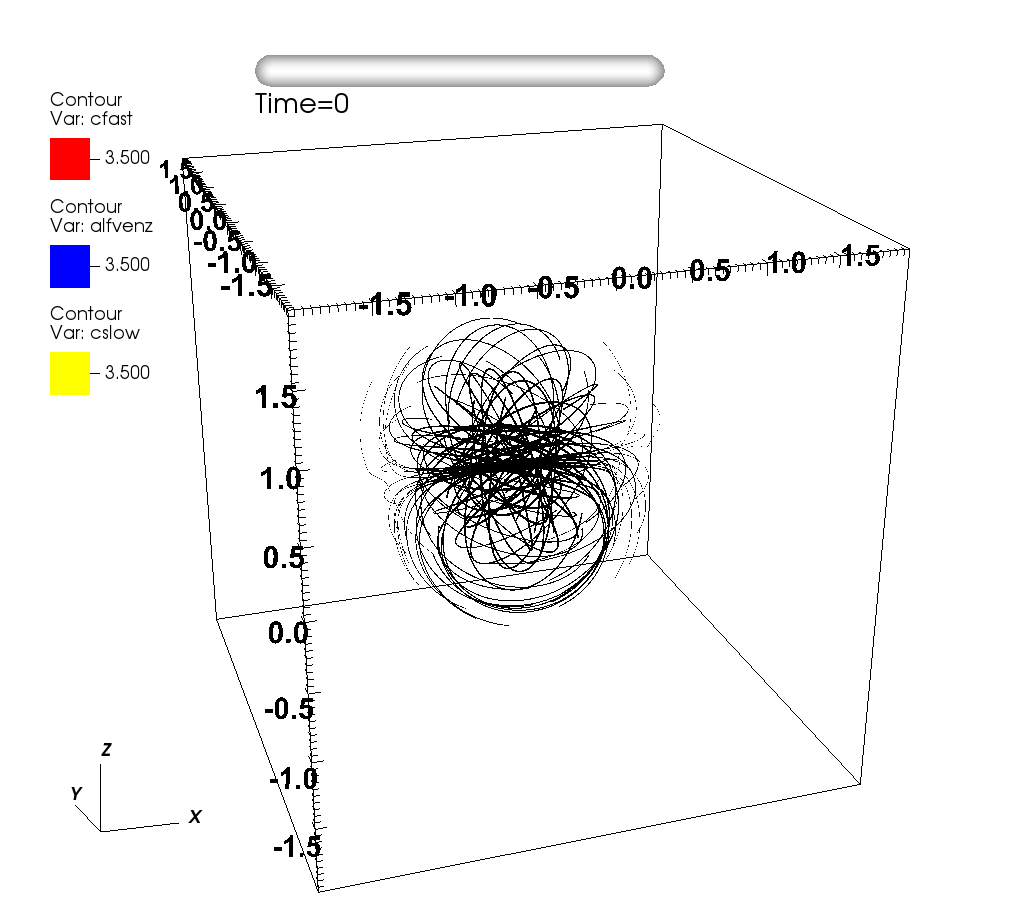}
    \includegraphics[width=0.45\textwidth]{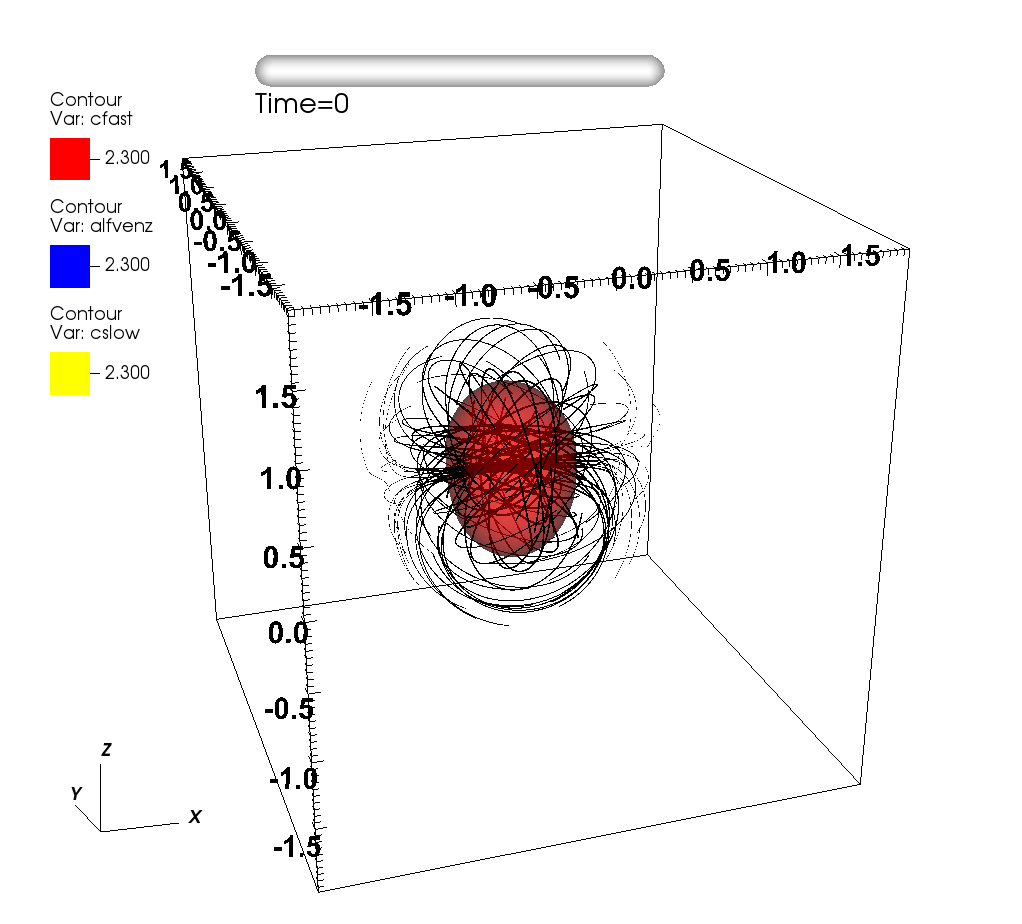}
    
    \includegraphics[width=0.45\textwidth]{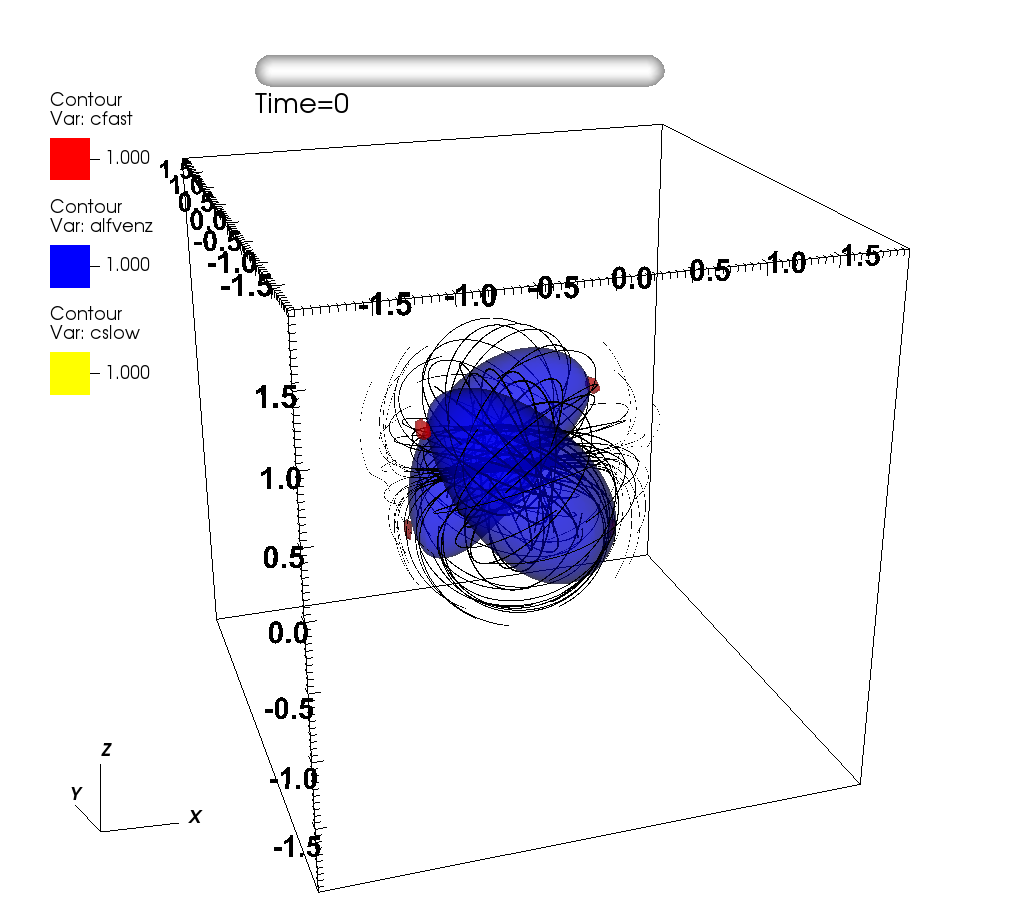}
    \includegraphics[width=0.45\textwidth]{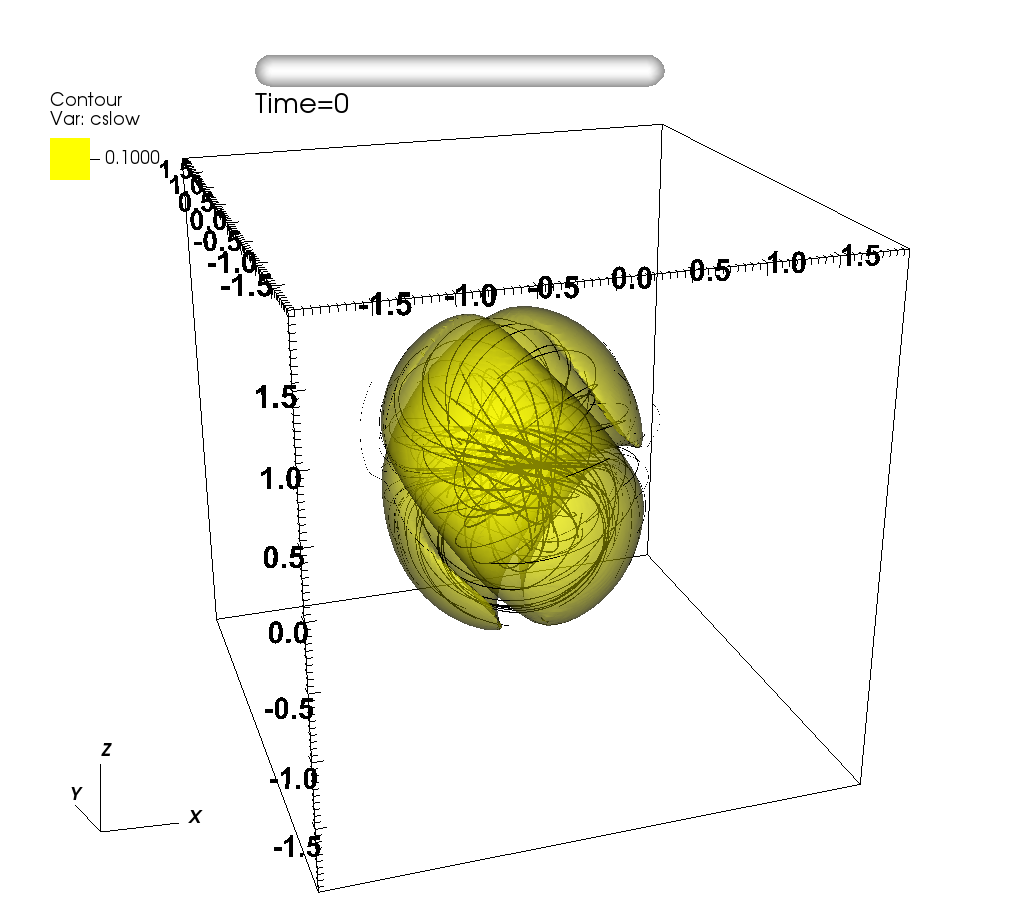}
    \caption{Contours of the characteristic velocities $c_f$ (red), $\alfven$ (blue), and $c_s$ (yellow) for each of the four selected values of $v_{ad}$.
    In each panel, $|v_{ad}|$ is less than the characteristic velocity inside the contour.
    The four cases are Case 1: $v_{\rm ad} < -\cfast$ (top left), Case 2: $-\cfast < v_{\rm ad} < -c_a$ (top right), Case 3: $-c_a < v_{\rm ad} < -\cslow$ (bottom left), and Case 4: $-\cslow < v_{\rm ad} < 0$ (bottom right).
    For Case 3 the contour of $\cfast$ is outside the box with the exception of small volumes where $\cfast < |v_{\rm ad}|$.
    For Case 4 only the slow speed is shown because the contour of $\cfast$ is fully outside the box and contours of $b_z$ and $c_s$ lie nearly on top of one another near the surface of the spheromak.}
    \label{fig:velocity_contours}
\end{figure*}

To select advection velocities, we consult Table \ref{tab:incoming}.
Our goal is to have each simulation highlight the behavior of the NRBC with a different number of incoming characteristics. 
Because all the characteristic speeds are functions of space in the simulation, we select advection velocities that put us in a single layer of Table \ref{tab:incoming} over a large volume of the interior of the spheromak.
Figure \ref{fig:velocity_contours} shows \change{a single contour of $c_f,$, $c_a$, and $c_s$ in red, blue, and yellow, respectively.  If $|v_{ad}|$ is selected to be equal to the value of a contour, then the associated characteristic will be incoming inside that contour.} 
The specific values selected for $v_{ad}$ are discussed further in the following section.
For brevity, going forward we refer to the set-ups with each of the advection velocities as Case 1: $v_z < -\cfast$, Case 2: $-\cfast < v_z < -\alfven$, Case 3: $-\alfven < v_z < -\cslow$, and Case 4: $-\cslow < v_z < 0$.

Note that part of our set-up requires allowing the simulation to relax to the presence of the spheromak before beginning the advection.
While the spheromak is a force-free field configuration in its interior, along its surface where the magnitude of $\vect{B}$ drops to zero, the spheromak does possess a surface current that produces an outwards directed Lorentz force.
This means that any spheromak will expand until its expansion generates a gas pressure gradient between the inside and outside of the spheromak large enough to cancel the non-zero Lorentz force.
As we want to be able to advect the spheromak at speeds corresponding to each layer of Table \ref{tab:incoming} with $v_z < 0$, this means that we need to make the magnetic field strong enough to meaningfully separate $\alfven$ and $\cslow$, i.e., we need to initialize the spheromak with plasma $\beta<1$.
As a result of this, $\alfven$ is much larger than the sound speed $a$, and balancing the non-zero Lorentz force across the surface of the spheromak requires an appreciable gas pressure gradient to be built up to put the setup into total force balance.
This is achieved by the spheromak expanding.
In fact, this argument holds independent of the value of $\beta$, with the amount of expansion required scaling inversely with $\beta$.
For this simulation suite, we set the magnetic field such that $\beta$ is of order 0.1 inside the spheromak, and then allow the simulation to relax for several average $\cslow$ crossing times through the spheromak before adding the advection.
The end result is that the spheromak is of order 10\% larger in radius when the advection begins than before the relaxation.

\subsection{Numerical specifications}\label{sec:validation_numerics}

Our simulations with a non-reflecting boundary condition are run on a grid with $N_{x} = N_{y} = N_{z} = 128$ grid cells in each direction.
The simulations span $\ell_x = \ell_y = \ell_z = 4$ pre-relaxation spheromak radii in each direction.
The spheromak is centered at $[x,y,z] = [0,0,0]$ and the simulation domain extends to $\pm 2$ in all directions.
At the maximum and minimum boundary in the $x$- and $y$-directions the simulation is periodic.
We place the NRBC at the minimum of the domain in the $z$-direction.
At the maximum of the domain in the $z$-direction, the simulation has a symmetric boundary, namely $U_\zeta(z_{\rm max} +\delta z) = U_\zeta(z_{\rm max} -\delta z)$ in all quantities except $v_z$, which is held fixed at the advection speed, and $B_z$, which is used to guarantee $\nabla \cdot B = 0$ at this boundary. 
The simulation is initialized with $v_x = v_y = v_z = 0$, $\rho=1$ and $\epsilon=1$ everywhere.
$B_x$, $B_y$, and $B_z$ are initialized according to the spheromak equations in \citetalias{TarKee24}, Appendix F with $B_0 = 2.0$ and $\kappa \approx 4.493$ to place the surface of the spheromak at $r=1$.
As discussed above, the simulation is then allowed to relax to the presence of the spheromak\change{, after which the characteristic speeds lie in the ranges} $\cfast \in [0.97,3.27]$, $\alfven \in [0,2.24]$, and $\cslow \in [0,0.95]$.
We then proceed to run the advection portion of the simulations imposing a uniform advection velocity $v_z = v_{\rm ad}$\change{ chosen to lie within each of those ranges.}

\change{To discuss our specific choices of advection speeds, we plot contours of $\cfast$, $\alfven$, and $\cslow$ in Figure \ref{fig:velocity_contours}. 
The four panels show these contours for values of 3.5 (top left), 2.3 (top right), 1.0 (bottom left), and 0.1 (bottom right).
In each panel $\cfast$ is the red contour, $\alfven$ is the blue contour, and $\cslow$ is the yellow contour.
}

\change{
Beginning in the upper left of Figure} \ref{fig:velocity_contours}\change{, 3.5 is greater than the maximum value of all the characteristic speeds, and as such no contours appear in this panel.
This means that advecting the spheromak at this speed would be expected to direct all the characteristic derivatives out of the box at all times.
To put this a different way, the number of $\Lsiginc$, hereafter denoted as $\# \Lsiginc$, would be expected to be zero at all times while the spheromak is advected through the boundary at this speed.
Therefore, we select $v_{\rm ad} = -3.5$ as our first advection speed, hereafter Case 1 which we denote as $v_{\rm ad} < - \cfast$.
}

\change{
Moving to the upper right panel, 2.3 is in the range of $\cfast$ but is still greater than the maximum of $\alfven$ or $\cslow$.
Therefore, we only see a contour of $\cfast$ (red) in this panel.
Advecting the spheromak downward at this speed would be expected to result in the characteristic mode with eigenvalue $v_{\rm ad} + \cfast$ pointed against the direction of advection inside this red contour, and as such $\# \Lsiginc = 1$ inside the red contour when this region is passing through the non-reflecting boundary condition during the advection.
As this isolates the effects of only this fast mode affecting the simulation, we select $v_{\rm ad} = -2.3$ as our second advection speed, hereafter Case 2 which we denote as $- \cfast < v_{\rm ad} < - \alfven$.
}

\change{The bottom left panel of Figure} \ref{fig:velocity_contours}\change{ shows contours of the characteristic velocities where they equal 1.0.
This is almost the minimum of $\cfast$ so only small red contours remain inside which no characteristic modes are incoming ($\# \Lsiginc = 0$ inside these very small regions).
Now, however, 1.0 is well within the range of $\alfven$ so we see substantial volumes of the spheromak encapsulated by blue contours.
1.0 is still larger than the maximum of $\cslow$, so no contours of $\cslow$ appear.
Choosing this advection speed, as we do in Case 3, is expected to result in at least one incoming characteristic with eigenvalue $v_{\rm ad} + \cfast$ at almost all locations and times, and two incoming characteristics with eigenvalues $v_{\rm ad} + \cfast$ and $v_{\rm ad} + \alfven$ ($\# \Lsiginc = 2$) inside the blue contours when they pass through the non-reflecting boundary.
Therefore we refer to Case 3 alternately as $- \alfven < v_{\rm ad} < - \cslow$ as this is satisfied over the majority of the interior of the spheromak.
}

\change{
Finally, the bottom right panel shows the contour where $\cslow = 0.1$ (yellow).
We omit the contour of $\cfast$ as 0.1 is less than the minimum of $\cfast$ and as such no red contour would appear.
We also omit the contour $\alfven$ for clarity.
This is because both $\alfven$ and $\cslow$ go to zero outside the spheromak and as the contour we plot is itself so close to zero, the blue contour of $\alfven$ would have lain very nearly on top of the yellow contour of $\cslow$, significantly muddying this panel.
As this is the only case we show which includes a contour of $\cslow$, using this as as our final advection speed will add a case where one additional characteristic mode can propagate into the box, namely the one with eigenvalue $v_{ad} + \cslow$ ($\# \Lsiginc = 3$) inside the yellow contour as it propagates across the NRBC.
This is the maximum number of characteristic modes we can get pointing into the simulation volume without the advection velocity being upwards, at which point we are in a regime likely more appropriately handled by a data-driven boundary condition as the advection could draw new structures and information from the external universe into the simulation.
Therefore, we make $v_{\rm ad} = -0.1$ our final choice of advection speed, namely Case 4, also denoted $-\cslow < v_{\rm ad} < 0$ as this is the condition satisfied by the majority of the volume of the spheromak.
}

\change{
To reiterate, we proceed running simulations with $v_{\rm ad} = $ -3.5, -2.3, -1.0, and -0.1 as our four advection speeds in Case 1-4, respectively.
As discussed above, these choices produce simulations that have largely distinct numbers of incoming characteristics once the spheromak reaches and passes through the NRBC, and as such allow us to analyze the effects of each characteristic mode on the simulation volume by comparing the four cases.
We will use the expected number of $\Lsiginc$ later in this section to test how well each NRBC is able to reproduce the known behavior of uniform, constant advection from the ground truth simulation, and what dependence any departures have on the number of incoming characteristics present in the boundary condition.
For ease of reference, we summarize that Case 1 is expected to have $\# \Lsiginc = 0$ at all times, Case 2 is expected to have $\# \Lsiginc = 1$ while the spheromak is passing through the boundary and $\# \Lsiginc = 0$ elsewise, Case 3 is expected to have $\# \Lsiginc = 2$ while the spheromak is passing through the boundary and $\# \Lsiginc = 1$ elsewise, and Case 4 is expected to have $\# \Lsiginc = 3$ while the spheromak is passing through the boundary and $\# \Lsiginc = 1$ elsewise as the region where $\cslow > |v_{ad}|$ is nearly identical to the region where $\alfven > |v_{ad}|$ for this case.
}

Factoring in the initial distance from the bottom of the spheromak to the NRBC boundary ($\delta \ell = \ell_z/4 = 1$), we run each simulation for $t=5/v_{\rm ad}$ such that the lower edge of the spheromak contacts the NRBC at $t \approx 1/v_{\rm ad}$, the spheromak fully exits the box at $t \approx 3/v_{\rm ad}$, and the remaining time allows us to examine whether remnants of the spheromak passing through the NRBC are visible in the simulations with NRBCs.
Based on these simulation durations, the ground truth ($GT$) simulations have $\ell_{z,GT} = 10$ with $-8\leq z\leq2$ and $N_{z,GT} = 320$.
\change{This puts} the $z_{\rm min}$ boundary in these simulations 2 spatial units beyond the lower edge of the spheromak at the end of the simulation such that the spheromak is fully contained within the $GT$ simulation at all times \change{and the impact of the ground truth boundary at its $z_{min}$ is negligible}.

\subsection{Comparison of Ground Truth and NRBC simulations}\label{sec:GTvsNRBC}
In comparing the ground truth (GT) and NRBC simulations, it is important to make a distinction between the types of differences that can occur between the simulations.
The first is changes in the MHD properties due to numerical reflections from the NRBC.
These are independent of the type of boundary condition being implemented, as they are numerical errors in the method which result in $\Linc$ being incorrectly defined at locations near to, or even immediately at, the NRBC.
These errors then impact the amplitude of the modes propagating away from the boundary, causing spurious information to enter the simulation.
We will refer to differences arising from these numerical errors as numerical reflections.

The second type of differences that can occur are due to the NRBC correctly fulfilling its designed purpose and setting $\Linc$ to have the prescribed value to numerical precision.
The explicit mathematical goal of non-reflecting boundary conditions is to either fully remove the evolution in time of incoming modes (as is the case when $\Linc$ is held at its initial value or zero) or to set this evolution to be based only on simulation interior properties (as is the case when $\Lsiginc = - \sum_\zeta S_{\sigma,\zeta}^{-1}C_\zeta$).
With the exception of some special cases, both are insufficient to provide what is needed to match the ground truth simulation.
This means that the evolution of a simulation with an NRBC is in general different from a larger simulation that includes a volume beyond the non-reflecting boundary, as information about the structure of the spheromak which has ``left the building'' through the non-reflecting boundary condition is essential to the integrity of the part of the spheromak still in the building.
\cite{Hed79} provides a simple example of why information which has already left the volume can be essential to the evolution of what is left behind by laying out the case of a slow mode leaving the volume, and then at a later time a fast mode.
The slow mode will be overtaken by the fast mode outside the volume, at which point it is possible for their interaction to result in a back-propagating mode that should re-enter the volume at a later time \citep{Hed79,Tho87}.
Put differently, it is in general not correct to assume that all the weighted sums of spatial derivatives of the MHD primitive variables present in the subsets of Eqn.~P1.14 remain constant in time as the simulation evolves (as is assumed by the boundary condition with $\Lsiginc$ constant or zero), nor is it correct to assume that these can be correctly recovered by information propagating perpendicular to the plane wherein these $\Lsiginc$ are defined (as is assumed by the boundary condition with $\Lsiginc = -\sum_\zeta S_{\sigma,\zeta}^{-1}C_\zeta$).
In point of fact, the information required to correctly specify these weighted sums of spatial derivatives in a way that would match the ground truth simulation has already been destroyed by the time we need it.
The direct, mathematically intended purpose of the non-reflecting boundary condition is to pass information out of the volume without reflection, which is explicitly without impact on the incoming characteristics.
The incoming characteristics encode all the information about the impact of the external universe on the simulation volume, so choosing to not allow outgoing characteristics to impact incoming characteristics means that all information about anything that leaves the box is irretrievably lost.

For both numerical reflections and physically meaningful information eliminated by the NRBC, the influence the boundary condition can have increases with the number of incoming characteristic derivatives defined by the boundary.
Therefore, these two types of differences are difficult to distinguish on a case-by-case basis, especially for complex simulations, as the advection of a spheromak presented here proves to be.
However, the NRBC implementation we present here is built on the same code foundation from \citetalias{TarKee24}, in which we present a test case of data-driving from this boundary with $\Lsiginc(x,y,t)$ computed from the full $\vect{U}$ at every time step.
Because we are defining numerical reflections as errors in the method, and at the core \citetalias{TarKee24} uses the same method as this paper, any numerical reflections as we discuss them here would manifest as a poor match between driven and ground truth simulations in \citetalias{TarKee24}.
The results presented in \citetalias{TarKee24} show excellent agreement between the ground truth simulation and the one driven with the full ground truth $\vect{U}$ at every time step (i.e. the case where we can solve for $\Linc$ at \change{every Courant step; see the blue line in their Figure 8 and discussion pertaining to it}), so we are confident that our method has negligible numerical reflections and discuss instead only the latter type of differences, namely those arising from the NRBC correctly doing its job and eliminating alteration of $\Linc$ by $\Lout$ (when $\Lsiginc = -\sum_\zeta S_{\sigma,\zeta}^{-1} C_\zeta$) or all alterations to $\Linc$ (when $\Lsiginc=0$).

\subsubsection{Overview of the general behavior of a spheromak advected through the NRBC}\label{sec:GTvsNRBC_general}

Before presenting the results of our investigation of the behavior of the NRBC in the following subsection (\S \ref{sec:GTvsNRBC_adv}), we first discuss the typical evolution of the spheromak as it passes through the NRBC.
As much of the general behavior of the simulations is the same until we go to the very slowest advection speed (see Sect. \ref{sec:GTvsNRBC_adv}), we here choose to present results from the simulation with $-\alfven < v_{\rm ad}~(= -1.0) < -\cslow$ only.
Comparable figures to those presented in this section are available for all four advection speeds as supplementary material (Supplementary Videos 2--5).

\begin{figure*}[!htbp]
    \centering
        \includegraphics[width=0.31\textwidth, viewport=0 0 926 900, clip=true]{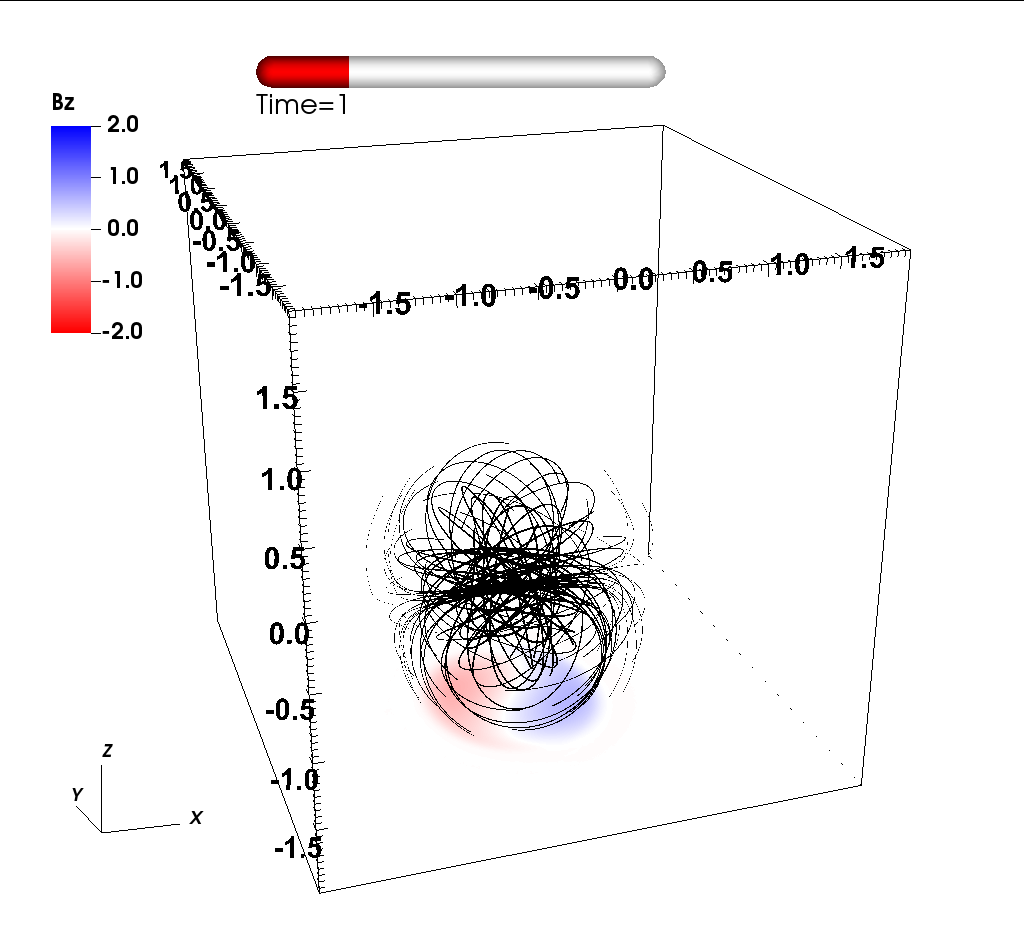}
        \includegraphics[width=0.31\textwidth, viewport=0 0 926 900, clip=true]{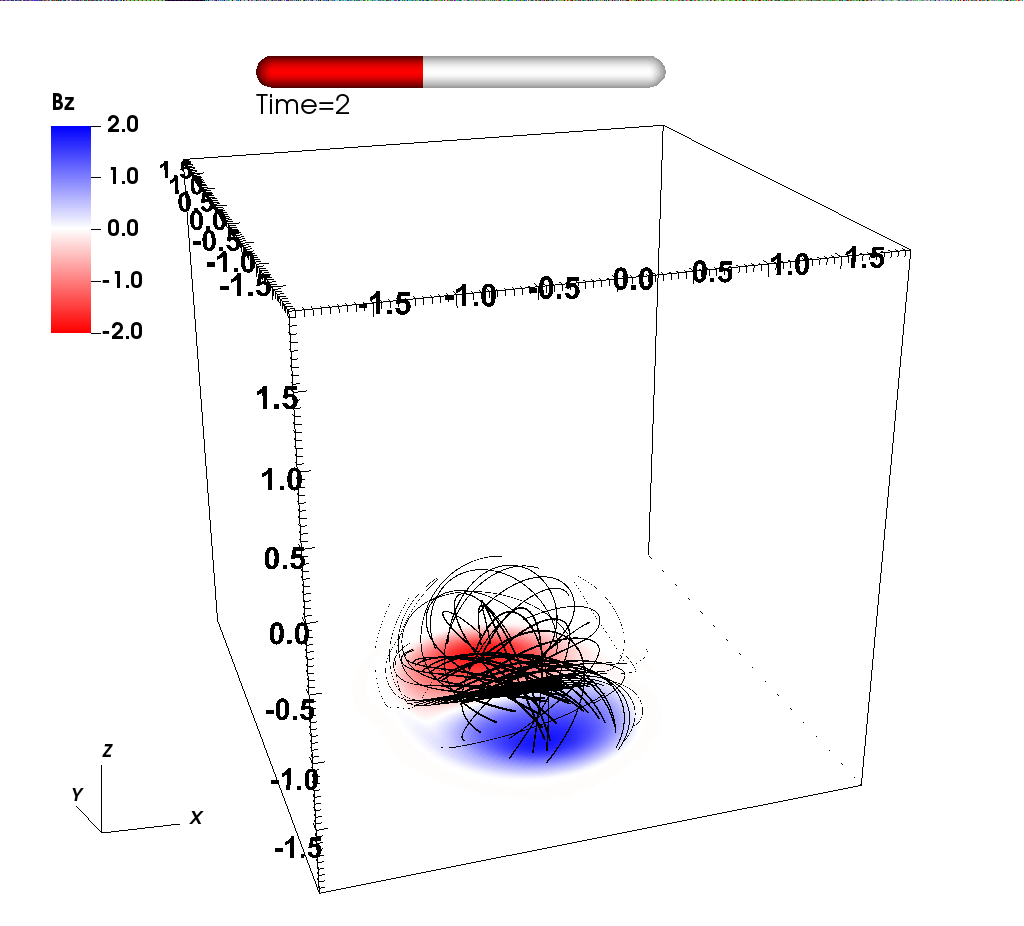}
        \includegraphics[width=0.31\textwidth, viewport=0 0 926 900, clip=true]{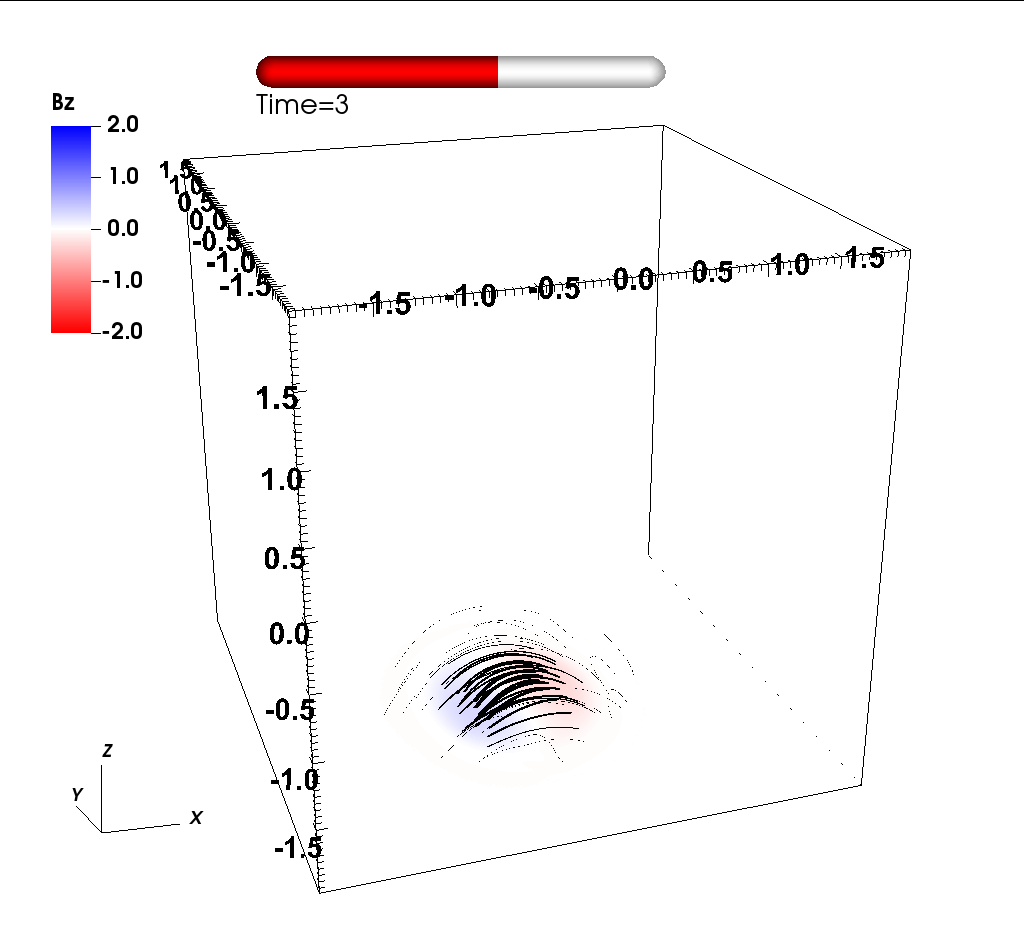}

        (a) Fixed NRBC
        \vspace{12pt}

        \includegraphics[width=0.31\textwidth, viewport=0 0 926 900, clip=true]{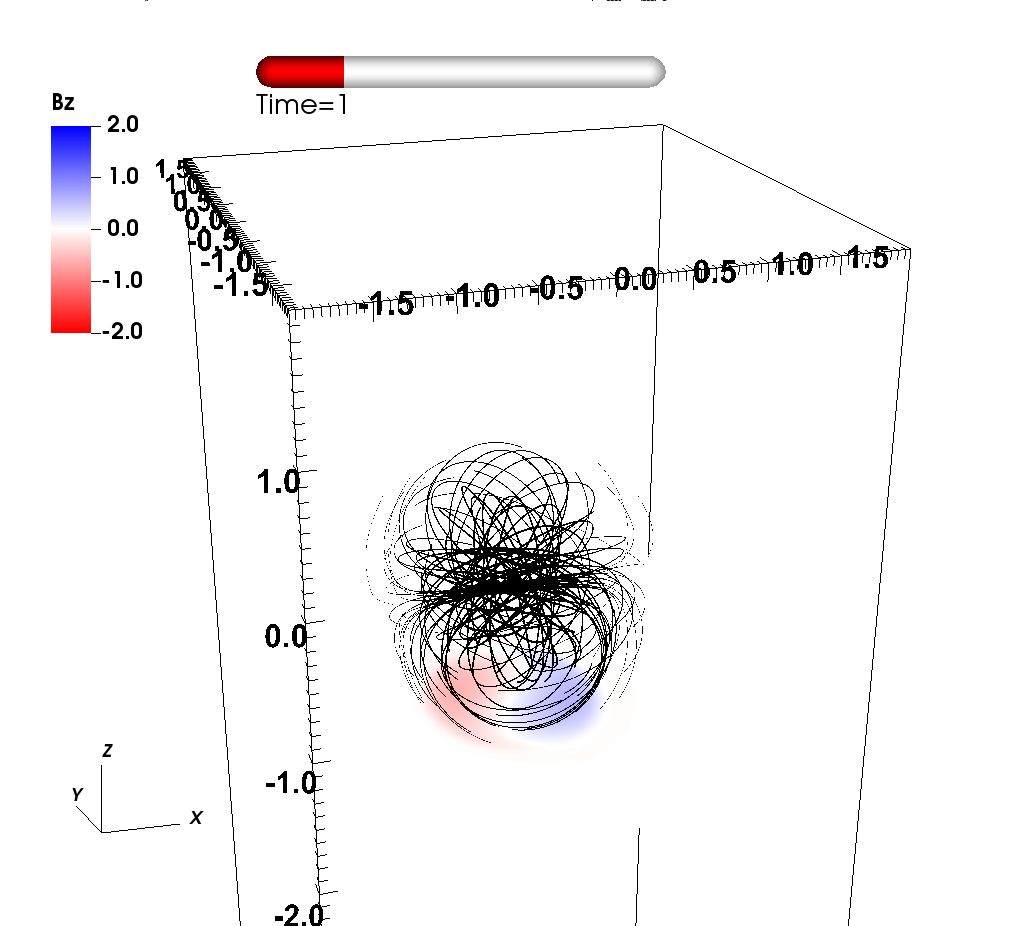}
        \includegraphics[width=0.31\textwidth, viewport=0 0 926 900, clip=true]{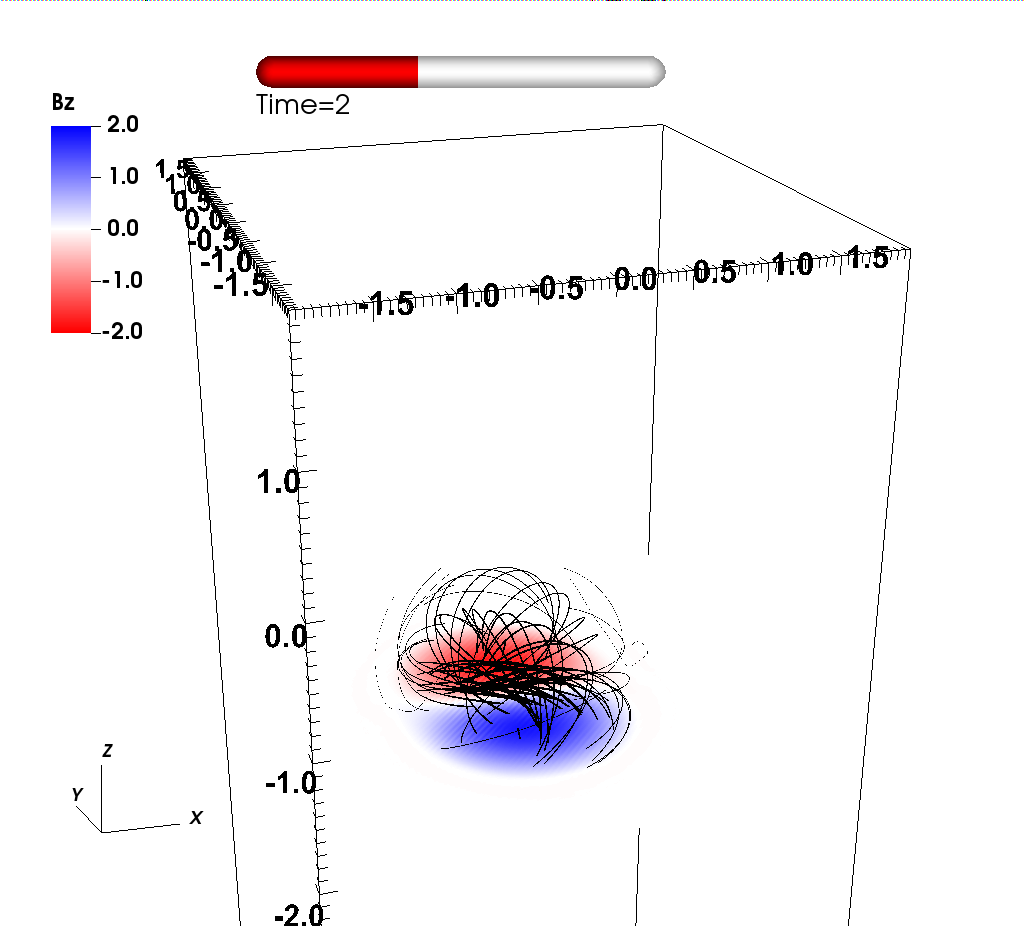}
        \includegraphics[width=0.31\textwidth, viewport=0 0 926 900, clip=true]{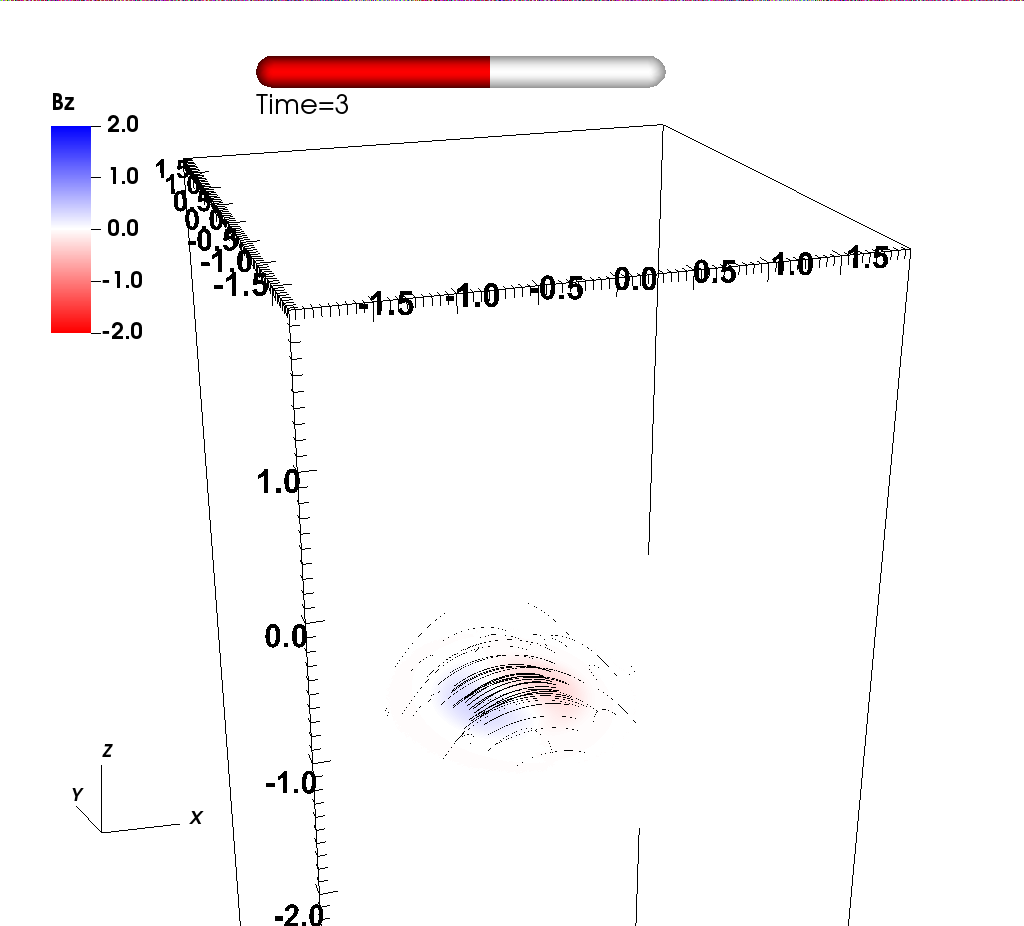}

        (b) Ground Truth
        \vspace{12pt}

        \includegraphics[width=0.31\textwidth, viewport=0 0 926 900, clip=true]{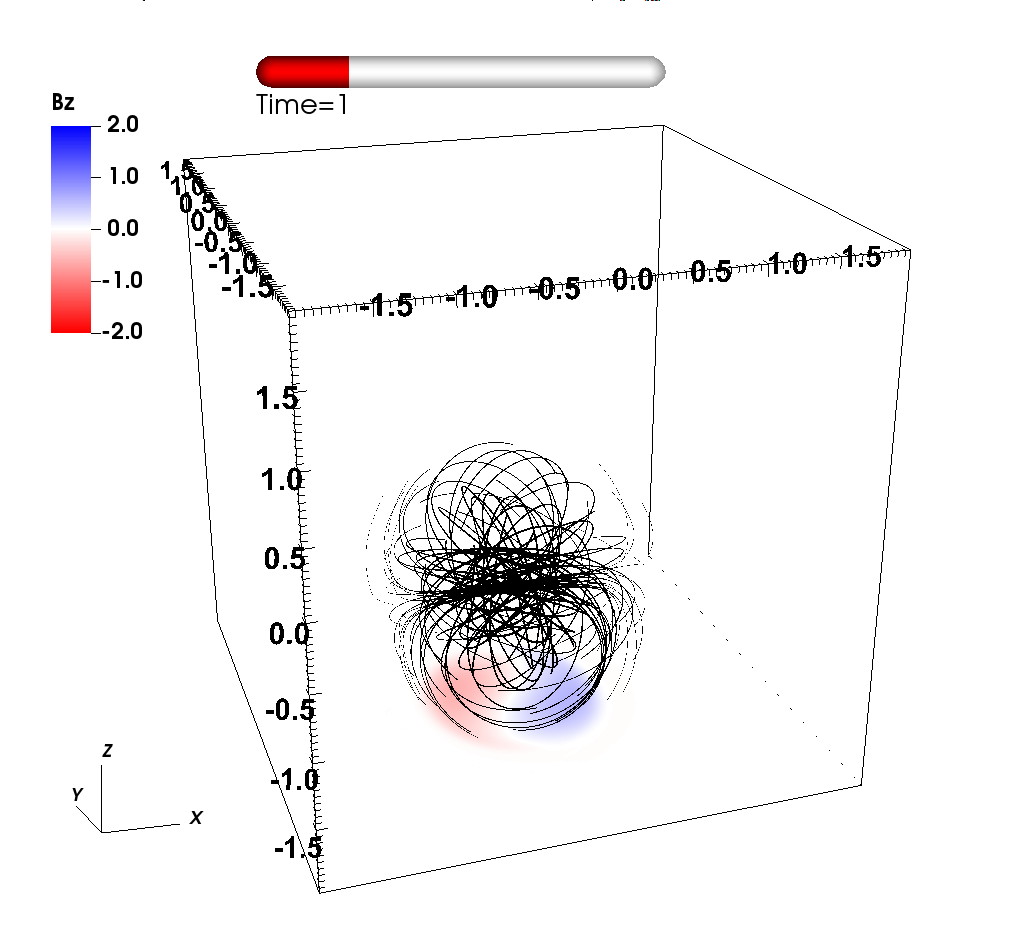}
        \includegraphics[width=0.31\textwidth, viewport=0 0 926 900, clip=true]{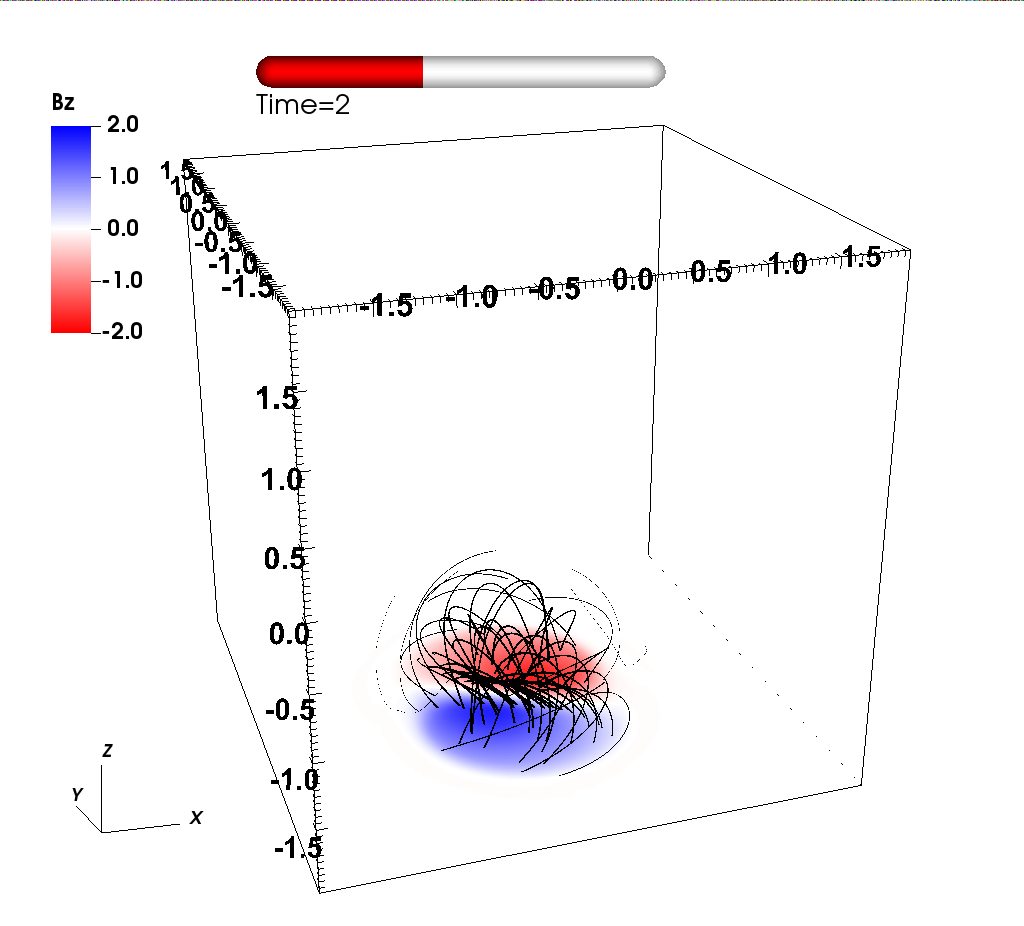}
        \includegraphics[width=0.31\textwidth, viewport=0 0 926 900, clip=true]{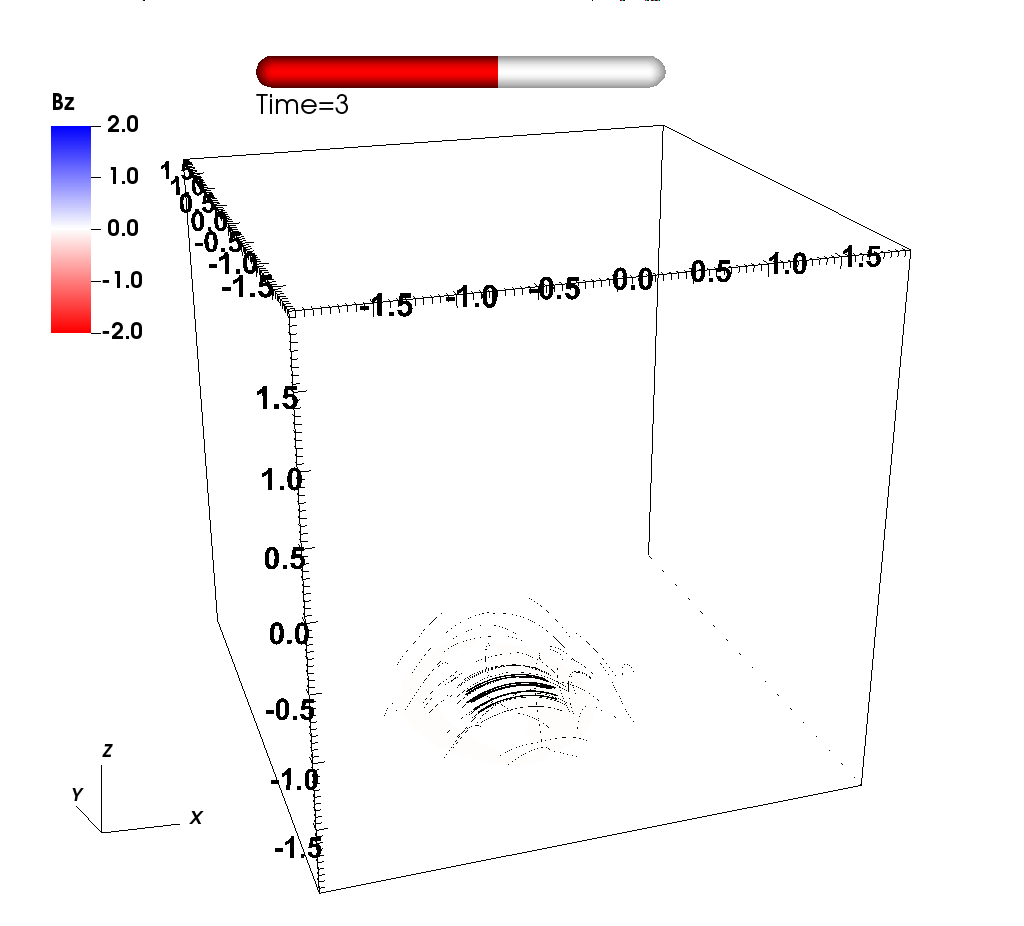}

        (c) Cancellation NRBC
	
	\caption{Magnetic field lines in the volume, and the normal component of the magnetic field, $B_z$, at the location of the non-reflecting boundary for the simulations with $-\alfven < v_{\rm ad}~(= -1.0) < -\cslow$. The simulation with a Fixed NRBC (here $\Lsiginc = 0$) is on the top and with a Cancellation NRBC ($\Lsiginc = -\sum_\zeta S_{\sigma,\zeta}^{-1} C_\zeta$) is on the bottom. The central row shows the same quantities in the overlapping volume and on the same plane for the ground truth simulation. Field lines were initiated at the same $x,y$ locations in the $z=-1.95$ plane for each column in all simulations. An animated version of this Figure is available as Supplementary Video 4.}
	\label{fig:halfSnap}
\end{figure*}

At a basic level, the configuration we are investigating is a Galilean transformation of a stationary, balanced spheromak.
Therefore, the desired result is that the spheromak should pass out of the volume without changes to any of the magnetohydrodynamic variables in the Lagrangian frame.
Importantly for investigating the behavior of the non-reflecting boundary, the simulation should begin with a topologically complex magnetic field configuration in the volume and end with $\vect{B}=\vect{0}$ after the spheromak passes out of the volume.
This critical result is recovered by all the simulations with non-reflecting boundaries which have $v_{\rm ad} < -\cslow$ (Cases 1, 2, and 3; rows 1, 2, and 3 in Table \ref{tab:incoming}).
To emphasize this point, in Figure \ref{fig:halfSnap} (an animated version of this Figure is available as Supplementary Video 4) we show the magnetic field lines in the volume and the normal component of the magnetic field, $B_{z}$, at the non-reflecting boundary (or in the case of the ground truth simulation, in a plane at the same height).
\change{
Figure \ref{fig:halfSnap} shows all three simulations at time $t=1$, when the spheromak has just contacted $z=-2$, the location of the non-reflecting boundary condition (left column); at time $t=2$ when the midplane of the spheromak reaches $z=-2$ in the ground truth simulation (middle column); and at time $t=3$ when the trailing edge of the spheromak reaches $z=-2$ in the ground truth simulation (right column).
}
The animation shows the spheromak in all three simulations advecting from its original location centered at $x=y=z=0$ toward the boundary condition at $z = -2$ (for the cases with NRBCs) or toward the extended lower volume of the simulation (for the GT case). 
Field lines in Figure \ref{fig:halfSnap} and in the animations are all initialized at the same $x,y$ locations in a plane that advects with the flow \change{to track the mid-plane of the spheromak,} and stops moving when it reaches $z = -1.95$ just inside the simulations with NRBCs. 
As the bottom (min($z$)) edge of the spheromak passes through the $z=-2$ layer where the NRBC is located in the simulations with such a boundary, distortions in the spheromak begin to occur as compared to the ground truth case. 
\change{These are visable at time $t=2$ in Figure \ref{fig:halfSnap}.}
At this speed and faster advection speeds, these distortions do not prevent the spheromak from passing through the simulation boundary or even drastically alter the advection speed of the spheromak.
\change{The small degree to which the advection time is altered can be seen at time $t=3$ in the right column of Figure \ref{fig:halfSnap}, as slightly different amounts of the spheromak remain in the volume.}
This reinforces the success of our implementation of a characteristics-based boundary condition into a code not based on a method of characteristics, and moreover its usage for non-reflecting boundaries.
This point was further emphasized in \citetalias{TarKee24}.

However, comparison of individual field lines, and even the orientation of the polarity inversion line (across which the normal component of the field changes sign) at the boundary of the simulation with a Cancellation NRBC, 
shows that the three simulations are not identical.
As discussed at the beginning of this section, this is to be expected as neither implementation of non-reflecting boundary conditions can in general be expected to correctly recover all $\Lsiginc$ as they would be computed in the ground truth simulation at the layer of the non-reflecting boundary condition at all times.
Therefore, we more deeply analyze these differences and their potential sources in the following subsection and in Section~\ref{sec:discussion}.



\subsubsection{Results at each advection speed} \label{sec:GTvsNRBC_adv}

\change{
The primary goal of this section is to quantify the difference between each NRBC simulation and the corresponding GT simulation in a simple way.
To do so, we begin with the weighted mean squared difference defined in Eqn. \ref{eq:wMSD}, but include one minor change.
The simulations of advected spheromaks have only very minor departures from uniform advection $\vect{v} = v_{\rm ad} \hat{z}$ throughout the full ground truth simulation, due to diffusion across the grid.
If we were to include $\vect{v}$ in the wMSD computation it would dominate the metric because this uniformity would make all elements in $\Matrix{K}^{-1}$ involving a component of velocity pathologically large.
Therefore, we here choose to omit $\vect{v}$ from the weighted mean squared difference such that $\vect{N}^T = (\rho,\epsilon,B_x,B_y,B_z)$ and $\chi$ and $\zeta$ in the covariance matrix vary over $\{\rho,\epsilon,B_x,B_y,B_z\}$.
Later in this section we break wMSD down and examine the impact of the boundary on each variable individually, at which point we reintroduce each of the velocity components on its own.}

\change{The expectation value $\langle\cdots\rangle$ in Eqn. \ref{eq:wMSD} is again taken over the total number of cells $N = N_x N_y N_z$ in the simulations with non-reflecting boundary conditions.
$\Matrix{K}$ is computed separately for each ground truth simulation at a time $\tau_{ad} = 1/v_{ad}$, namely when the spheromak has advected one spatial unit.
Computing $\Matrix{K}$ at this time allows the variations in the ground truth simulation to develop due to diffusion from advecting over several tens of grid cells, while also keeping the majority of the spheromak inside the domain of overlap with the NRBC simulations to keep the variance and covariance of the magnetic field components physically meaningful.
$\langle G_\chi \rangle$ and $\langle G_\zeta \rangle$ in Eqn. \ref{eq:covariance}, as well as $\Matrix{K}$ as a whole, are computed over the region of the ground truth simulation with $\|B\|>10^{-10}$, while wMSD is computed over the full overlapping region of both simulations.
}

\begin{figure*}[!htbp]
	\centering
	\includegraphics[width=0.4\textwidth]{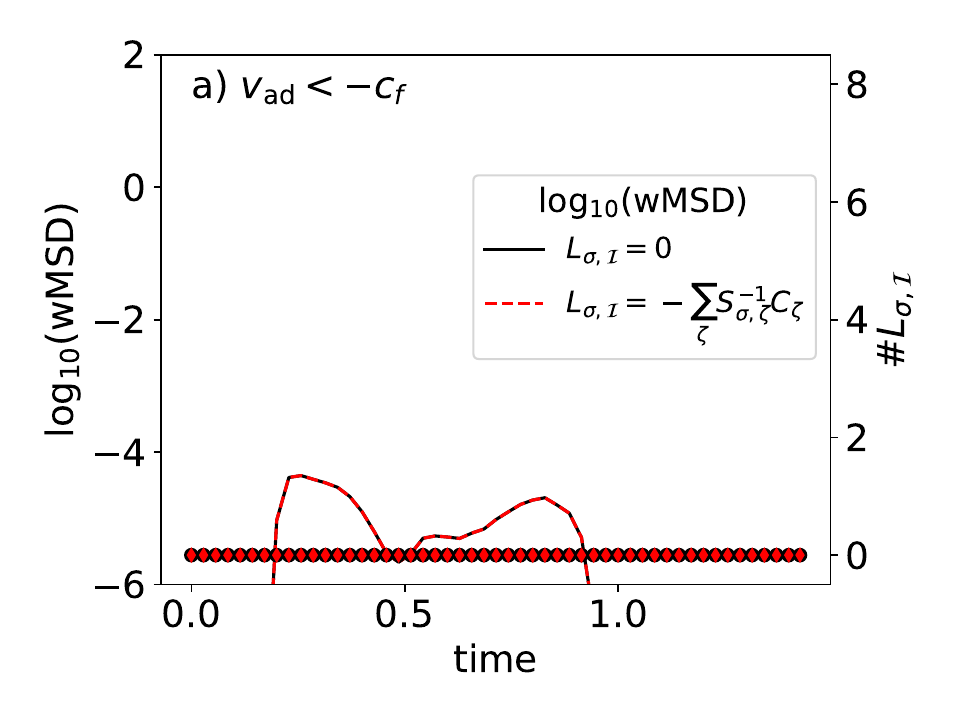}
	\includegraphics[width=0.4\textwidth]{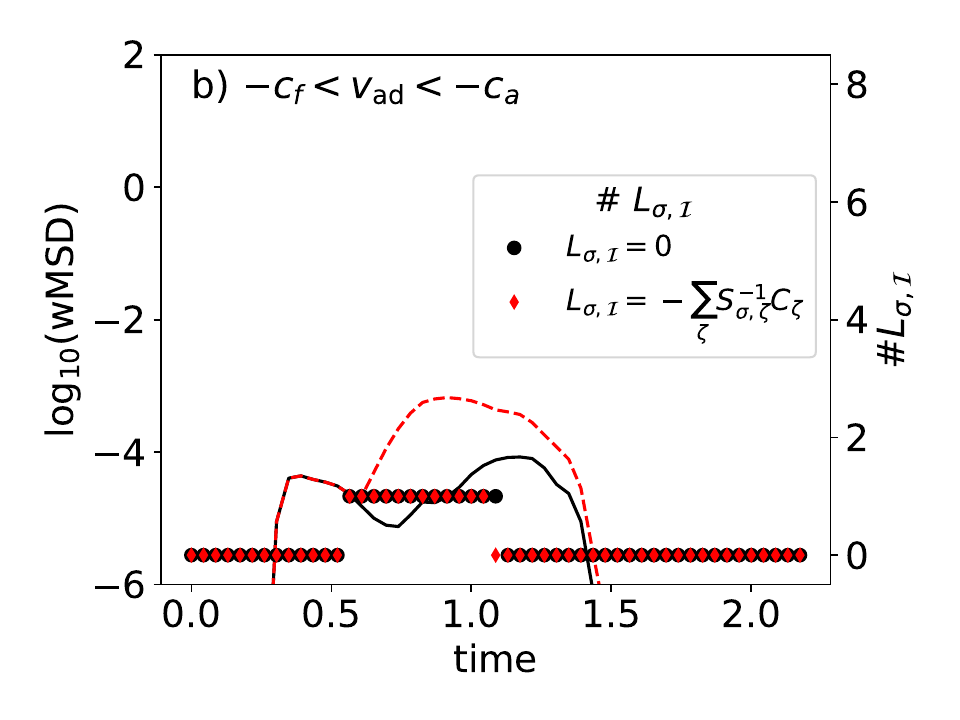}
	
	\includegraphics[width=0.4\textwidth]{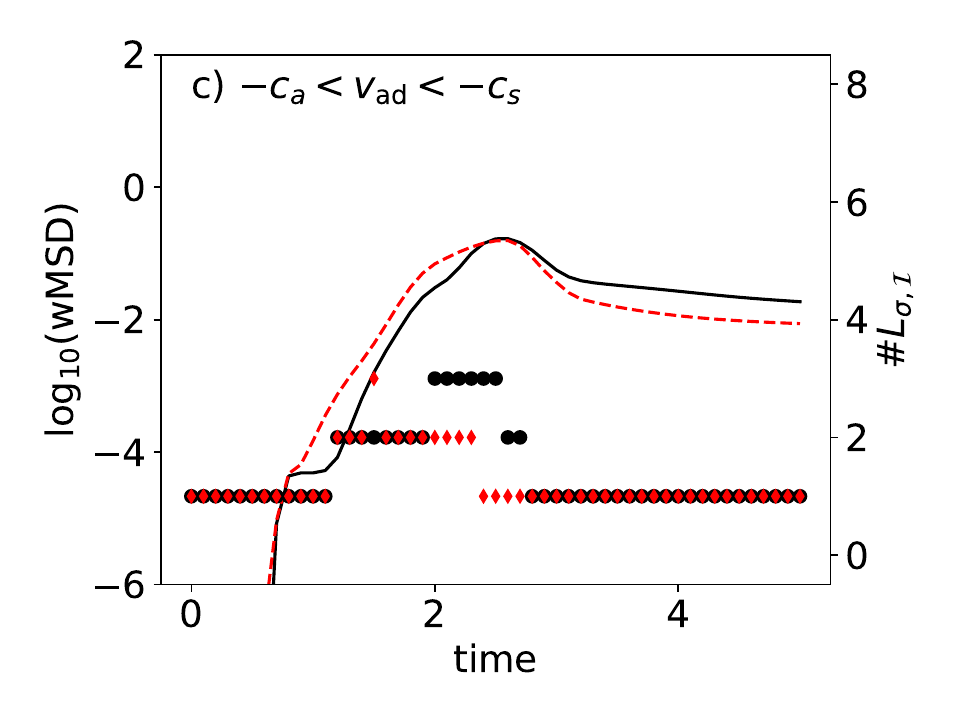}
	\includegraphics[width=0.4\textwidth]{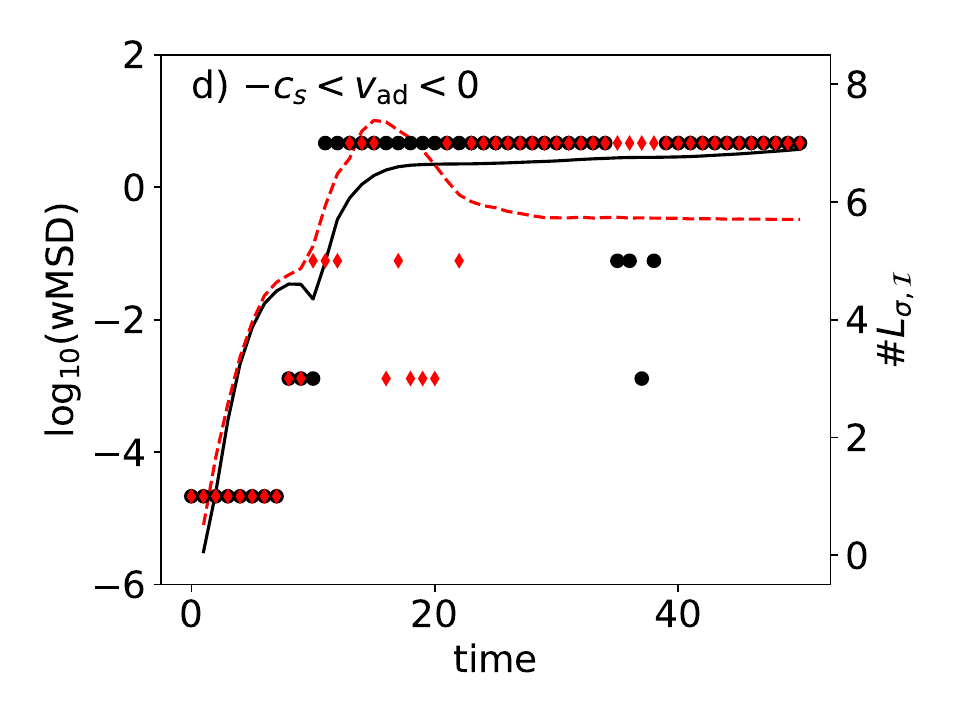}

	\caption{\change{Lines plot the wMSD for each simulation (left axis), and symbols plot the maximum number of incoming characteristics, $\# \Lsiginc$, within the boundary layer (right axis), as a function of time, for}  Case 1 (a),  Case 2 (b), Case 3 (c), and Case 4 (d). Solid black curves \change{and black circles correspond to} simulations with a Fixed NRBC (\change{labeled} $\Lsiginc = 0$) and dashed red curves \change{and red diamonds} to simulations with a Cancellation NRBC ($\Lsiginc = -\sum_\zeta S_{\sigma,\zeta}^{-1} C_\zeta$).  \change{Both legends refer to all four panels.}}
	\label{fig:mean_sq_error}
\end{figure*}

In Figure \ref{fig:mean_sq_error}, we plot the wMSD for each of the simulations advecting a spheromak through a non-reflecting boundary, where the difference is computed against the ground truth simulation with the same advection velocity.
In all four panels solid black lines denote $\rm wMSD$ for simulations with a Fixed NRBC (here $\Lsiginc = 0$) and dashed red lines denote $\rm wMSD$ for simulations with a Cancellation NRBC ($\Lsiginc = -\sum_\zeta S_{\sigma,\zeta}^{-1} C_\zeta$).
To augment these plots of the mean squared difference, we also plot the maximum number of incoming characteristic derivatives found in any cell in the boundary layer as a function of time, using the axis on the right.
For this portion of each panel black circles show the maximum number of incoming characteristics for simulations with a Fixed NRBC (here $\Lsiginc = 0$) and the red diamonds show this quantity for simulations with a Cancellation NRBC ($\Lsiginc = -\sum_\zeta S_{\sigma,\zeta}^{-1} C_\zeta$).
\change{Note that at many times, and especially for Cases 1 (panel a) and 2 (panel b), both the Fixed and Cancellation NRBCs produce the same maximum number of incoming characteristics for a given Case, so their symbols overlap.}

\begin{figure*}[!htbp]
	\centering
	\includegraphics[width=0.32\textwidth]{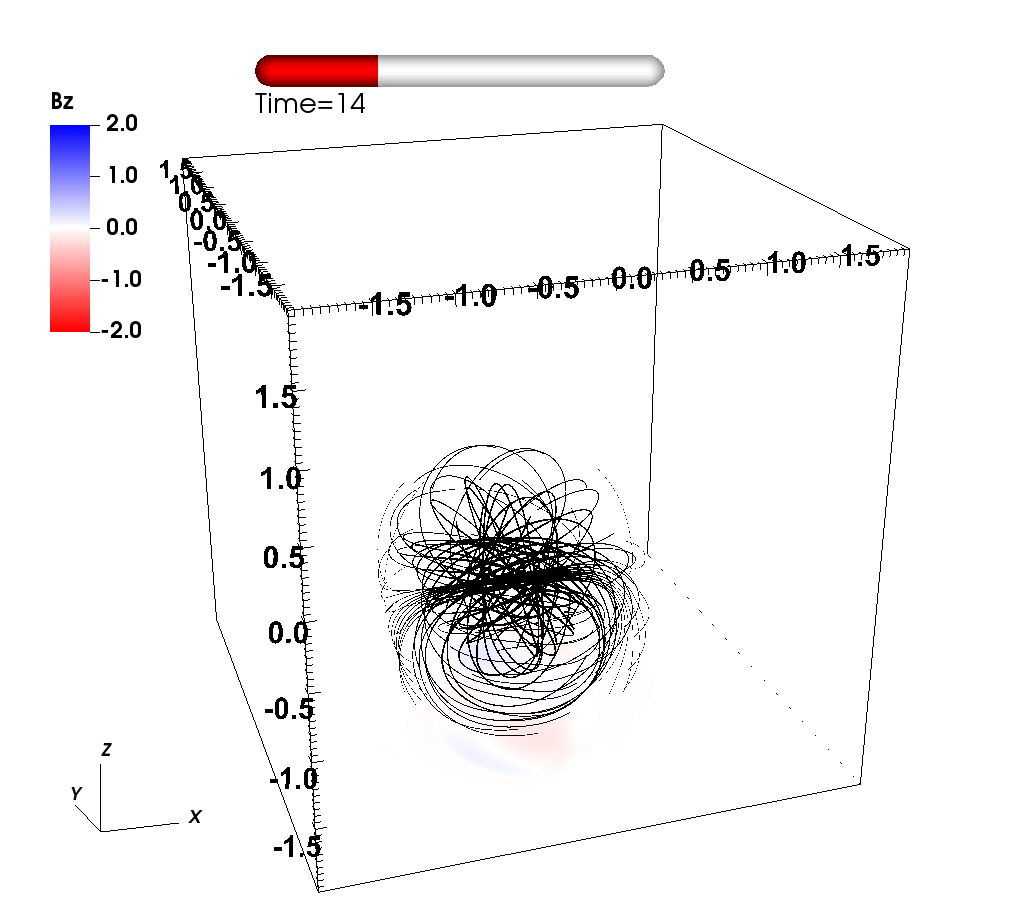}
	\includegraphics[width=0.32\textwidth]{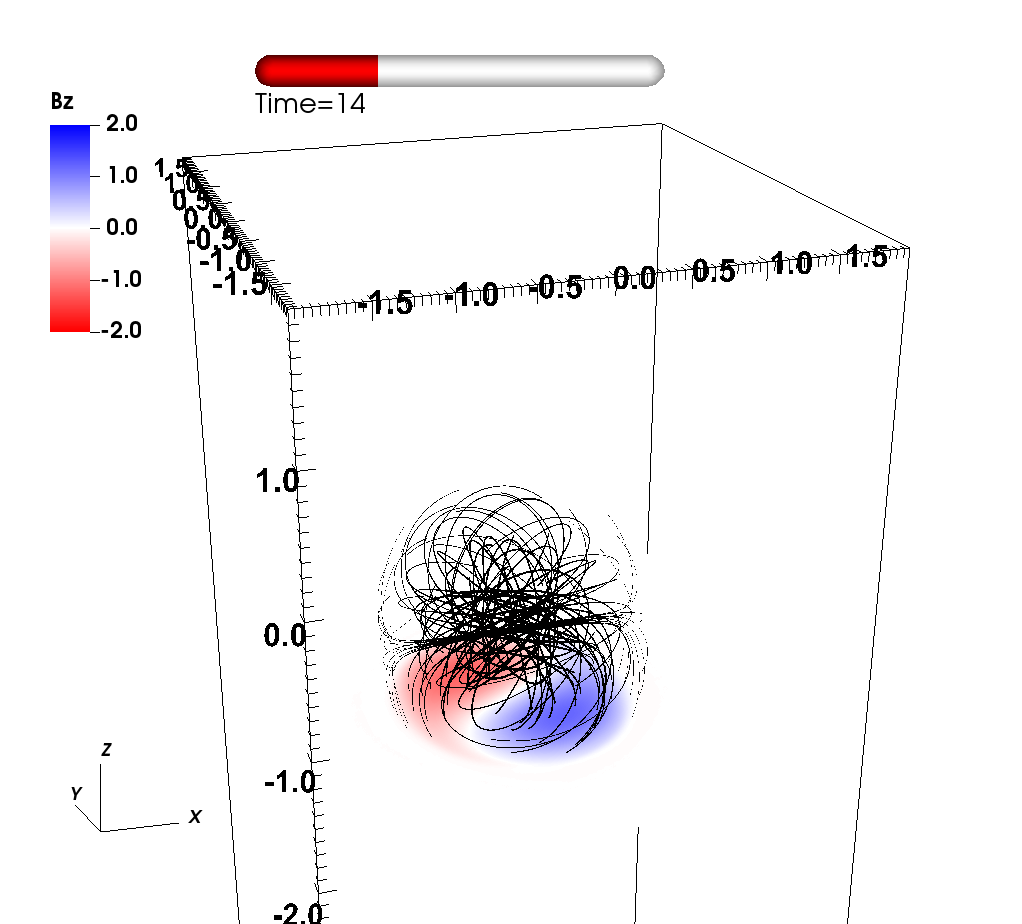}
	\includegraphics[width=0.32\textwidth]{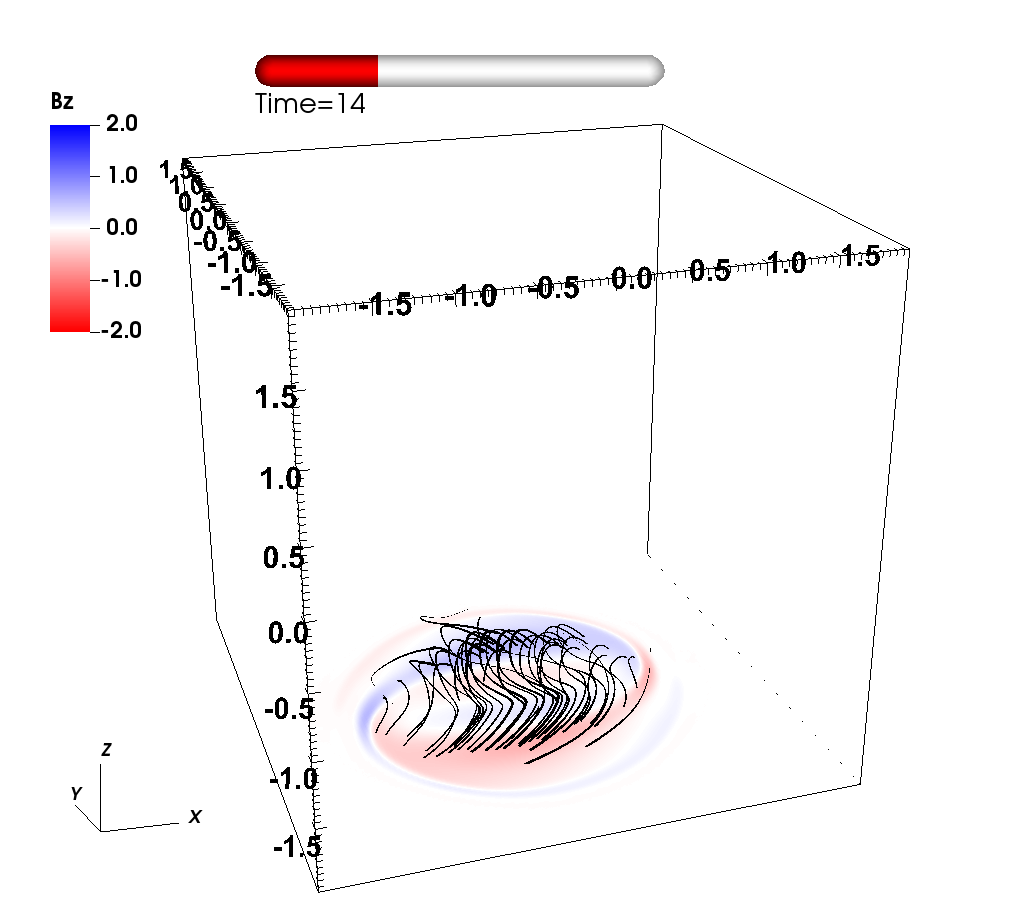}
	
	\caption{Magnetic field lines in the volume, and the normal component of the magnetic field, $B_z$, at the location of the non-reflecting boundary for the simulations with $-\cslow < v_{\rm ad}~(= -0.1) < 0$ (Case 4). 
    The simulation with a Fixed NRBC (here $\Lsiginc = 0$) is on the left and with a Cancellation NRBC ($\Lsiginc = -\sum_\zeta S_{\sigma,\zeta}^{-1} C_\zeta$) is on the right. The central panel shows the same quantities in the overlapping volume and on the same plane for the ground truth simulation. Magnetic field lines are initiated from $x,y$ positions which track the motion of the midplane of the spheromak as it advects and ultimately accelerates in the cases with NRBCs. An animated version of this Figure is available as Supplementary Video 5. The animation shows the spheromak in each simulation beginning by advecting from its initial position at $x=y=z=0$ toward the $z_{min}$ boundary condition. Once the lower edge of the spheromak encounters the NRBC in the simulations which have one, it becomes notably distorted and the motion of the spheromak deviates from constant speed advection. For the simulation with a Fixed NRBC (here $\Lsiginc = 0$) the spheromak bounces off the boundary condition twice before the animation ends. For the simulation with a Cancellation NRBC ($\Lsiginc = -\sum_\zeta S_{\sigma,\zeta}^{-1} C_\zeta$), the spheromak markedly accelerates, exiting the simulation domain in approximately half the time it would have taken under constant speed advection.}
	\label{fig:halfSnap_vslow_0}
\end{figure*}

Comparing the panels of Figure~\ref{fig:mean_sq_error} shows a clear trend where the slower the advection is, the more different the simulation with non-reflecting boundaries are from the ground truth.
This trend is independent of which implementation of non-reflecting boundaries is chosen, and is a direct result of the increasing number of incoming characteristic derivatives from a simulation in Case 1 ($v_{ad} < -\cfast$), which we can see from Table \ref{tab:incoming} is tuned to have no incoming characteristics, to a simulation in Case 4 ($-\cslow < v_{ad} < 0$), which we can see from Table \ref{tab:incoming} \change{is} tuned to have three incoming characteristics over the bulk of the spheromak volume.

In the former case, the two simulations with non-reflecting boundary conditions are identical, and the negligible differences between them and the ground truth simulation are due to numerical diffusion from coupling two MHD codes together at the boundary layer.
Additionally, it is important to note how extremely similar the two simulations are, as wMSD $<10^{-4}$ at all times for both types of non-reflecting boundary.
To give a sense of just how remarkably similar wMSD = $10^{-4}$ is, if the MHD variables were distributed as Gaussians and were uncorrelated, this would be equivalent to one variable being different by an average of one percent of a standard deviation, or all five variables being different by an average $0.2\%$ of a standard deviation.
This reinforces that our method has coupled simulations using fundamentally different methods for solving MHD without introducing any appreciable difference compared to a simulation using a single method for solving MHD.

Meanwhile, when we consider simulations in Case 4, snapshots of which are shown in Figure \ref{fig:halfSnap_vslow_0}, the evolutions of both simulations are visibly dissimilar for all times after the spheromak encounters the non-reflecting boundary and, as we discuss in more depth below, these simulations produce drastically wrong numbers of incoming characteristics.
In this slowest advection case, the two versions of NRBCs produce wildly different, and both wildly incorrect, behaviors.
As can be seen from Figure \ref{fig:halfSnap_vslow_0} and its animation in Supplemental Video 5, the simulation with a Cancellation NRBC ($\Lsiginc = -\sum_\zeta S_{\sigma,\zeta}^{-1} C_\zeta$) pulls the spheromak through the boundary faster than the advection, while part of the spheromak bounces off the ``non-reflecting'' boundary when a Fixed NRBC (in this case $\Lsiginc = 0$) is used.

The largest jump in wMSD occurs between Case 2 and Case 3.
This is the step between super-Alfv\`enic outflow through the NRBC and predominantly sub-Alfv\`enic outflow.
For simulations with super-Alfv\`enic outflow, namely Cases 1 and 2, it is not possible to discern any differences from the ground truth by eye.
\change{wMSD remains correspondingly low in these cases.  In contrast,}
for simulations with sub-Alfv\`enic inflow, namely Cases 3 and 4, the impact of the misrepresented incoming characteristic derivatives is severe enough to cause by-eye differences and a wMSD which becomes of order unity or higher.
\change{At that level, wMSD} indicates typical differences between simulations with NRBCs and the ground truth on the order of the scale of the distribution of primitive variables in the ground truth spheromak itself.

Beyond this, comparing the number of incoming characteristic derivatives in the simulations in Cases 3 and 4 with what would be expected in these cases in Table \ref{tab:incoming} shows that a larger number of characteristic derivatives end up being incoming than what the simulation is tuned to have.
This highlights that the impact of the non-reflecting boundary can be severe enough that the characteristic speeds and the local velocity normal to the boundary are both altered, and, in tandem, the number of characteristic derivatives the boundary sets can increase in a positive feedback loop.
In fact, close examination of the simulations in Case 2 shows that even in this case the impact of the non-reflecting boundary is sufficient to make simulations with Fixed (here $\Lsiginc = 0$) and Cancellation ($\Lsiginc = -\sum_\zeta S_{\sigma,\zeta}^{-1} C_\zeta$) NRBCs have different numbers of incoming characteristic derivatives from one another for one of the outputs near the end of the time when the spheromak is passing through the non-reflecting boundary.

\begin{figure*}
	\centering
	\includegraphics[width=0.35\textwidth]{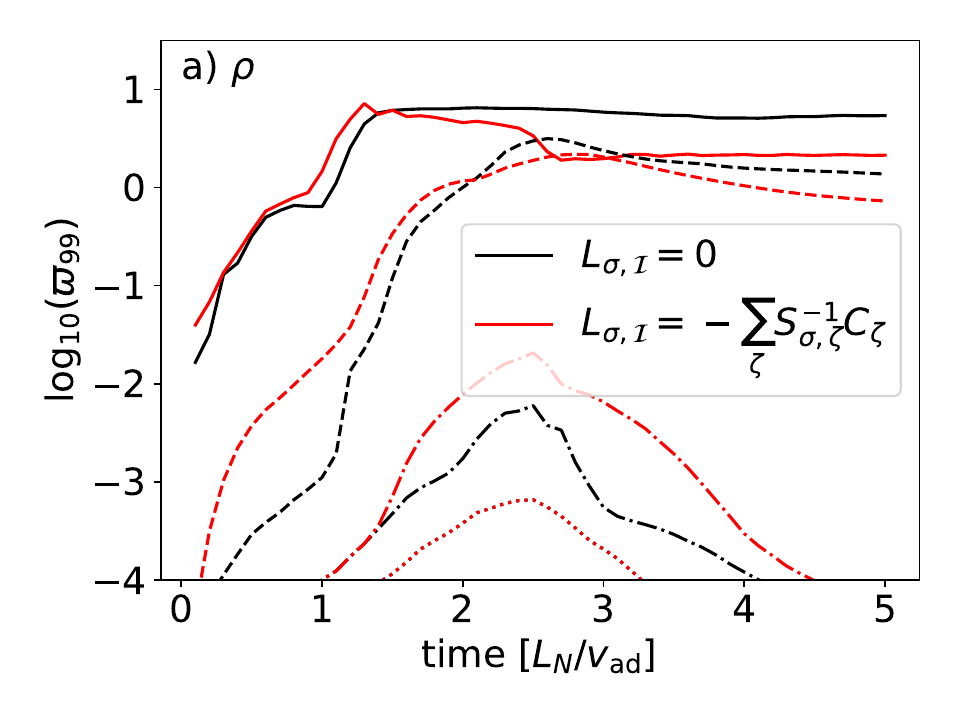}
	\includegraphics[width=0.35\textwidth]{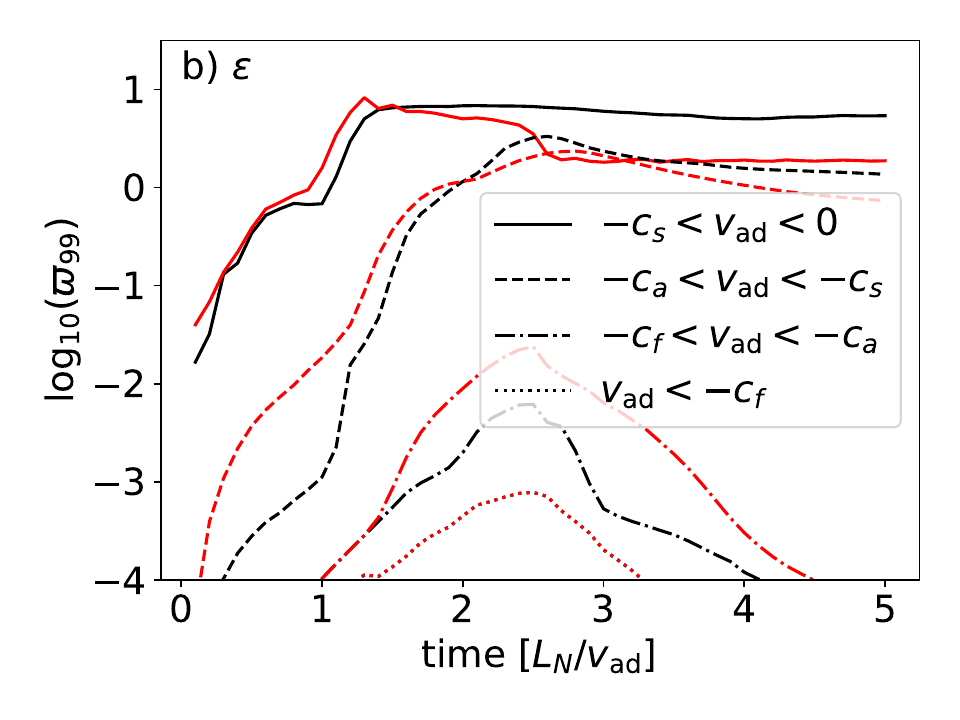}
	
	\includegraphics[width=0.35\textwidth]{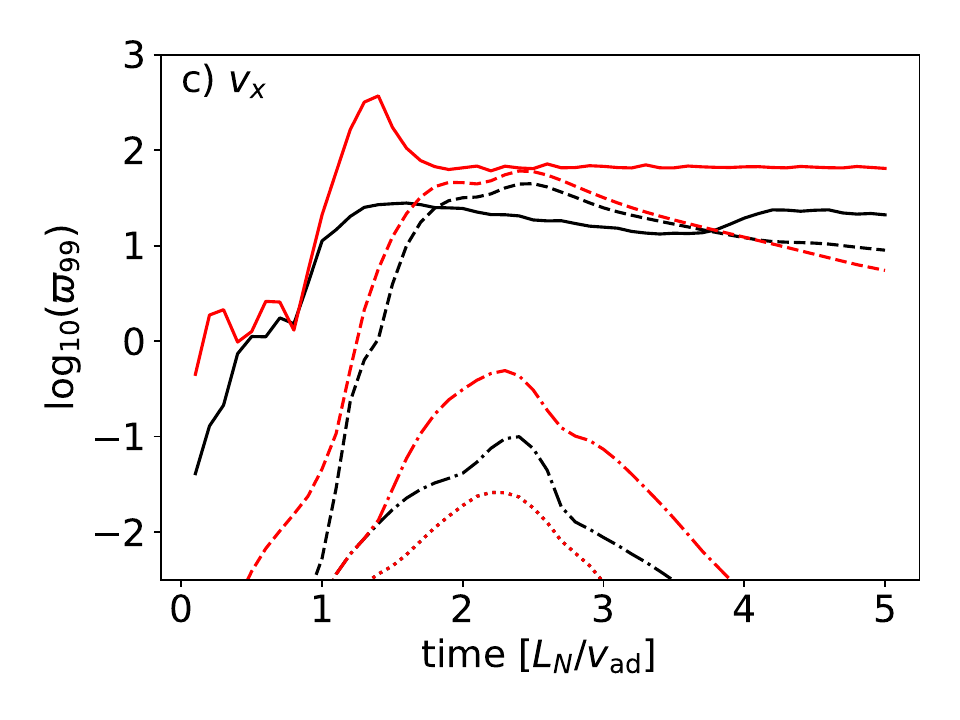}
	\includegraphics[width=0.35\textwidth]{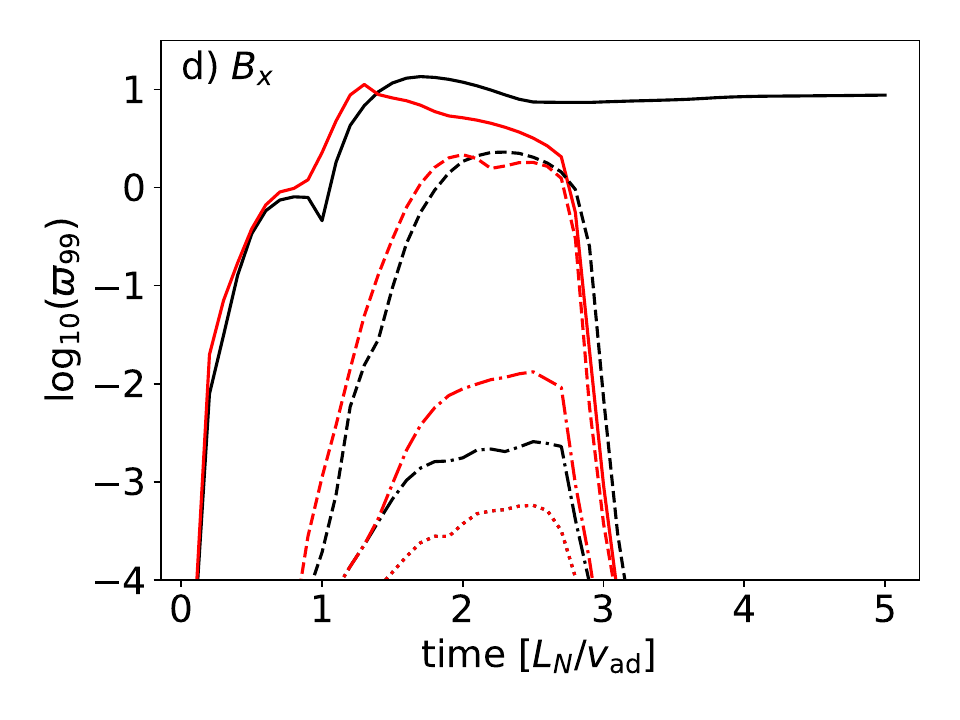}

 	\includegraphics[width=0.35\textwidth]{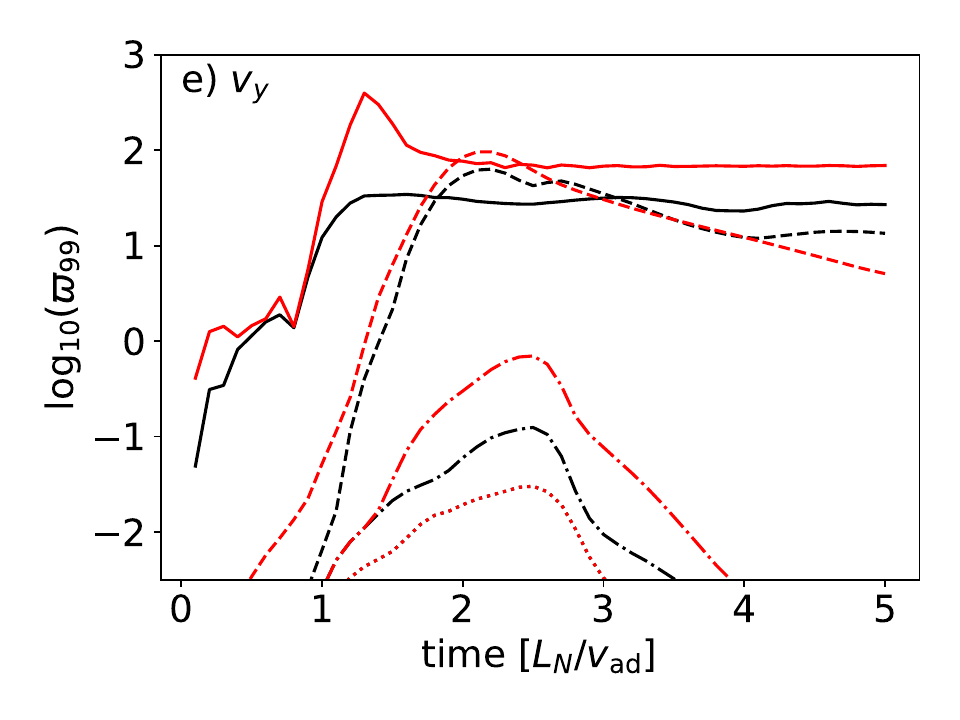}
	\includegraphics[width=0.35\textwidth]{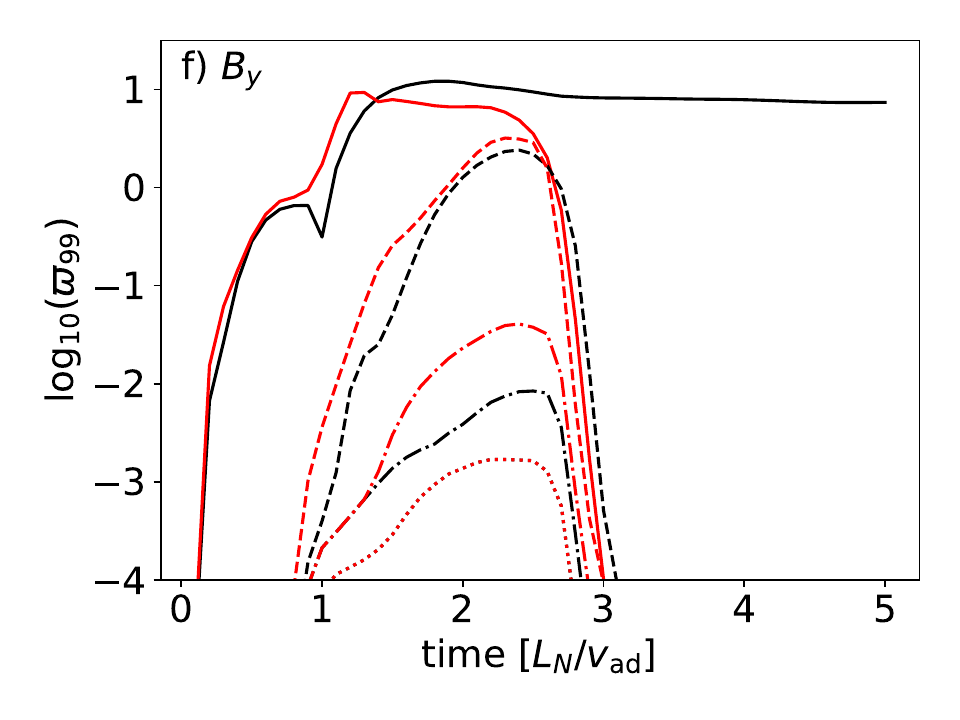}

 	\includegraphics[width=0.35\textwidth]{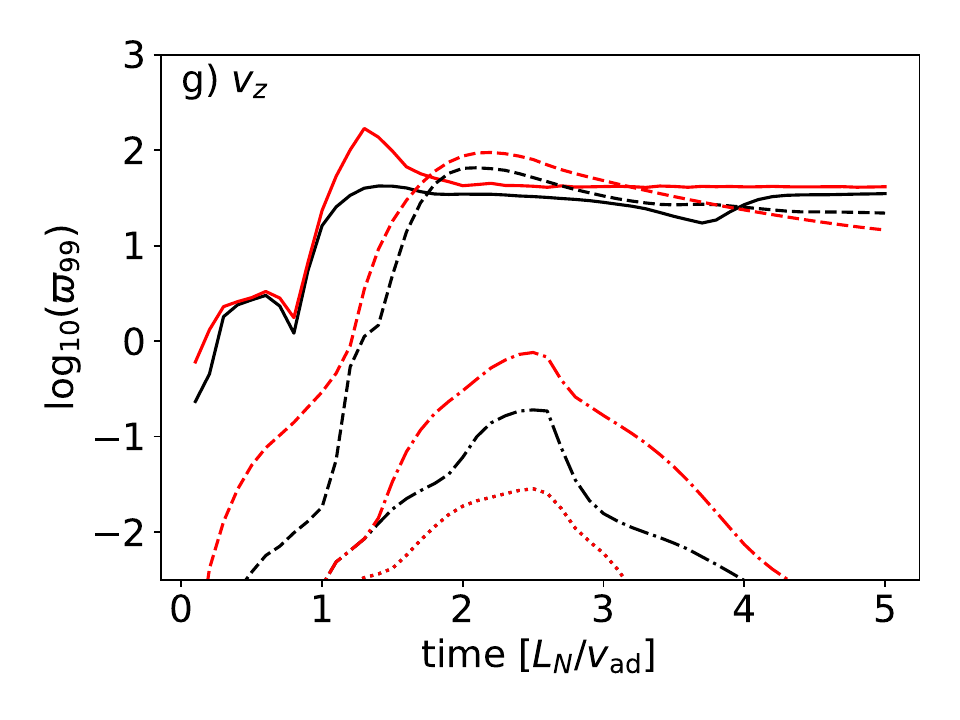}
	\includegraphics[width=0.35\textwidth]{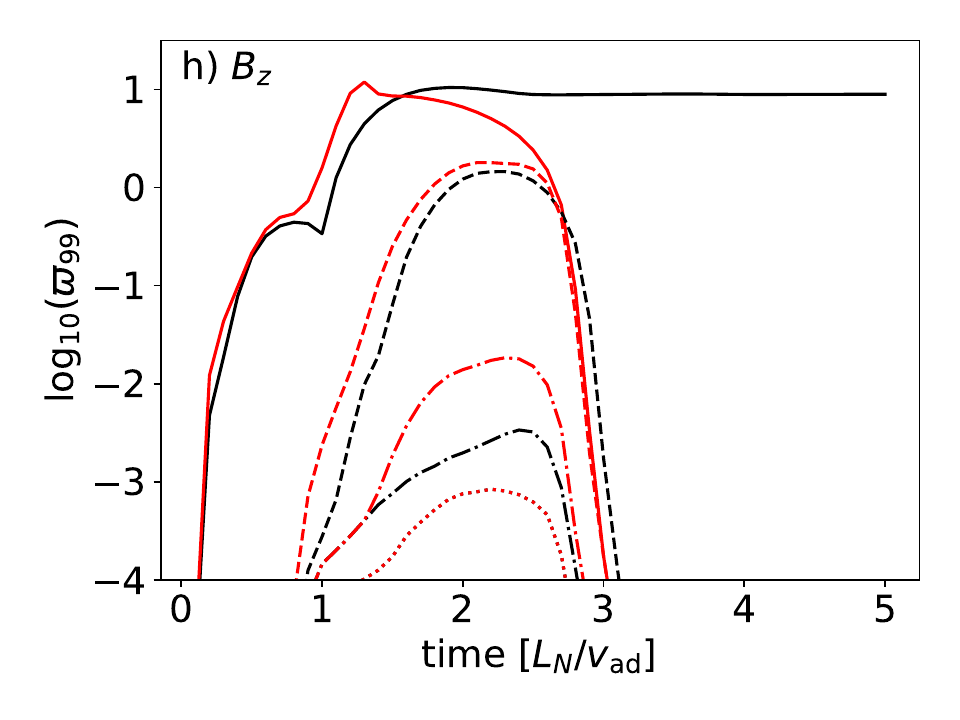}
		
	\caption{Threshold $\varpi_{99}$ for density (a), energy (b), $v_x$ (c), $v_y$ (e), $v_z$ (g), $B_x$ (d), $B_y$ (f), and $B_z$ (h). In each panel, black curves denote simulations with a Fixed NRBC (here $\Lsiginc = 0$) and red curves simulations with a Cancellation NRBC ($\Lsiginc = -\sum_\zeta S_{\sigma,\zeta}^{-1} C_\zeta$). The four cases are represented by different line styles with Case 1 dotted, Case 2 dash-dotted, Case 3 dashed, and Case 4 solid. As was the case in Figure \ref{fig:mean_sq_error}, the curves for Case 1 lie directly on top of one another.}
	\label{fig:sig99}
\end{figure*}

\subsubsection{Behavior of each variable} \label{sec:individualwMSD}
To examine the impact of the boundary condition on each variable individually, we split up the wMSD metric and calculate the quantity $\varpi_{99}(\zeta)$, which is the 99th percentile of the difference distribution for variable $U_\zeta$.  That is, 99 percent of the cells in a simulation with a NRBC have a weighted difference less than $\varpi_{99}(\zeta)$: 
\begin{equation}
    \varpi_{99}(\zeta) \geq \frac{|U_\zeta - G_\zeta|}{\varpi_{\zeta,GT}}\;,
\end{equation}
where $U_\zeta$ is one of the primitive MHD variables in the full MHD state vector (\change{i.e.,} including velocity \change{once again}) for the simulation with a non-reflecting boundary, $G_{\zeta}$ is the same variable in the ground truth simulation, and $\varpi_{\zeta,GT} \equiv \sqrt{\langle(G_\zeta - \langle G_\zeta\rangle)^2\rangle}$ is the standard deviation of $G_{\zeta}$ which, like $\Matrix{K}$, is computed over the region of the ground truth simulation with $\|B\|>10^{-10}$.
This quantity is plotted for all eight primitive variables and in all cases for simulations with non-reflecting boundaries in Figure \ref{fig:sig99}.
For all panels of Fig. \ref{fig:sig99} time is measured in unit length over advection speed, $L_N/v_{\rm ad}$, to synchronize the simulations with different durations and advection speeds.
Note that $\varpi_{99}$ for the components of $\vect{v}$ is between one and two orders of magnitude higher than the other MHD variables.
As mentioned previously, this is because the velocity in the ground truth simulation is nearly constant in time and uniform in space, so the normalizing standard deviation is much smaller than for the other variables, and not necessarily because the difference in velocity is intrinsically much larger.
This is the reason we omitted $\vect{v}$ from wMSD.

With this in mind, comparing the panels with one another points out that the final state of some variables can be very well represented even when the evolution of the simulations overall are dramatically different.
For instance, the simulation in Case 4 (solid) with a Cancellation NRBC ($\Lsiginc = -\sum_\zeta S_{\sigma,\zeta}^{-1} C_\zeta$) (red) ends up with remarkably small differences in the magnetic field.
This is because this simulation, like the ground truth simulation, ends with the spheromak having exited the volume of interest even though the way in which this occurs is drastically different.
Meanwhile, the simulation in Case 4 with a Fixed NRBC (in this case $\Lsiginc = 0$) is the only one which does not go to low error in magnetic field by the end of the simulation, as it is the only one that does not clear the magnetic structure (see the solid black curves).
Additionally considering the errors in $\rho$, $\epsilon$, and the components of $\vect{v}$ reinforces that neither simulation in Case 4 is actually adequately reproducing the ground truth, which in turn underlines the importance of detailed comparisons of all MHD quantities in addition to holistic properties like wMSD which may mask some interesting and important differences between individual simulations.

Within each of the panels, as was the case in Figure \ref{fig:mean_sq_error}, there is again a clear trend of decreasing quality of agreement with increasing number of incoming characteristic derivatives as the advection speed is lowered.
As mentioned above, there is a particularly notable jump in how well the simulations with non-reflecting boundaries do at reproducing the ground truth when we go from Case 2 to Case 3 when advection switches from super- to predominantly sub-Alfv\`enic, as this results in $\varpi_{99}$ increasing by around two orders of magnitude for every variable.
Comparing with Figure \ref{fig:mean_sq_error}, going from Case 2 to Case 3 was also when we began getting a different number of incoming characteristic derivatives than the simulations were tuned to have, emphasizing the compounding effect of each additional incoming characteristic in allowing simulations to diverge in behavior.

As alluded to in the introduction to this section, even the initially alarming result of a spheromak bouncing off a nominally ``non-reflecting'' boundary is actually the non-reflecting boundary condition operating as it is mathematically designed to.
This draws attention to an important difference between what non-reflecting boundaries are designed to do and what they sometimes are asked and expected to do.
In the following section we will discuss both why the non-reflecting boundary behaves as it does for these test cases in more detail, as well as discussing how one could formulate a boundary condition that behaves more like we would wish in this case.

\section{Discussion}\label{sec:discussion}

To dig further into why the simulations we present with non-reflecting boundary conditions do not evolve like the corresponding ground truth simulations, it is instructive to examine the MHD equations, Eqn.~P1.1, and especially how they are formulated in terms of characteristic derivatives in Eqn.~P1.18.
The problem we are considering, that of an advected spheromak, is a force-balanced setup.
This means that the gradient of gas pressure and the Lorentz force exactly cancel out with one another in the momentum equation.
Moreover, because this setup has uniform velocity, $(\vect{v}\cdot\nabla)\vect{v} = \vect{0}$.
Taken together, this means the problem we are considering has $\partial_t \vect{v} = \vect{0}$.
Clearly this is not preserved in the simulations we present with non-reflecting boundary conditions, most obviously when $-\cslow < v_{\rm ad} < 0$, so the goal of this section is to elucidate why this is.

To simplify the discussion somewhat, let us first consider a steady-state, force-free, and zero velocity scenario.
In such a scenario, we again have the cancellation of the gradient of gas pressure and the Lorentz force, and now $\vect{v} = \vect{0}$ such that $\partial_t \vect{v}$ remains zero.
With the exception of the trivial case of uniform plasma, the spatial gradients present in such a set up mean that $\vect{L} \neq \vect{0}$ and $\vect{C} \neq \vect{0}$.

If we wanted to achieve $\partial_t\vect{v}=\vect{0}$ in terms of the characteristic derivatives in the momentum equation (Eqns.~\ref{eq:vx_char}, \ref{eq:vy_char}, and \ref{eq:vz_char}), we might be tempted to simply set $\Lsiginc$ to be directly equal to the associated $\Lsigout$, namely $L_4 = L_3$, $L_6 = L_5$, and $L_8 = L_7$.
Because there are no inhomogeneous terms, $\vect{C}$ is built up purely from a comparable inter-combination of characteristic derivatives in the $x$ and $y$ directions (see \citetalias{TarKee24}, Appendices B and C), so we could assume the same pairwise cancellation occurs in the transverse directions resulting in $C_{v_{x}} = C_{v_{y}} = C_{v_z} = 0$.
This then indeed does satisfy $\partial_t\vect{v} =\vect{0}$.

However, because $\vect{v}=\vect{0}$ for this simplified scenario, the state of the system that we are trying to maintain with the characteristics must also have $\partial_t \rho = 0$, $\partial_t \epsilon = 0$, and $\partial_t \vect{B} = \vect{0}$, as can be seen easily from Eqn.~P1.1.
From Eqn.~P1.18, the simplest solution providing no time evolution\footnote{The time evolution of $B_z$ in Eqn.~\ref{eq:Bz_char} does not provide any constraints here as it is governed by only $L_1$ in the boundary-normal direction, which is zero because $v_z=0$.} of $\rho$ (Eqn.~\ref{eq:rho_char}), $\epsilon$ (Eqn.~\ref{eq:eps_char}), $B_x$ (Eqn.~\ref{eq:Bx_char}), and $B_y$ (Eqn.~\ref{eq:By_char}) is that $L_2 =0$ and each of the remaining $\Lsiginc$ has equal magnitude and opposite sign to the associated $\Lsigout$, namely $L_4 = -L_3$, $L_6 = -L_5$, and $L_8 = -L_7$.
As in the prior paragraph, due to the structure of the $\vect{C}$ and the transverse terms, imposing the same cancellation in the transverse directions would result in $C_\rho = C_\epsilon = C_{B_x} = C_{B_y} = C_{B_z} = 0$ which would indeed result in no time evolution of mass density, energy density, and magnetic field.

Since $\vect{L} \neq 0$, the two simple approaches one might want to attempt to set $L_\sigma$ are not mutually compatible with one another.
Moreover, both methods result in non-zero elements of $\vect{C}$ for the other equations they are not considering.
Specifically, the first method which satisfies $\partial_t\vect{v} = \vect{0}$ in a simple way generates non-zero $C_\rho$, $C_\epsilon$, $C_{B_x}$, $C_{B_y}$, and $C_{B_z}$, while the second method satisfying $\partial_t \rho = 0$, $\partial_t \epsilon = 0$, and $\partial_t \vect{B} = \vect{0}$ generates non-zero $C_{v_{x}}$, $C_{v_{y}}$, and $C_{v_z}$.
This points us toward the true solution; either the momentum equations or everything except the momentum equations must be satisfied by a \emph{non-zero} weighted sum of the components of $\vect{L}$ and $\vect{C}$.

Additionally examining Eqn.~P1.14 resolves the conflicting requirements set forth by these two simple methods for determining each $L_\sigma$.
$v_z = 0$ and $\partial_z \vect{v} = 0$ in this simplified case, which reduces Eqns.~\ref{eqn:li3} and \ref{eqn:li4} to $L_4 = - L_3$, Eqns.~\ref{eqn:li5} and \ref{eqn:li6} to $L_6 = - L_5$, and Eqns.~\ref{eqn:li7} and \ref{eqn:li8} to $L_8 = - L_7$.
This means that for this case, the simple solution for time evolution of $\rho$, $\epsilon$, and $\vect{B}$ is the correct one.
As such, it is the momentum equations which must be satisfied by the weighted sum of $\Lsiginc$ and $\Lsigout$ with the associated also non-zero transverse terms $C_{v_x}$, $C_{v_y}$, and $C_{v_z}$.

In general, neither method of defining a non-reflecting boundary correctly recovers even this simplified case.
A Cancellation NRBC ($\Lsiginc = -\sum_\sigma S^{-1}_{\sigma,\zeta} C_\zeta$) attempts to handle some of the 3D balance of forces by setting up incoming characteristic derivatives to cancel subsets of the characteristic derivatives transverse to the boundary, but does not account for the contribution of $\Lsigout$ to the force balance.
A Fixed NRBC accounts for the 3D balance marginally better, as the incoming characteristics are calculated from the initial configuration, and therefore will accommodate the original force balance.
However, even the most minute adjustment in the plasma properties in the simulation near the boundary means that the force balance will be adjusted, and the incoming characteristics which were correct for the initial conditions will not preserve the required 3D cancellation of forces for any slight deviation.
Put a different way, in general for the non-reflecting boundary conditions that we consider here, the external universe exerts a force at the simulation boundary. 


To investigate how the non-reflecting boundary condition specifically impacts the case of an advecting spheromak, it is useful to focus on terms representing the Lorentz force in the characteristic derivatives, as the contribution of gas pressure to the force balance is negligible in the spheromak interior.
Doing so, we see that terms which contribute to the the Lorentz force (terms with $\beta_x B_y^\prime$, $\beta_y B_x^\prime$, $\beta_x B_x^\prime$, and $\beta_y B_y^\prime$) appear in $L_3$, $L_4$, $L_5$, $L_6$, $L_7$, and $L_8$.
This means that appropriately representing the balanced Lorentz force in the spheromak interior requires appropriately representing six of the eight characteristic derivatives in all three directions (i.e., 18 characteristic derivatives in total).
Once some portion of the spheromak has advected through the non-reflecting boundary, depending on the advection speed, a subset of the boundary-normal Alfv\'en, magnetosonic slow, and magnetosonic fast modes that carry information about the Lorentz force are incorrectly represented by the way the boundary condition sets the incoming subset of $L_3 \cdots L_8$.
Therefore, it is ultimately unsurprising that the simulations with non-reflecting boundary conditions go through substantially different evolutions than the larger, ground truth simulations.

Based on this information, initially one may be tempted to simply move the boundaries of the simulation far away from the region of interest and then to discard the evolution of the simulation near the boundaries as ``contaminated'' by the boundary condition.
The results shown in Sec.~\ref{sec:GTvsNRBC} demonstrate the shortcomings of such a choice, however.
Once the boundary condition fails to reproduce the impact of the external universe, that information begins propagating away from the boundary at all the incoming characteristic speeds simultaneously. 
This means that successfully removing the impact of the boundary conditions from the simulation volume of interest requires that the boundaries must all be a distance $v_\perp + \cfast$ times the total simulation run time away from any region being analyzed, where $v_\perp$ is the velocity normal to the boundary.
If one wishes to push this as far as possible, it may be feasible to relax this constraint and continue trusting the simulation contents until a time equal to the distance to the boundary condition divided by $v_\perp + \cfast$ after the boundary condition fails to reproduce the impact of the external universe.
Identifying when such a failure occurs without a comparison ground truth simulation is, however, fraught at best, and more likely simply not possible for arbitrarily complex simulations.

As a brief aside, the same type of issue exists for simulations which are driven incorrectly.
Here ``driven incorrectly'' may be read as driven using poorly constrained or internally inconsistent data series (e.g., a series of velocities and magnetic field configurations which violate the induction equation), or as driven in a way that is not magnetohydrodynamically consistent with the contents of the simulation.
In either case, information about the incorrect driving layer will propagate into the simulation, contaminating the contents of the simulation volume.
As it is the information which is propagating away from the driving layer which is most interesting in a driven case, this issue is particularly pronounced in driven simulations.
Removing the possibility for the boundary driving to be inconsistent with the simulation interior, and using a minimization method to solve for the best incoming characteristics in the case of driving with bad data, are major successes of the method we have described in \citetalias{TarKee24}.

Therefore, a natural next step would be to want some set of ``minimum impact'' boundary conditions which do what we wish non-reflecting boundary conditions would do, namely they would allow not only outgoing waves, but also complex outgoing structures to pass through the boundary condition with minimal impact to the simulation interior.
Developing such a boundary condition is a future goal of the current research line, but is beyond the scope of this paper.
Nevertheless, a brief enumeration of the properties such a boundary condition would need to reproduce is illustrative and worth undertaking before we conclude.
Specifically, such a ``minimum impact'' boundary condition would need to allow waves \emph{and} topologically complex structures to pass out of the boundary at a rate only determined by the dynamics of the simulation interior.
Ideally, this would mean that a feature passing through the boundary would decelerate, accelerate, or coast at constant velocity as prescribed by the dynamics established in the simulation interior without the boundary condition contributing to the dynamics.
This is, however, already a highly non-trivial ask as it requires the boundary condition to react and retain the dynamics established in the simulation interior in a way that preserves the back reaction of the un-simulated external universe on the simulation interior.

The \code{Bifrost} \citep{GudCar11} boundary conditions provide an interesting suggestion of a stepping stone between the non-reflecting boundary conditions that we have examined in this paper and our goal of a ``minimum impact'' boundary condition.
As we mentioned in a footnote earlier in this paper, the \code{Bifrost} boundary conditions use $\Lsiginc = -\sum_\sigma S^{-1}_{\sigma,\zeta} C_\zeta|_{\vect{v}=\vect{0}}$, namely the zero velocity limit of our boundary condition $\Lsiginc = -\sum_\sigma S^{-1}_{\sigma,\zeta} C_\zeta$.
Constructing $C_\zeta|_{\vect{v}=\vect{0}}$ (see \citetalias{TarKee24}, Appendix D) only retains non-zero $C_{v_x}$, $C_{v_y}$, and $C_{v_z}$, each of which represents forces tangential to the boundary.
Therefore, the \code{Bifrost} boundary conditions use $\Lsiginc$ to cancel out transverse forces, while leaving transverse advection terms alone.
This is intriguingly close to our desired ``minimal impact'' boundary condition in that it allows advection initiated in the simulation interior to persist while removing forces at the simulation boundary.
However, like our boundary condition with $\Lsiginc = -\sum_\sigma S^{-1}_{\sigma,\zeta} C_\zeta$, the \code{Bifrost} boundary conditions do not account for the contribution of $\Lsigout$ to the force balance, and they do not treat any of the non-advective contributions to the temporal evolution of $\rho$, $\epsilon$, or $\vect{B}$ which depend on non-zero velocity, for instance compressive heat and expansion cooling in the energy equation, Eqn. \ref{eq:eps_char}, through $(P/\rho)\nabla\cdot\vect{v}$.
As such, even these more sophisticated boundary conditions fall short of our goal of generalized ``minimal impact'' boundary conditions.

\section{Conclusions}\label{sec:conclusions}

We have presented our implementation of non-reflecting boundary conditions built on a characteristics-based decomposition of the MHD equations.
This implementation is part of our ongoing implementation of characteristics-based boundary conditions for data-driven simulations, more results of which can be found in \citetalias{TarKee24}.
The first key result of the current paper is that we have successfully numerically implemented two varieties of non-reflecting boundary conditions into the Lagrangian remap code \lare{}.
We demonstrated that our implementations are successful by presenting a test where perturbations propagate through the boundary without reflection.
The second key result of this paper is that we have found that both standard implementations of non-reflecting boundary conditions yield unintended consequences when presented with complex MHD features, and moreover that these unintended consequences render non-reflecting boundary conditions as they are typically defined a poor choice for our goal of developing ``minimal impact'' boundary conditions.
Specifically, the implementations of non-reflecting boundary conditions we present in this paper yielded physically alarming but mathematically correct results like an advecting spheromak bouncing off a non-reflecting boundary condition with a Fixed NRBC (here $\Lsiginc = 0$) or getting sucked through a non-reflecting boundary with a Cancellation NRBC ($\Lsiginc = -\sum_\zeta S_{\sigma,\zeta}^{-1} C_\zeta$) instead of continuing to advect at a constant velocity as it does in a larger ground truth simulation.

Ultimately these departures of simulations with non-reflecting boundaries from comparable ground truth simulations come down to the intrinsic loss of information inherent to a non-reflecting boundary condition.
A non-reflecting boundary condition attempts to either remove the incoming characteristic derivatives entirely (Fixed, in this case $\Lsiginc = 0$), or to set them entirely in terms of characteristic derivatives transverse to the boundary and inhomogeneous terms (Cancellation, $\Lsiginc = -\sum_\zeta S_{\sigma,\zeta}^{-1} C_\zeta$).
Neither of these methods correctly represents the behavior of the external universe in general, as information from the external universe must be allowed to propagate into the simulation volume.
Moreover, the differences introduced by the non-reflecting boundary are not confined to regions close to the boundaries, as information about any change in the MHD state at the boundary propagates into the volume at all the characteristic speeds of MHD.
In these cases the non-reflecting boundary condition has done exactly what it is \emph{constructed} to do, even though this is not what one may \emph{want} it to do.

Motivated by the results here, future work will focus on the development and implementation of more generally usable boundary conditions for MHD simulations.
As discussed in Section~\ref{sec:discussion}, an ideal outcome of this future work would be the development of a ``minimum impact'' boundary condition which would behave in the way a non-reflecting boundary condition is often hoped to, namely that any features propagating toward the boundary would continue propagating as though the simulation boundary were not there.
This is particularly important for the top boundary of simulations of coronal mass ejections, as a common goal of such simulations is to distinguish between scenarios which lead to eruptions and those where the dynamics are confined.
If either non-reflecting boundary discussed here were to be used, it could substantially bias the results by leading to eruptions being easier, in the case of $\Lsiginc = -\sum_\zeta \Matrix{S}_{\sigma, \zeta}^{-1} C_\zeta$, or harder, in the case of $\Lsiginc = 0$, to trigger.
Therefore, the development of such a ``minimum impact'' boundary is a core goal of this research line and one which we hope to present in a future paper in this series.

As a final note, as briefly touched on at various points in this paper and as discussed in depth in \citetalias{TarKee24}, the non-reflecting boundary conditions which we have investigated here and the ``minimum impact'' boundary conditions we hope to develop in the future represent only a small subset of the cases where characteristic-based boundary conditions are useful.
Future work in this research line will therefore also pursue the use of characteristics-based boundary conditions for other classes of problems.
\change{One application of particular interest to us is of course further developing} the data-driven boundaries set out in \citetalias{TarKee24}.
\change{
Another is the use of characteristics-based boundary conditions for code coupling.
In brief, such an application builds on the power of an implementation of characteristics-based boundary conditions in a code which is itself \emph{not} based on the method of characteristics.
This means that this boundary condition method could be well suited to coupling simulations at different scales or in different sub-volumes where the dominant physical effects are not necessarily the same.
However, in practice there are numerous caveats to such code coupling.
For instance, one must ensure that the region near the boundaries of both numerical simulations are in a regime where any non-ideal effects that the two simulations do not share are not dominant.
Additionally, it will be challenging to separate numerical and physical effects when validating such an implementation.
As such, the development of such code coupling constitutes a research and code development line in and of itself.}



\begin{acknowledgments}
This work is supported by the National Solar Observatory, the Office of Naval Research, and the NASA HSR and LWS Programs. 
JEL, MGL, PWS, LAT, and NDK acknowledge support from NASA grants NNH16ZDA001N-LWS “Implementing and Evaluating a Vector-Magnetogram-Driven Magnetohydrodynamic Model of the Magnetic Field in the Low Solar Atmosphere”, NNH21ZDA001N-LWS “The Origin of the Photospheric
Magnetic Field: Mapping Currents in the Chromosphere
and Corona”, and NNH17ZDA001N-LWS “Investigating Magnetic Flux Emergence with Modeling and Observations to Understand the Onset of Major Solar Eruptions”. 
MGL additionally acknowledges support from the Office of Naval Research. 
PWS and JEL additionally acknowledge support from the NASA Internal Science Funding Model (H-ISFM) program “Magnetic Energy Buildup and Explosive Release in the Solar Atmosphere.”
This research has made use of NASA's Astrophysics Data System.
\end{acknowledgments}

\software{
  Correctness of the MHD eigensystem was verified using \textsl{Mathematica} \citep{Mathematica}.
  The primary 3D MHD code, \lare{} v.2.10 \citep{ArbLon01}, is written in Fortran 2003 and parallelized with MPI.
  3D renderings were generated using the \code{VisIt} visualization software \citep{HPV:VisIt}.
  The majority of the numerical analysis was carried out using the \code{NumPy} package \citep{HarMil20} and figures were prepared using the \code{Matplotlib} library \citep{Hun07} in \code{python3}.
}

\appendix

\section{Simple, 1D tests} \label{app:waves}
\change{
As mentioned in Section \ref{sec:validation}, in addition to the hot sphere in an angled magnetic field, we also ran a variety of additional test problems to verify our implementation of non-reflecting boundary conditions into the \lare{} code.
We here present some of the more familiar 1D test cases.
All test problems are presented on a grid with 1024 numerical cells per unit length.
As \lare{} is an inherently 3D code, these 1D tests are run with 8 grid cells in both directions perpendicular to the direction of interest and periodic boundaries on these transverse directions.
The simulations are invariant in these transverse directions.
The domain for the ground truth simulations extends over $-1 < z < 2$ while the simulations with non-reflecting boundary conditions extend over $0 < z < 2$.
The boundary at $z = 2$ in both simulations imposes zero gradient in all MHD properties across the boundary.
We use this same boundary at $z = -1$ in the ground truth simulation.
As there is no variation in the transverse directions, nor is there initially any variation near the boundary in the boundary normal direction, Cancellation and Fixed NRBCs are identical, and both reduce to the simple case of $\Lsiginc=0$.
For simplicity, all simulations presented here with non-reflecting boundary conditions use the Fixed NRBC at $z = 0$.
We have verified that the results for these 1D test cases are identical if we instead use Cancellation NRBCs.
The use of zero-gradient boundaries generates substantial reflections, so all simulations are terminated before reflections from $z_{min}$ in the ground truth simulations, or $z_{max}$ in either simulation reach $z=0$.
All figures zoom in to the region near $z=0$ to emphasize any numerical reflections from the NRBC.
All figures plot an absolute error in each MHD property with a gradient in the $z-direction$.
}

\subsection{Linear Alfv\'en Wave}

\begin{figure*}
    \centering
    \includegraphics[width=\linewidth]{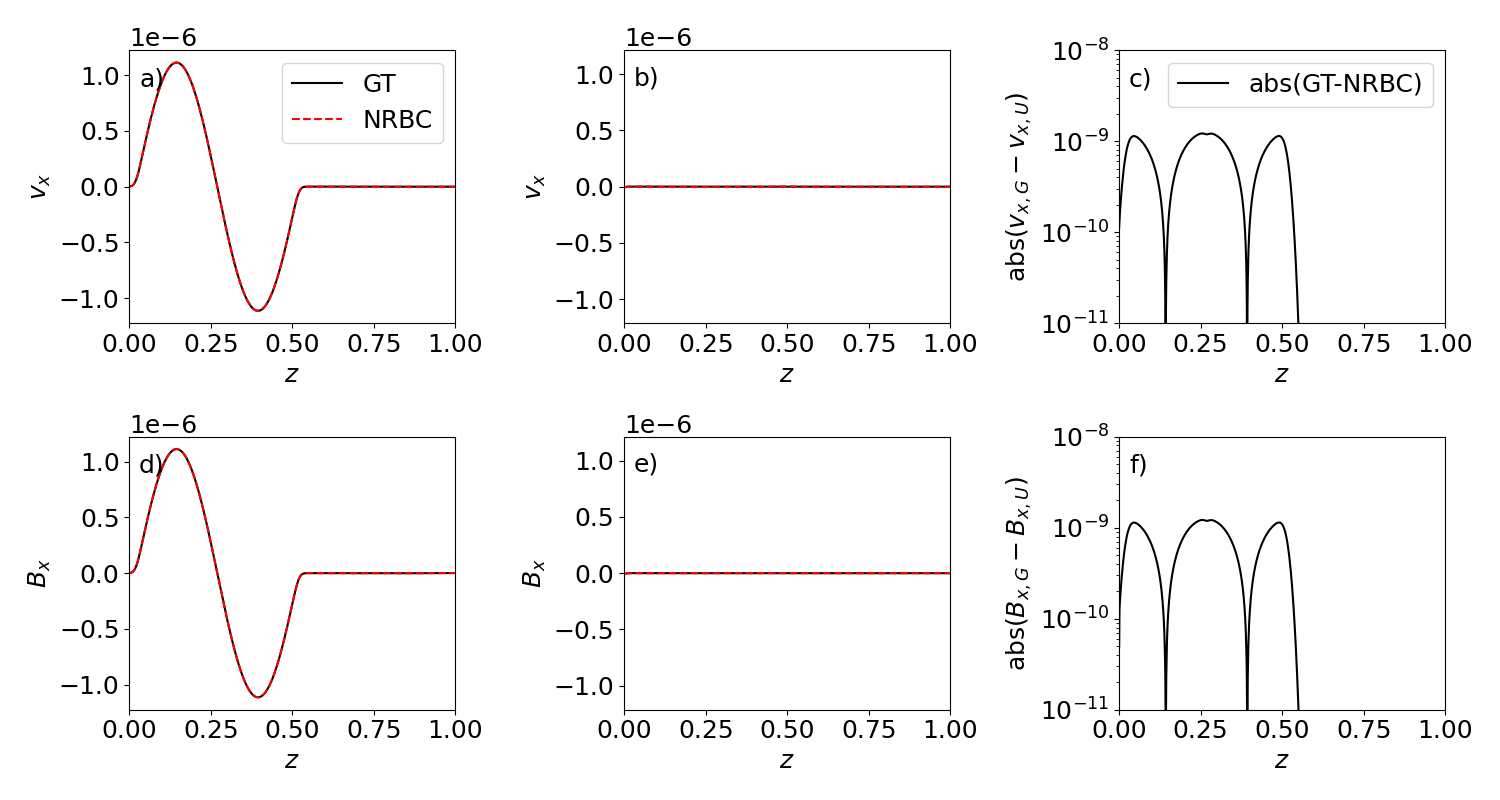}
    \caption{1D test case of an Alfv\'en wave with perturbations in the $x$-direction propagating in the negative $z$-direction. On the left, panel a) shows $v_x$ and panel d) shows $B_x$ for a time immediately before the wave reaches the non-reflecting boundary at $z=0$. Panels b) and e) respectively show $v_x$ and $B_x$ again, but now for a time immediately after the wave has passed the non-reflecting boundary condition location. In both columns the solid, black line represents the ground truth simulation and the dashed, red line represents the simulation with a non-reflecting boundary condition. Panels c) and e) on the right show the absolute value of the difference between the ground truth simulation, denoted $G$, and the simulation with the NRBC, denoted $U$, in $v_x$ and $B_x$ respectively. An animated version of this figure is available as Supplementary Video 6.}
    \label{fig:alfven}
\end{figure*}


\change{
Our first sample test problem is that of a linear Alfv\'en wave propagating toward the non-reflecting boundary.
The background state is $\rho = 1.0$, $\epsilon = 1.0$, $B_z = 1.0$ and the Alfv\'en wave is imposed in the $x$-direction with amplitude $B_x = 1e-6$ to ensure that the Alfv\'en wave will not steepen due to gradients in the magnetic field magnitude and the Alfv\'en speed.
Figure \ref{fig:alfven} shows the snapshot immediately before the full Alfv\'en wave has passed the non-reflecting boundary condition in the left column and the snapshot immediately after in the middle column.
The right column shows the absolute difference between the simulation with a non-reflecting boundary and one one which extends beyond $z=0$, demonstrating that the NRBC passes all but about one part in one thousand of the linear Alfv\'en wave.
It is important to note that even for this extremely simple case where we have a pure wave, we can not uniquely identify this reflection as a failure of the NRBC to do what we have designed it to do.
A careful examination of the terms in Eqn. \ref{eqn:li} shows that the linear ``Alfv\'en" wave actually has non-zero amplitude only in the fast mode characteristics, with both the forward and backward propagating fast mode having non-zero amplitude.
The implementation of a non-reflecting boundary condition explicitly sets the incoming fast mode to have zero amplitude, which in turn by definition incorrectly represents the impact of the portion of the Alfv\'en wave which has left the simulation volume on the portion of the Alfv\'en wave remaining in the volume, such that even this seemingly obvious ``reflection" is the NRBC working as it is designed to, even though it is certainly not how one would reasonably expect such a boundary condition to behave.
Equivalent arguments apply to all the remaining test cases, so we omit the comparable discussion in the remaining subsections of this appendix.
}

\subsection{Sod Shock Tube}

\begin{figure*}[htb!]
    \centering
    \includegraphics[width=\linewidth]{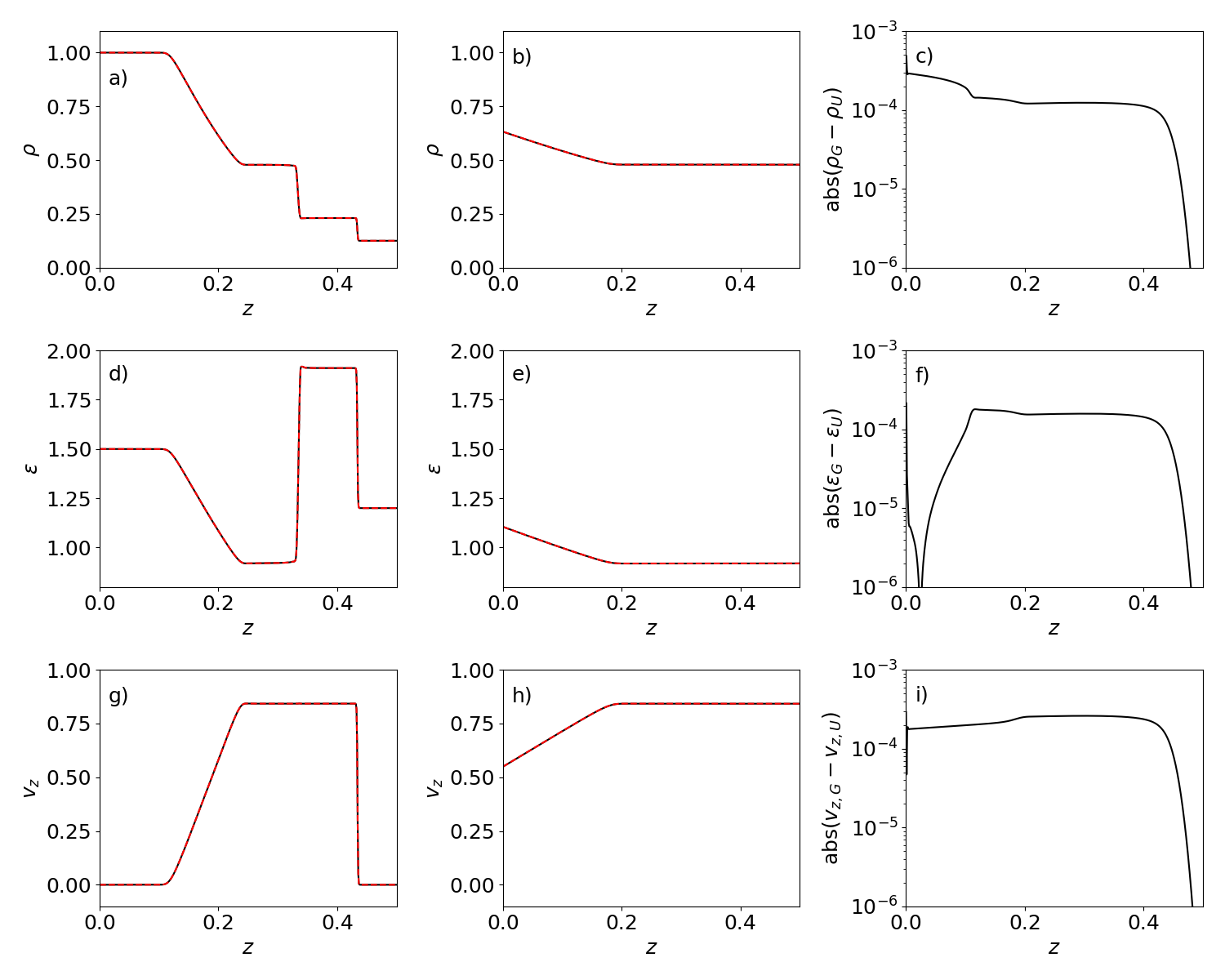}
    \caption{The 1D Sod shock tube problem. The left column shows the conditions after the solution has had time to develop from the initial conditions and the middle column shows a time after the left propagating sonic rarefaction wave has begun interacting with the non-reflecting boundary condition. From top to bottom, the panels show density (a and b), energy density (d and e), and $v_z$ (g and h). In both the left and middle columns the solid, black line represents the ground truth simulation and the dashed, red line represents the simulation with a non-reflecting boundary condition. The right panels show the absolute value of the difference between the ground truth simulation, denoted $G$, and the simulation with the NRBC, denoted $U$, for the same time as in the middle column. An animated version of this figure is available as Supplementary Video 7.}
    \label{fig:sod}
\end{figure*}


\begin{sidewaysfigure*}[htb!]
    \centering
    \includegraphics[width=\linewidth]{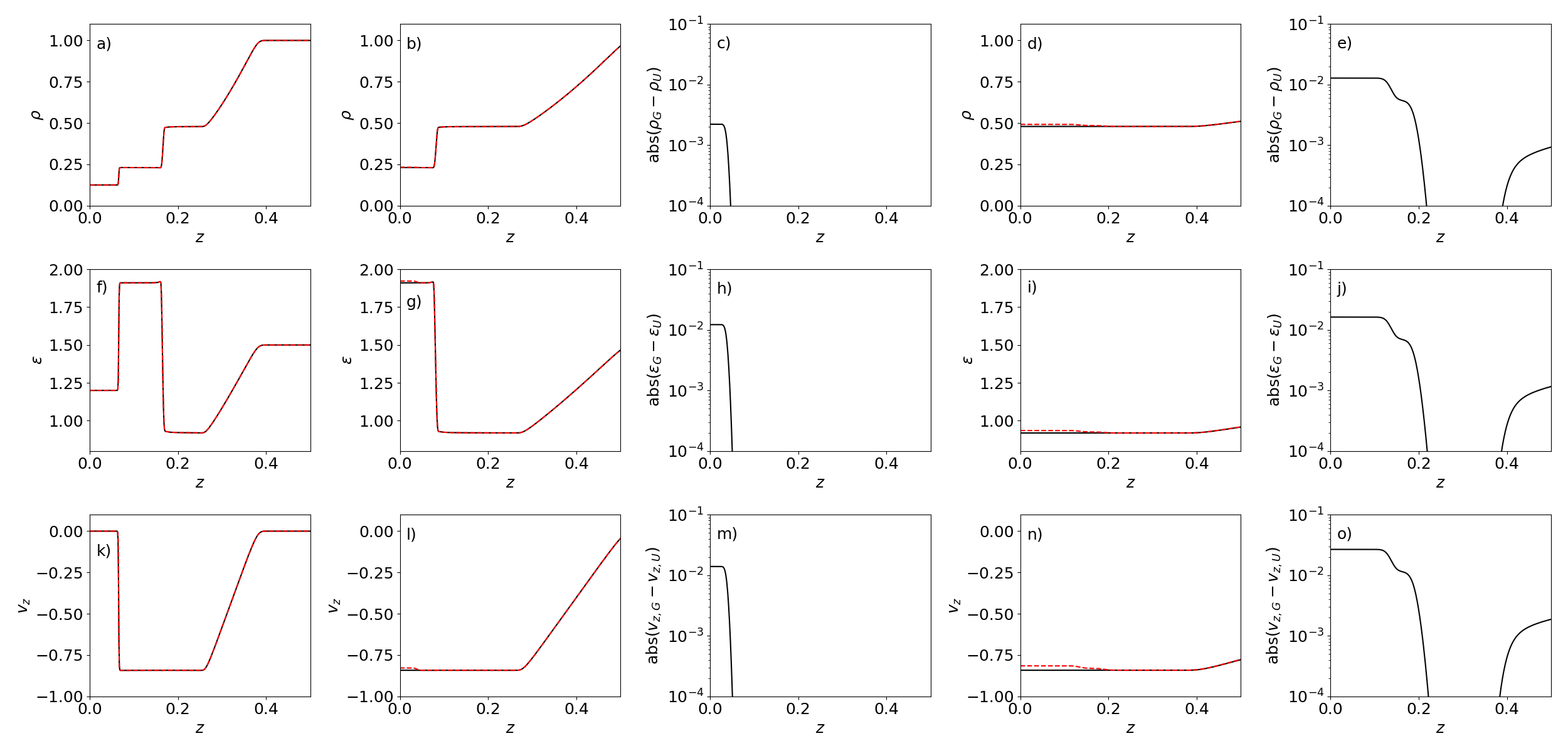}
    \caption{The reversed 1D Sod shock tube. The left column shows a time after the solution has had time to develop from the initial conditions. The second and third columns respectively show the MHD variables and the absolute value of the difference between the ground truth simulation, denoted $G$, and the simulation with the NRBC, denoted $U$ at a time after the sonic shock has passed the location of the NRBC. The fourth and fifth columns show the same quantities for a time after the contact discontinuity has passed the NRBC. From top to bottom, the rows show (a-e) density and differences therein, (f-j) energy density and differences therein, and (k-o) $v_z$ and differences therein. An animated version of this figure is available as Supplementary Video 8.}
    \label{fig:sod_reverse}
\end{sidewaysfigure*}



\change{
Our second simple 1D test case is the Sod shock tube \citep{Sod78}.
For purposes of this discussion, we use the convention that the Sod shock tube has left state $\rho = 1.0$, $\epsilon = 1/(\rho(\gamma-1))$ and right state $\rho = 0.125$, $\epsilon = 0.1/(\rho(\gamma-1))$  and the reversed Sod shock tube flips these states.
Figure \ref{fig:sod} shows the Sod shock tube after the solution has had time to develop from the initial conditions in the left column.
From left to right, each panel in this column depicts the sonic rarefaction wave, a contact discontinuity, and a sonic shock.
Moving forward to a time when the rarefaction has progressed roughly halfway through the NRBC at $z=0$, shown in the middle panel of Figure \ref{fig:sod} with errors plotted in the right column, we see minimal reflection from the NRBC with the departures of the simulation with a NRBC from the ground truth simulations on the level of $0.01-0.1\%$.
The animated version of Figure \ref{fig:sod}, Supplementary video 7, shows that this is the typical, low magnitude of error achieved for the full passage of the rarefaction through the NRBC.
}

\change{
Examining the reversed Sod shock instead directs the sonic shock and contact discontinuity toward the NRBC, as can be seen from Figure \ref{fig:sod_reverse}.
These are significantly more difficult for the boundary conditions to handle as they represent departures from purely linear MHD.
Nevertheless, the passage of the shock and contact discontinuity both result in order $1\%$ departures between the ground truth and NRBC simulations, as highlighted by the remaining panels of Figure \ref{fig:sod_reverse}, with columns two and three focusing on a time just after the shock passes the boundary and columns four and five focusing on a time after the passage of the contact discontinuity.
}

\subsection{Brio-Wu Shock Tube}

\begin{sidewaysfigure*}
    \centering
    \includegraphics[width=0.8\linewidth]{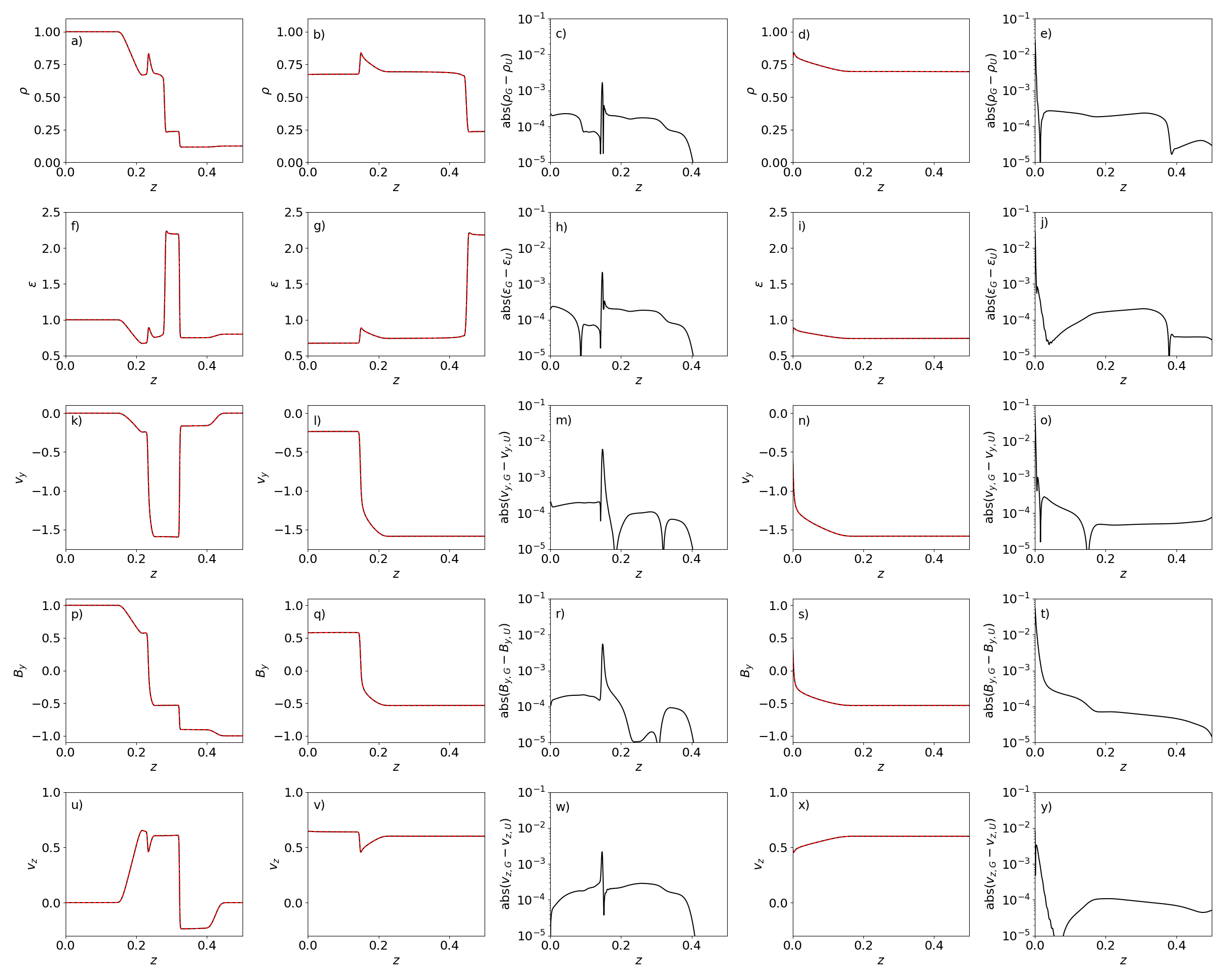}
    \caption{The 1D Brio-Wu shock tube. The left column shows the MHD variables at a time after the solution has developed from the initial conditions. Columns two and four show times after the left propagating fast rarefaction has passed the location of the NRBC and after the front of the left propagating compound wave has passed the location of the NRBC. Columns three and five show the absolute value of the difference between the ground truth simulation, denoted $G$, and the simulation with the NRBC, denoted $U$, at the same times. From top to bottom, the rows show the variable and the differences in (a-e) density, (f-j) energy density, (k-o) $v_y$, (p-t) $B_y$, and (u-y) $v_z$. An animated version of this figure is available as Supplementary Video 9.}
    \label{fig:brioWu}
\end{sidewaysfigure*}



\begin{sidewaysfigure*}
    \centering
    \includegraphics[width=0.8\linewidth]{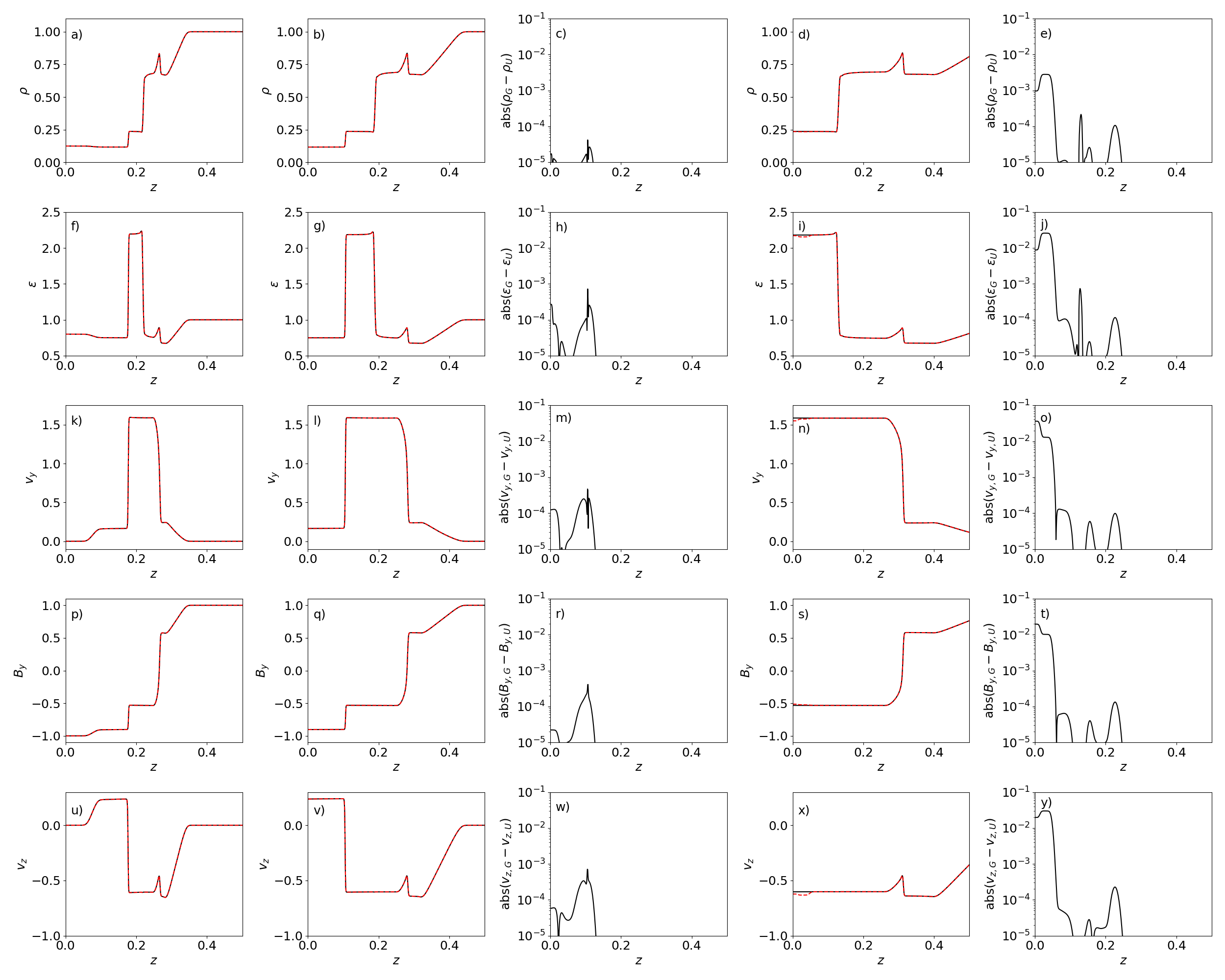}
    \caption{The reversed 1D Brio-Wu shock tube. The left column shows the MHD variables at a time after the solution has developed from the initial conditions. Columns two and four show times after the left propagating fast rarefaction has passed the location of the NRBC and after the slow shock has passed the location of the NRBC. Columns three and five show the absolute value of the difference between the ground truth simulation, denoted $G$, and the simulation with the NRBC, denoted $U$, at the same times. From top to bottom, the rows show the variable and the differences in (a-e) density, (f-j) energy density, (k-o) $v_y$, (p-t) $B_y$, and (u-y) $v_z$. An animated version of this figure and Figure \ref{fig:brioWu_reverse_2} is available as Supplementary Video 10.}
    \label{fig:brioWu_reverse_1}
\end{sidewaysfigure*}

\begin{figure*}
    \centering
    \includegraphics[width=0.85\linewidth]{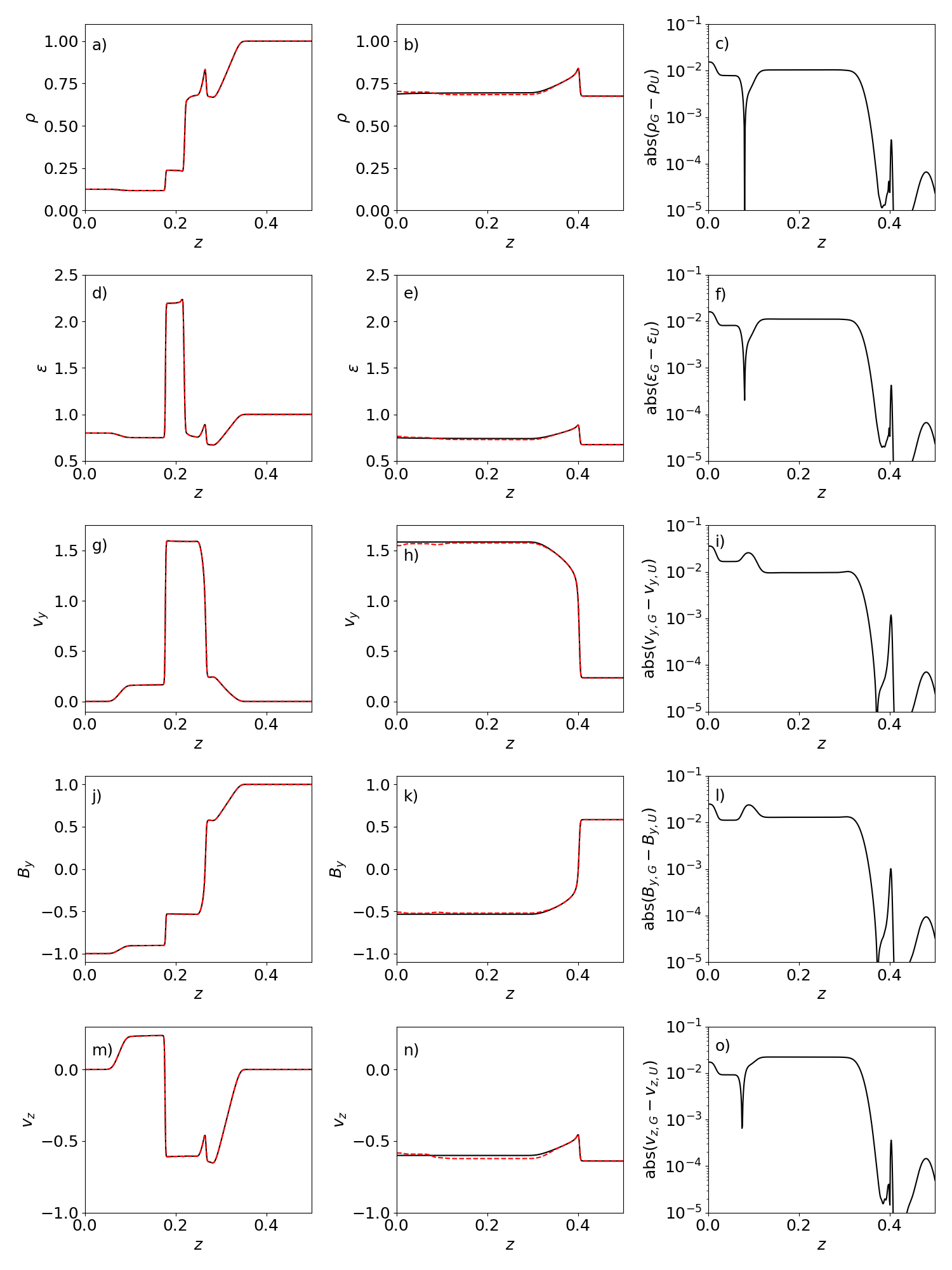}
    \caption{As Figure \ref{fig:brioWu_reverse_1}. The left column still shows the time after the solution has developed from the initial conditions to facilitate comparison. Columns two and three now show a time after the contact discontinuity has passed the location of the NRBC.}
    \label{fig:brioWu_reverse_2}
\end{figure*}




\change{As the final 1D test problem we highlight in this appendix, we use the Brio-Wu shock tube \citep{BriWu88}.
As we did with the Sod shock tube, we test the NRBC against cases with the Brio-Wu shock tube oriented both ways.
We first examine the case with the left state given by $\rho = 1.0$, $\epsilon = 1.0/\rho$, $B_y =1.0$, and $B_z = 0.75$ and the right state by $\rho = 0.125$, $\epsilon = 0.1/\rho$, $B_y =-1.0$, and $B_z = 0.75$.
Note the omission of the factor $(\gamma-1)$ in the definition of $\epsilon$, which arises from the Brio-Wu shock tube defining $\gamma \equiv 2.0$, such that $(\gamma - 1) = 1.0$.
Allowing this to evolve in the simulation generates, as shown from left to right in the left column of Figure \ref{fig:brioWu}, a fast rarefaction, a compound wave, a contact discontinuity, a slow shock, and a fast rarefaction.
The first two are propagating left and the last three are propagating right.
We highlight the differences between ground truth and NRBC simulations introduced by the passage of the fast rarefaction in the second and third columns of Figure \ref{fig:brioWu} and the front of the compound wave in the fourth and fifth columns of Figure \ref{fig:brioWu}.
This shows very comparable behaviors to the Sod shock tube with the linear rarefaction introducing discrepancies at the level of $<0.1\%$ and the sharp leading edge of the compound wave introducing discrepancies at the level of a few percent.
An animated version of Figure \ref{fig:brioWu} is available as Supplementary Figure 9.
}

\change{
Finally, we present the reversed Brio-Wu shock tube.
To fully compare the ground truth solution and the simulation with a non-reflecting boundary condition, the results of the reversed Brio-Wu shock tube are split over Figures \ref{fig:brioWu_reverse_1} and \ref{fig:brioWu_reverse_2}.
For completeness, we show the reversed solution in the left panels of both Figure \ref{fig:brioWu_reverse_1} and \ref{fig:brioWu_reverse_2}, with a fast rarefaction, slow shock, contact discontinuity, compound wave, and fast rarefaction from left to right.
Here the amplitude of the fast rarefaction is itself minimal, such that it introduces discrepancies in the solutions at the level of $\lesssim 0.01\%$, as shown in columns two and three of Figure \ref{fig:brioWu_reverse_1}.
Meanwhile, the discontinuities passing the boundary again introduce reflections at the level of a few percent, as shown for the slow shock in columns four and five of Figure \ref{fig:brioWu_reverse_1} and for the contact discontinuity in the second and third columns of Figure \ref{fig:brioWu_reverse_2}.
An animated version of Figures \ref{fig:brioWu_reverse_1} and \ref{fig:brioWu_reverse_2} is available as Supplementary Figure 10.
}

\bibliography{biblio}{}
\bibliographystyle{aasjournal}



\end{document}